%% file: M8.tex
\documentclass[preprint,prd,showpacs,preprintnumbers,nofootinbib,fleqn,floatfix]{revtex4}

\usepackage{amsmath}
\usepackage{amssymb}
\usepackage{graphicx}
\usepackage{dcolumn}
\usepackage{enumerate}

\begin{document}


\title{Revised value of the eighth-order QED contribution to the 
anomalous magnetic moment of the electron}

\author{T. Aoyama}
\affiliation{Institute of Particle and Nuclear Studies,
High Energy Accelerator Research Organization (KEK), Tsukuba, Ibaraki 305-0801, Japan}

\author{M. Hayakawa}
\affiliation{Department of Physics, Nagoya University, Nagoya, 
Aichi 464-8602, Japan}

\author{T. Kinoshita}
\affiliation{Newman Laboratory for Elementary-Particle Physics, 
Cornell University, Ithaca, New York 14853, U.S.A.}

\author{M. Nio}
\affiliation{Theoretical Physics Laboratory,
Nishina Center, RIKEN, Wako, Saitama 351-0198, Japan }




\begin{abstract}
We have carried out a new 
evaluation of the eighth-order contribution to the electron
$g\!-\!2$ using FORTRAN codes generated by an automatic code 
generator {\sc gencode}{\it N}.
Comparison of the ``new"  result with the ``old" one
has revealed an inconsistency in the treatment of the infrared
divergences in the latter.
With this error corrected we now have two independent determinations
of the eighth-order term.
This leads to the revised value 
$1~159~652~182.79~(7.71) \times 10^{-12}$ of the electron $g\! - \! 2$, 
where the uncertainty comes mostly from that of the
best non-QED value of the fine structure constant $\alpha$.
The new value of $\alpha$ derived from the revised theory and 
the latest experiment is
$\alpha^{-1} = 137.035~999~084~(51)~[0.37~\text{ppb}]~$, 
which is about 4.7 ppb smaller than the previous $\alpha^{-1}$.
\end{abstract}


\newcommand{\FourthFigures}[1]{
\begin{figure}[tb]
\includegraphics*[width=14cm]{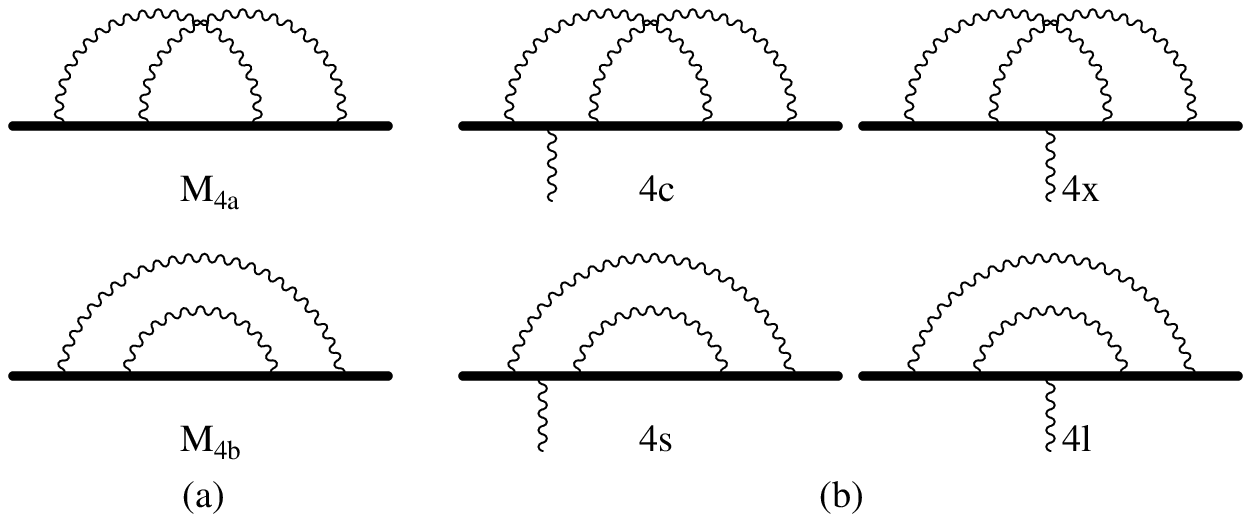}
\caption{Fourth-order $q$-type diagrams. (a)  
Self-energy-like diagrams $M_{4a}$ and $M_{4b}$. ~~~(b) Vertex diagrams $4c, 4x, 4s,$
and $4l$.
Their contributions to the magnetic moment 
are related to $M_{4a}=2 M_{4c}+M_{4x}$, and  $M_{4b}=2 M_{4s}+M_{4l}$.
}
\label{4th-figures}
\end{figure}
}
\newcommand{\SixthFigures}[1]{
\begin{figure}[tb]
\includegraphics*[width=15cm]{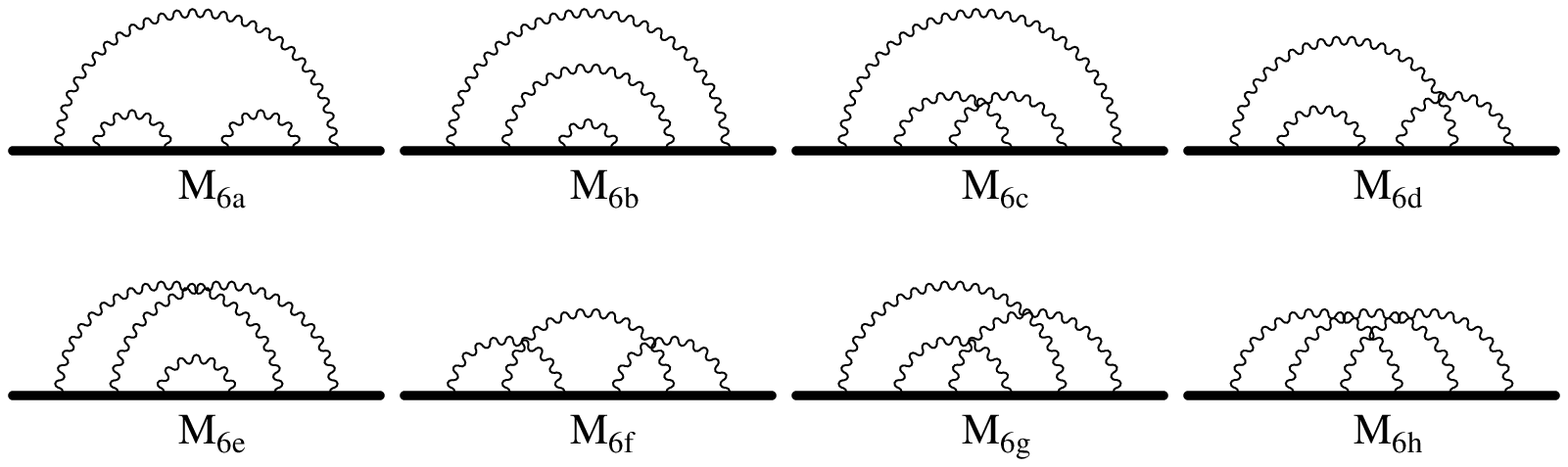}
\caption{Sixth-order $q$-type self-energy-like diagrams $M_{6x},~x=a,\ldots,h$.
The time reversal diagrams of $M_{6d}$ and $M_{6g}$ are not shown here.
The fermion lines of a diagram is named  1 to 5 from left to right.
The vertex diagram  obtained by inserting an external photon vertex into the
fermion line $i$ of the self-energy diagram $6x$ is named $6x i$.
}
\label{6th-figures}
\end{figure}
}
\newcommand{\AllFiguresOfEighthOrderSetV}[1]{
\begin{figure}[tb]
\includegraphics*[width=15cm]{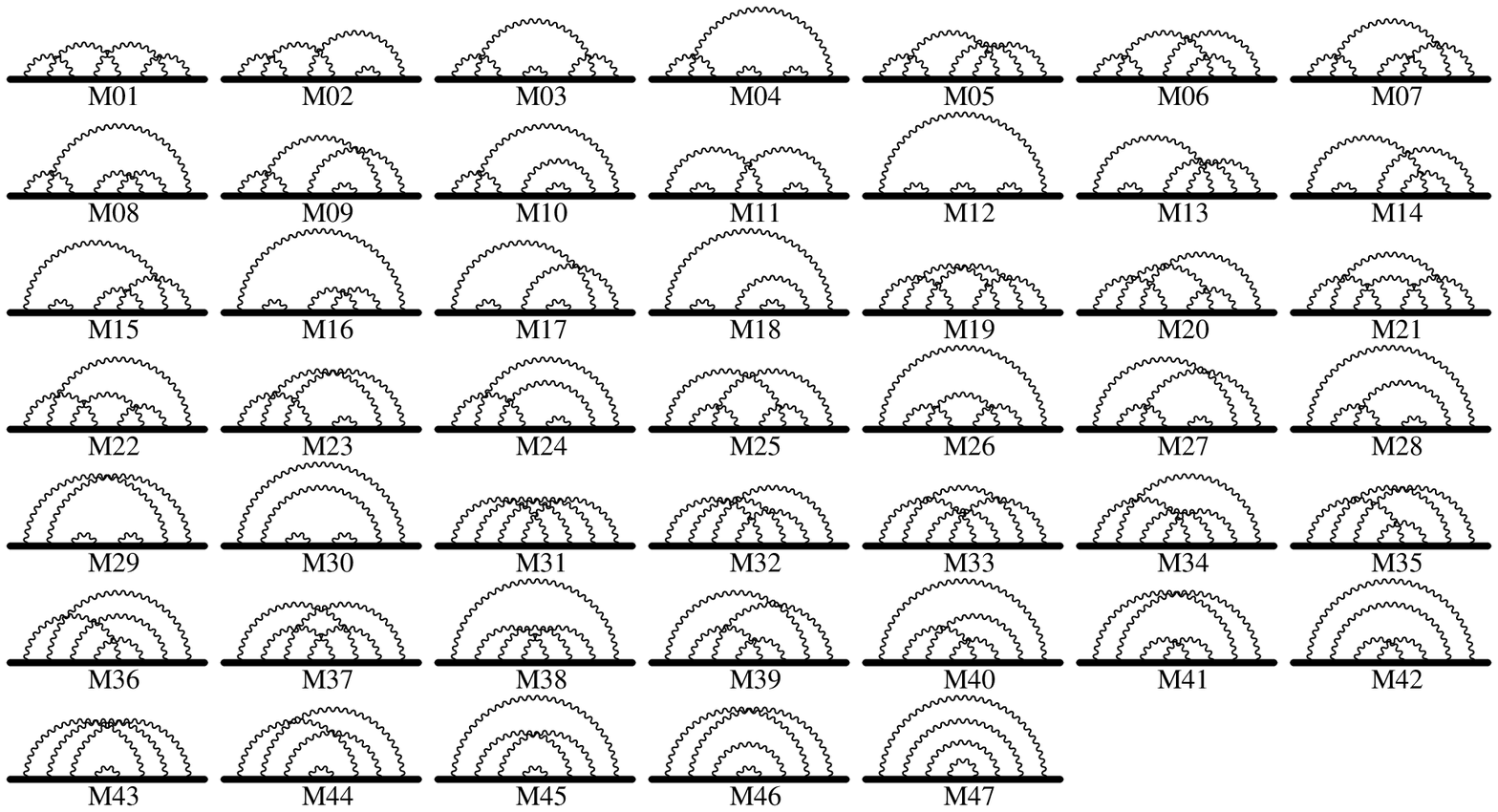}
\caption{Eighth-order Group V diagrams represented by 47 self-energy-like diagrams 
$M_{01}$--$M_{47}$.} 
\label{fig:EighthOrderSetV}
\end{figure}
}
\newcommand{\MsixteenMeighteenFigure}[1]{
\begin{figure}[tb]
\includegraphics*[width=4.25in]{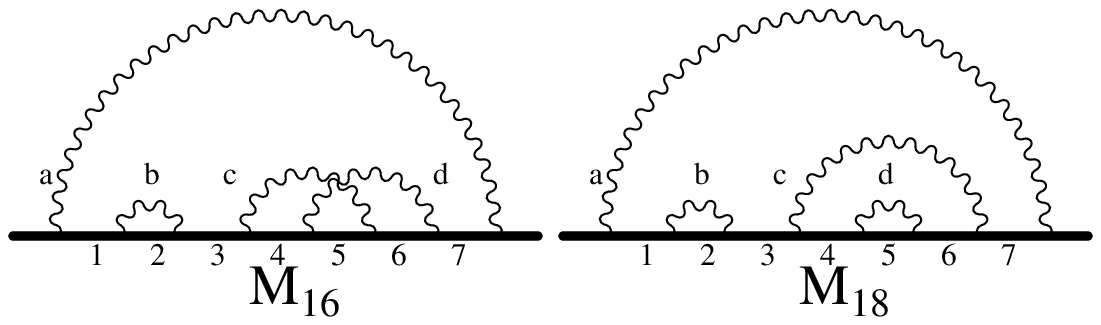}
\caption{Self-energy-like diagrams of $M_{16}$ and $M_{18}$. Feynman parameters
assigned to the electron lines are $z_1 \sim z_7$ and those to the photon lines
are $z_a \sim z_d$. 
} 
\label{fig:M16andM18}
\end{figure}
}
\newcommand{\MfourteenFigure}[1]{
\begin{figure}[tb]
\includegraphics*[width=5.3in]{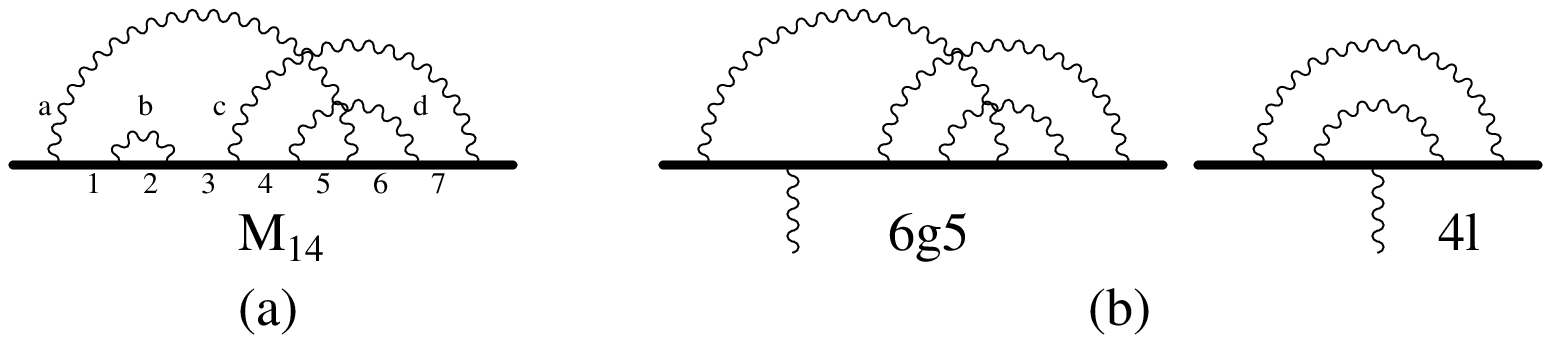}
\caption{(a) Self-energy-like diagram of $M_{14}$. ~~(b)Vertex diagrams 
$6g5$ and $4l$ from which the vertex renormalization constants $L_{6g5}$ and
$L_{4l}$, respectively,  are derived.} 
\label{fig:M14andL6g5}
\end{figure}
}
\newcommand{\ComparFirstTable}{
\begin{table*}
\caption{
 Comparison of the numerical calculation of $M_{01}$--$M_{15}$ of 
the eighth-order Group V diagrams. 
 The second column shows the analytic expression 
for $\Delta M_i^{\rm old} - \Delta M_i^{\rm new}$ 
for each diagram $M_i$ in terms of lower-order finite quantities multiplied by the
multiplicity. 
 The value $A$ in the third column is obtained 
by substituting the values of lower-order renormalization constants, 
such as $\Delta M_{4a}, \Delta L_{4s}$ and $\Delta M_{4a(1^\star)}$, 
for the corresponding expression in the second column.  See Appendices \ref{appendixM} and \ref{appendixRconst}
for the detail.
 In contrast, the value 
for $\Delta M_i^{\rm old} - \Delta M_i^{\rm new}$ in the fourth column, 
denoted by value $B$, is obtained by 
taking the direct difference between  
the value of $\Delta M_i^{\rm old}$ quoted from Ref.~\cite{Kinoshita:2005zr}, 
and the one of $\Delta M_i^{\rm new}$ 
calculated via {\sc gencode}{\it N} in the ``new"  IR subtraction procedure 
\cite{Aoyama:2005kf,Aoyama:2007IR}. 
 The fifth column lists up the differences $A-B$. 
 If the whole calculation is done correctly,
$A-B$ must vanish within the numerical uncertainty.
 In evaluating $\Delta M^{\rm new}$ the double precision is used for the 
diagrams without a self-energy subdiagram, while 
the quadruple precision is used for the remainder.
\label{Table:M01M15}
}
\newcolumntype{.}{D{.}{.}{8}}
\begin{ruledtabular}
\begin{tabular}{cl...}
\multicolumn{1}{c}{{\rm Diagram}} &
\multicolumn{1}{c}{{\rm difference}}&
\multicolumn{1}{c}{{\rm value}~$A$} &
\multicolumn{1}{c}{{\rm value}~$B$} &
\multicolumn{1}{c}{{$A-B$}} \\
\hline
$M_{01}$  &  0              
        &  0              &  -0.0129( 47)  &    0.0129( 47)  \\
$M_{02}$  & $2\Delta L_{6f1} M_2$                 
        &   -0.0063(  2)  &   0.0060(110)  &   -0.0124(110)  \\
$M_{03}$  & $\Delta L_{6f3} M_2$  
        &   -0.1133(  1)  &  -0.1055(100)  &   -0.0078(100)\\
$M_{04}$  & $2(\Delta L_{6d1} + \Delta L_{6d3} )M_2 $                 
        &    0.3350(  2)  &   0.3408(175)  &   -0.0058(175)  \\
$M_{05}$  &  0              
        &  0              &   0.0020( 28)  &   -0.0020( 28)  \\
$M_{06}$  &  0                      
        &  0              &  -0.0223( 61)  &    0.0223( 61)  \\
$M_{07}$  &  0                      
        &  0              &  -0.0102( 40)  &    0.0102( 40)  \\
$M_{08}$  & $2 (\Delta \delta m_{4a}\Delta M_{4a(1^\star)}+\Delta L_{4c} 
   \Delta M_{4a}) $            
        &   -2.1809(  7)  &  -2.1790(121)  &   -0.0019(121)  \\
$M_{09}$  & $2\Delta L_{6f2} M_2 $                
        &    0.0806(  1)  &   0.0894(109)  &   -0.0088(109)  \\
$M_{10}$  & $2(\Delta \delta m_{4b}\Delta M_{4a(1^\star)}+\Delta L_{6d2} M_2+  
                \Delta L_{4c}\Delta M_{4b}) $                   
        &   15.8898( 49)  &  15.8795(147)  &    0.0103(155)  \\
$M_{11}$  & $2\Delta L_{6d5} M_2 $              
        &    0.6949(  2)  &   0.6827(112)  &    0.0122(112)  \\
$M_{12}$  & $( 2\Delta L_{6a1} + \Delta L_{6a3}) M_2 $                 
        &    1.2842(  0)  &   1.2875( 74)  &   -0.0034( 74)  \\
$M_{13}$  & $2\Delta L_{6h1} M_2 $                
        &   -0.4211(  2)  &  -0.4238( 48)  &    0.0027( 48)  \\
$M_{14}$  & $2\Delta L_{6g5} M_2 $                
        &    0.0892(  2)  &   0.0960( 95)  &   -0.0068( 95)  \\
$M_{15}$  & $2\Delta L_{6g1} M_2 $                
        &    0.0883(  2)  &   0.0893( 71)  &   -0.0009( 71)\\
\end{tabular}
\end{ruledtabular}
\end{table*}
}
\newcommand{\ComparSecondTable}{
\begin{table*}
\caption{
Comparison of the numerical calculations of $M_{16}$--$M_{30}$  of the 
eighth-order Group V diagrams. 
\label{Table:M16M30}
}
\newcolumntype{.}{D{.}{.}{8}}
\begin{ruledtabular}
\begin{tabular}{cl...}
\multicolumn{1}{c}{{\rm Diagram}} &
\multicolumn{1}{c}{{\rm difference}}&
\multicolumn{1}{c}{{\rm value}~$A$} &
\multicolumn{1}{c}{{\rm value}~$B$} &
\multicolumn{1}{c}{{$A-B$}} \\
\hline
$M_{16}$  & $2(\Delta \delta m_{4a}\Delta M_{4b(1^\star)} 
       +\Delta L_{6c1}M_2 + \Delta L_{4s}\Delta M_{4a})$                  
        &   -2.6042(  6)  &  -2.6316(235)  &    0.0274(235)  \\
$M_{17}$  & $2(\Delta L_{6e1}  + \Delta L_{6d4}) M_2$                  
        &   -2.1201(  2)  &  -2.1010(189)  &   -0.0173(189)  \\
$M_{18}$  & $2 \{ \Delta \delta m_{4b}\Delta M_{4b(1^\star)} 
      +\Delta L_{4s}\Delta M_{4b} $
          &   16.9686( 39)  &  17.1897(206)  &   -0.2207(210)  \\
        &   $+(\Delta L_{6b1} +\Delta L_{6a2}) M_2 \}$
          &                 &                &                 \\
$M_{19}$  &  0                      
        &    0            &   0.0002(  3)  &   -0.0002(  3)  \\
$M_{20}$  &  0                      
        &    0            &   0.0010( 17)  &   -0.0010( 17)  \\
$M_{21}$  &  0                      
        &    0            &   0.0003(  3)  &   -0.0003(  3)  \\
$M_{22}$  &  0                      
        &    0            &  -0.0090( 25)  &    0.0090( 25)  \\
$M_{23}$  & $2\Delta L_{6h2} M_2$                 
        &    0.0501(  2)  &   0.0438( 59)  &    0.0064( 59)  \\
$M_{24}$  & $2\Delta L_{6g2} M_2$                 
        &    0.0789(  2)  &   0.0945( 61)  &   -0.0155( 61)  \\ 

$M_{25}$  &  0                      
        &    0            &  -0.0031( 20)  &    0.0031( 20)  \\
$M_{26}$  & $\Delta \delta m_{6f}( M_{2^\star} - M_{2^\star}[I] ) $             
        &    2.5119(  3)  &   2.5369( 95)  &   -0.0250( 95)  \\
$M_{27}$  & $2\Delta L_{6g4} M_2                 $
        &   -0.0630(  1)  &  -0.0459( 90)  &   -0.0171( 90)  \\
$M_{28}$  & $2\{\Delta \delta m_{6d}( M_{2^\star} - M_{2^\star}[I] ) 
                          +\Delta L_{6c2} M_2 \} $              
        &   -7.5332(  5)  &  -7.5307(153)  &   -0.0025(153)  \\
$M_{29}$  & $2\Delta L_{6e2} M_2                $
        &   -0.2857(  2)  &  -0.2809(109)  &   -0.0048(109)  \\
$M_{30}$  & $\Delta \delta m_{6a}( M_{2^\star} - M_{2^\star}[I] ) 
                                 + 2\Delta L_{6b2} M_2   $          
        &    0.2763(  6)  &   0.2675(153)  &    0.0088(153) \\
\end{tabular}
\end{ruledtabular}
\end{table*}
}
\newcommand{\ComparThirdTable}{
\begin{table*}
\caption{
Comparison of the numerical calculations of $M_{31}$--$M_{47}$  of the 
eighth-order Group V diagrams. 
\label{Table:M31M47}
}
\newcolumntype{.}{D{.}{.}{8}}
\begin{ruledtabular}
\begin{tabular}{cl...}
\multicolumn{1}{c}{{\rm Diagram}} &
\multicolumn{1}{c}{{\rm difference} }&
\multicolumn{1}{c}{{\rm value}~$A$} &
\multicolumn{1}{c}{{\rm value}~$B$} &
\multicolumn{1}{c}{$A-B$} \\
\hline 
$M_{31}$  &  0                      
        &    0            &   0.0007(  5)  &   -0.0007(  5)  \\
$M_{32}$  &  0                      
        &    0            &  -0.0024( 10)  &    0.0024( 10)  \\
$M_{33}$  &  0                      
        &    0            &   0.0001(  3)  &   -0.0001(  3)  \\
$M_{34}$  &  0                      
        &    0            &  -0.0010( 13)  &    0.0010( 13)  \\
$M_{35}$  &  0                      
        &    0            &   0.0001( 13)  &   -0.0001( 13)  \\
$M_{36}$  &  0                      
        &    0            &  -0.0027( 22)  &    0.0027( 22)  \\
$M_{37}$  &  0                      
        &    0            &   0.0004(  5)  &   -0.0004(  5)  \\
$M_{38}$  & $\Delta \delta m_{6h} ( M_{2^\star} - M_{2^\star}[I]$ )                       
        &   -0.9088(  3)  &  -0.9112( 40)  &    0.0024( 40)  \\
$M_{39}$  &  0                      
        &    0            &  -0.0031( 18)  &    0.0031( 18)  \\
$M_{40}$  & $2\Delta \delta m_{6g} ( M_{2^\star} - M_{2^\star}[I]$ )                       
        &    3.8281(  3)  &   3.8326( 71)  &   -0.0045( 71)  \\
$M_{41}$  & $\Delta \delta m_{4a} ( \Delta M_{4a(2^\star)} )
                          +    \Delta L_{4x}\Delta M_{4a}     $           
        &    0.9809(  3)  &   0.9713( 83)  &    0.0096( 83)  \\
$M_{42}$  &  $ \Delta \delta m_{6c}( M_{2^\star} - M_{2^\star}[I] ) 
                                + \Delta L_{4l}\Delta M_{4a} $  
        &   -7.0218(  4)  &  -7.0202(114)  &   -0.0016(114)  \\
     &       $+ \Delta \delta m_{4a}      
                      \{\Delta M_{4b(2^\star)}
             -\Delta \delta m_{2^\star}(M_{2^\star}-M_{2^\star}[I])\}  $ 
                                                        \\       
$M_{43}$  & $\Delta L_{6h3} M_2 $                 
        &    0.4724(  1)  &   0.4703( 42)  &    0.0022( 42)  \\
$M_{44}$  & $2\Delta L_{6g3} M_2$                 
        &   -0.0748(  1)  &  -0.0499( 69)  &   -0.0250( 69)  \\
$M_{45}$  & $\Delta \delta m_{6e} ( M_{2^\star} - M_{2^\star}[I] )  
+ \Delta L_{6c3} M_2 $ 
        &   -0.0523(  3)  &  -0.0498( 90)  &   -0.0025( 90)  \\
$M_{46}$  &
         $\Delta\delta m_{4b}\Delta M_{4a(2^\star)}+\Delta L_{6e3}M_2 
+ \Delta L_{4x}\Delta M_{4b}  $                 
        &   -7.9339( 22)  &  -7.9232( 86)  &   -0.0107(89)  \\
$M_{47}$ & $\Delta \delta m_{6b} ( M_{2^\star} - M_{2^\star}[I] )+ \Delta L_{6b3} M_2 + 
       \Delta L_{4l}\Delta M_{4b} $
        &   10.5872( 15)  &  10.5864(102)  &    0.0008(103)  \\ 
      & $+ \Delta\delta m_{4b} \{ \Delta M_{4b(2^\star)}-\Delta\delta m_{2^\star}
          (M_{2^\star} - M_{2^\star}[I] ) \} $ & & &  
\end{tabular}
\end{ruledtabular}
\end{table*}
}
\newcommand{\RconstSixthTable}{
\begin{table}
\caption{Finite renormalization constants used in Table \ref{Table:M01M15}, \ref{Table:M16M30}, and \ref{Table:M31M47}.
Sixth-order vertex renormalization constants are shown in this table. Their validity is checked
by comparing the sum $X_{LBD}\equiv \sum_{i=1}^5 \Delta L_{6xi} + \frac{1}{2} \Delta B_{6x} + 2 \Delta \delta m_{6x},\quad
x=a,\cdots h$
to the previous $X_{LBD}$ values listed in Ref.~\cite{Kinoshita:2005zr}.
\label{Table:L6}
}
\newcolumntype{.}{D{.}{.}{10}}
\begin{ruledtabular}
\begin{tabular}{l.@{\hskip 2em}l.@{\hskip 2em}l.}
      $\Delta L_{6a1}$&    0.539604
          (~45)&
      $\Delta L_{6a2}$&   -0.167211
          (~81)&
      $\Delta L_{6a3}$&    1.489159
          (~98)  \\
      $\Delta L_{6b1}$&   -1.479745
          (109)&
      $\Delta L_{6b2}$&    0.582944
          (106)&
      $\Delta L_{6b3}$&   -0.016344
          (~73)  \\
      $\Delta L_{6c1}$&   -0.219365
          (~98)&
      $\Delta L_{6c2}$&    0.071504
          (~87)&
      $\Delta L_{6c3}$&   -0.552261
          (107)  \\
      $\Delta L_{6d1}$&    0.834949
          (~96)&
      $\Delta L_{6d2}$&   -0.090796
          (~92)&
      $\Delta L_{6d3}$&   -0.499995
          (~97)  \\
      $\Delta L_{6d4}$&   -1.378190
          (109)&
      $\Delta L_{6d5}$&    0.694916
          (101) & &  \\
      $\Delta L_{6e1}$&   -0.741904
          (144)&
      $\Delta L_{6e2}$&   -0.285670
          (108)&
      $\Delta L_{6e3}$&   -0.141787
          (122)  \\
      $\Delta L_{6f1}$&   -0.006322
          (114)&
      $\Delta L_{6f2}$&    0.080648
          (~97)&
      $\Delta L_{6f3}$&   -0.226693
          (106)   \\
      $\Delta L_{6g1}$&    0.088204
          (~70)&
      $\Delta L_{6g2}$&    0.078922
          (103)&
      $\Delta L_{6g3}$&   -0.074834
          (~92)  \\
      $\Delta L_{6g4}$&   -0.062995
          (~85)&
      $\Delta L_{6g5}$&    0.089213
          (~69) & &  \\
      $\Delta L_{6h1}$&   -0.421132
          (108)&
      $\Delta L_{6h2}$&    0.050140
          (108)&
      $\Delta L_{6h3}$&    0.944887
          (116)   \\
      $\Delta \delta m_{6a}$  & -0.15331(26) &
      $\Delta \delta m_{6b}$  &  1.83795(19) &
      $\Delta \delta m_{6c}$  & -3.05047(17)  \\
      $\Delta \delta m_{6d}$  & -1.90117(11) &
      $\Delta \delta m_{6e}$  &  0.11193(13) &
      $\Delta \delta m_{6f}$  &  1.25594(10)  \\
      $\Delta \delta m_{6g}$  &  0.95702(6) &
      $\Delta \delta m_{6h}$  & -0.45441(5) & &  
\end{tabular}
\end{ruledtabular}
\end{table}
}
\newcommand{\RconstFourthTable}
{
\begin{table}
\caption{Finite renormalization constants used in Table \ref{Table:M01M15}, \ref{Table:M16M30} and \ref{Table:M31M47}.
Fourth-order and second-order quantities are given here.
\label{Table:L4}
}
\newcolumntype{.}{D{.}{.}{8}}
\begin{ruledtabular}
\begin{tabular}{l.@{\hskip 4em}l.}
$\Delta L_{4c}$ &     0.003387(16) &
$\Delta L_{4x}$ &    -0.481834(54)  \\
$\Delta L_{4s}$ &     0.407633(20) &
$\Delta L_{4l}$ &     0.124796(67)  \\
$\Delta B_{4a}$ &    -0.039811(15)  &
$\Delta B_{4b}$ &    -0.397283(15)    \\
$\Delta \delta m_{4a}$ &  -0.301485(61)&
$\Delta \delta m_{4b}$ &   2.20777 (44)  \\  
$\Delta M_{4a}$ &     0.218359(39) &
$\Delta M_{4b}$ &    -0.187526(39)       \\
$\Delta M_{4a(1^\star)}$ &       3.6192(31) &
$\Delta M_{4a(2^\star)}$ &      -3.6003(19)  \\
$\Delta M_{4b(1^\star)}$ &       4.2486(15) &
$\Delta M_{4b(2^\star)}$ &       1.6432(15)  \\
$\Delta M_2$         &        0.5  &
$\Delta M_{2^\star}$     &        1     \\
$\Delta M_{2^\star}[I]$  &       -1    &
$\Delta \delta m_{2^\star}$ &     -0.75  \\
$\Delta B_2$         &        0.75   &
$\Delta B_{2^\star}[I]$   &       -0.5     \\
$\Delta L_{2^\star}$        &       -0.75      &
$\Delta B_{2^\star}$         &      1.5 
\end{tabular}
\end{ruledtabular}
\end{table}
}

\maketitle

\section{Introduction and Summary}

The anomalous magnetic moment of the electron 
has played a central role in testing the validity of QED 
\cite{Kusch:1947,Schwinger:1948iu}. 
The test became very stringent when the precision of measurement of
the electron and positron was improved 
by three orders of magnitude over the best earlier result \cite{Rich:1972}
by the University of Washington group in 
the Penning trap experiment \cite{VanDyck:1987ay} 
\begin{align}
  a_{e^-} &= 1~159~652~188.4~(4.3) \times 10^{-12}
  \quad
  [3.7 \text{ppb}]
  \,, 
\nonumber \\
  a_{e^+} &= 1~159~652~187.9~(4.3) \times 10^{-12}
  \quad
  [3.7 \text{ppb}]
  \,,
\label{aeUW87value}
\end{align}
where $a_e \equiv (g\! - \! 2)/2$ 
and $g$ is the $g$-factor of electron. 
 The main source of the remaining uncertainty in Eq.~(\ref{aeUW87value}) 
is the uncontrolled shift of the frequency 
due to the resonance between the electron and 
the metal cavity of hyperbolic shape.
 Brown {\it et al.} \cite{Brown:1985sa} 
showed that this source of uncertainty
can be reduced significantly
using a metal trap 
with the cylindrical cavity whose resonance structure 
can be calculated analytically.

 The recent Harvard measurement is based on the cylindrical cavity. 
 Their value 
announced in 2006 is \cite{Odom:2006gg} 
\begin{equation}
  a_e({\rm HV}06) = 1~159~652~180.85~(0.76) \times 10^{-12}
  \quad
  [0.66\text{ppb}]
  \,, 
\label{aeHV06value}
\end{equation} 
which has a 5.5 times smaller uncertainty than 
the previous measurements listed in Eq.~(\ref{aeUW87value}).
Very recently,  the same Harvard group has succeeded in reducing 
the uncertainty further by a factor 2.7 \cite{Hanneke:2008}:
\begin{equation}
  a_e({\rm HV}08)= 1~159~652~180.73~(0.28) \times 10^{-12} 
\quad [0.24\text{ppb}]
  \,.
\label{aeHV08value}
\end{equation}

 To match the precision of the measurement
the theory of $a_e$ must include radiative corrections 
of up to the eighth-order of QED perturbation theory 
as well as the hadronic and weak contributions 
\begin{equation}
  a_e = a_e({\rm QED}) + a_e({\rm hadron}) + a_e({\rm weak}).
\label{ae_theory} 
\end{equation} 
 The hadronic 
\cite{Davier:1998si,Krause:1996rf,Melnikov:2003xd,Bijnens:2007pz}
and weak contributions \cite{Czarnecki:1996ww} 
to $a_e$ are very small, but not entirely negligible relative to the measurement
uncertainties  (\ref{aeHV06value}) or (\ref{aeHV08value}): 
\begin{gather}
  a_e({\rm hadron}) = 1.682~(20) \times 10^{-12}, \\
  a_e({\rm weak } ) = 0.0297~(5) \times 10^{-12}.
\label{hadweak}
\end{gather}

 The QED contribution $a_e({\rm QED})$ can be divided further into four 
parts taking account of the presence of other leptons: 
\begin{equation}
  a_e({\rm QED})
  =  
  A_1
  + A_2(m_e/m_\mu) 
  + A_2(m_e/m_\tau) 
  + A_3(m_e/m_\mu, m_e/m_\tau) \,, 
\label{ae_qed} 
\end{equation} 
where $m_e, m_\mu,$ and $m_\tau$ are masses of the electron 
($e$), muon ($\mu$) and tau-lepton ($\tau$), respectively. 
$A_1$, being dimensionless, depends only on the fine structure constant $\alpha$. 
$A_2$ denotes contributions from the Feynman diagrams which 
have closed loops of either muon or tau-lepton. 
 $A_3$ stands for the contributions 
of the Feynman diagrams which contain 
both $\mu$ loop and $\tau$ loop. 
 Each $A_i$ can be calculated by the QED perturbation theory 
\begin{equation} 
  A_i 
  =
    A_i^{(2)}\left( \frac{\alpha}{\pi} \right)
  + A_i^{(4)}\left( \frac{\alpha}{\pi} \right)^2
  + A_i^{(6)}\left( \frac{\alpha}{\pi} \right)^3
  + \cdots\,.
\end{equation}

 The purpose of this paper is to give a detailed account of derivation
of the revised value of the eighth-order coefficient of $A_1$ reported recently
\cite{Aoyama:2007PRL}
\begin{equation}
  A_1^{(8)} = -1.914~4~(35)~.     
\label{newa8} 
\end{equation} 

 Making use of our automating algorithms 
in handling ultraviolet (UV) 
and infrared (IR) divergences \cite{Aoyama:2005kf, Aoyama:2007IR}, 
we are now able to generate
the eighth-order FORTRAN codes very easily and swiftly.
 However, numerical evaluation of these codes is 
still nontrivial and requires a huge computational resource. 
 Thus far the ``new" calculation has achieved 
a relative uncertainty of about 3 \% . 
 Although this is still more than an order of magnitude 
less accurate than that of Ref.~\cite{Kinoshita:2005zr}, 
it is good enough for the purpose of checking the old calculation.

 Comparison of the ``new" 
numerical result 
with the old one 
has revealed an inconsistency in the treatment of 
the IR divergence in the latter.
 With this error of the old calculation corrected,
we now have two independent determinations of $A_1^{(8)}$.
 Of course, precise evaluation of all terms of ``new" $A_1^{(8)}$
by the integration routine VEGAS \cite{vegas}
requires an enormous amount of computation.
 Fortunately, as is described in Sec. \ref{reconst},
the correction term itself can be evaluated 
easily and very precisely. 
 This is why we are able to give the uncertainty in Eq.~(\ref{newa8})
which is essentially identical with that of the previous calculation 
\cite{Kinoshita:2005zr}. 

 Besides $A_1^{(8)}$
the known terms of Eq. (\ref{ae_qed}) are as follows
\cite{%
Schwinger:1948iu,%
Petermann:1957,Sommerfield:1957,%
Kinoshita:1995,%
Laporta:1996mq,
Samuel:1990qf,Li:1992xf,Czarnecki:1998rc,%
Laporta:1993ju,Laporta:1992pa,%
Lautrup:1977hs,%
Passera:2006gc,%
CODATA:2002}: 
\begin{gather}
  A_1^{(2)}  = 0.5, \nonumber \\
  A_1^{(4)}  =-0.328~478~965~579 \cdots, \nonumber \\
  A_1^{(6)}  = 1.181~241~456~587 \cdots, \nonumber \\
  A_1^{(10)} = 0.0~(4.6), \nonumber \\ 
  A_2^{(4)}(m_e/m_\mu )  = 5.197~386~70~(27)\times 10^{-7}, \nonumber \\
  A_2^{(4)}(m_e/m_\tau ) = 1.837~63~(60)\times 10^{-9}, \nonumber \\
  A_2^{(6)}(m_e/m_\mu )  =-7.373~941~58~(28)\times 10^{-6}, \nonumber \\
  A_2^{(6)}(m_e/m_\tau ) =-6.581~9~(19)\times 10^{-8}, \nonumber \\
  A_3^{(6)}(m_e/m_\mu, m_e/m_\tau) = 0.190~945~(62)\times 10^{-12}
  \,.
\label{knownterms}
\end{gather}
Here, $A_1^{(2)}$, $A_1^{(4)}$, and $A_1^{(6)}$ are known analytically. 
$A_2^{(4)}$, $A_2^{(6)}$ and $A_3^{(6)}$ are known analytically as functions of mass ratios 
so that their uncertainties are due to those of measured lepton masses only.
 Note that $A_1^{(10)}$ is actually unknown 
and the value listed above is
an educated guess 
calculated by the recipe 
proposed in Ref.~\cite{CODATA:2002}
to indicate a likely range of the value taken by $A_1^{(10)}$.
 This will soon be replaced by a real number,
which is being evaluated by FORTRAN codes generated
with the help of the automatic code generator {\sc gencode}{\it N} 
\cite{Aoyama:2005kf,Aoyama:2007IR}. Until then $A_1^{(10)}$ 
in  Eq.~(\ref{knownterms})
is the largest source of theoretical uncertainty.

 In order to obtain the numerical value of the theoretical $g\! - \! 2$,
an explicit value of the fine structure constant $\alpha$, 
which is determined by the physical phenomena other than $g\! - \! 2$, 
is required. 
 At present the best values of $\alpha$ available in the
literature are from the Cesium atom experiments 
\cite{Wicht:2002,Gerginov:2006} and 
the Rubidium atom experiment \cite{Clade:2006PRA} 
\begin{align}
  \alpha^{-1}({\rm Cs06}) &= 137.036~000~00~(110)
  \quad
  [8.0 {\rm ppb}]\,,
\label{alphaCs}
\\
  \alpha^{-1}({\rm Rb06}) &= 137.035~998~84~(91)\phantom{0}
  \quad
  [6.7 {\rm ppb}]\,.
\label{alphaRb}
\end{align}
They lead to the theoretical predictions of $a_e$: 
\begin{align}
  a_e({\rm Cs}) &= 1~159~652~172.99~(0.10)(0.31)(9.32) \times 10^{-12},
\nonumber \\
  a_e({\rm Rb}) &= 1~159~652~182.79~(0.10)(0.31)(7.71) \times 10^{-12},
\label{newae}
\end{align}
respectively,
where the uncertainty $0.10$ comes from the eighth-order result (\ref{newa8}),
$0.31$ is an estimated uncertainty of the tenth-order term,
and $9.32$ and $7.71$ come from the uncertainties of the input values
of the fine structure constants 
given in Eqs.~(\ref{alphaCs}) and (\ref{alphaRb}).
The uncertainty  due to the hadronic 
and weak contributions (\ref{hadweak}) is  $0.02\times 10^{-12}$. 
 The revised theoretical anomaly $a_e$ is in closer agreement 
with the experimental 
values
(\ref{aeHV06value})
and (\ref{aeHV08value}) 
than the old value \cite{Kinoshita:2005zr}.

 Unfortunately, the precision of  $a_e$ given in Eq.~(\ref{newae}) is 
not high enough
for direct confrontation between  the experimental and theoretical
$a_e$'s. 
 This is because the uncertainties in $a_e$ due to these $\alpha$'s amount 
to $9.3 \times 10^{-12}$ for $\alpha({\rm Cs06})$
and $7.7 \times 10^{-12}$ for $\alpha({\rm Rb06})$, respectively,  
which are an order of magnitude 
larger than the experimental uncertainty $0.76 \times 10^{-12}$ and 
the theoretical uncertainty $ 0.28 \times 10^{-12}$ of $a_e$.

 This implies that,
assuming the validity of QED, 
 the electron $g\! - \! 2$ is in fact the best source 
of the fine structure constant $\alpha$, 
an order of magnitude better than any alternative.
 Because of high precision of the experiments (\ref{aeHV06value}) and (\ref{aeHV08value})
the fine structure constant $\alpha$ determined from $a_e$ 
is rather sensitive to the revision of the theoretical prediction. 
 Equating the Harvard measurements (\ref{aeHV06value}) or (\ref{aeHV08value}), and 
the theory (\ref{ae_theory}), we obtain \cite{Gabrielse:2006gg,Gabrielse:2007prl, Hanneke:2008}
\begin{align}
 & 
\alpha^{-1}(a_e ({\rm HV06=Th07}) ) 
  =
  137.035~999~070~(12)(37)(90)
  \quad
  [0.71 {\rm ppb}]
  \,,
\label{newalpha} \\
&
 \alpha^{-1}(a_e ({\rm HV08=Th07}) ) 
  =
  137.035~999~084~(12)(37)(33)
  \quad
  [0.37 {\rm ppb}]
  \,,
\label{newalpha08}
\end{align} 
where the first and second uncertainties 
come from the numerical uncertainties 
of $A_1^{(8)}$ and $A_1^{(10)}$, respectively,
and the third in Eq. (\ref{newalpha}) or Eq. (\ref{newalpha08}) comes from the experiment 
(\ref{aeHV06value}) or (\ref{aeHV08value}), respectively.

 These values of $\alpha^{-1}$ are
smaller than the old $\alpha^{-1} (a_e ({\rm HV06=Th06}))$
by $-6.411~80(73) \times 10^{-7}$  
which is about 4.7 ppb (or about 7 s.~d.), 
but are still in good agreement 
with $\alpha^{-1}({\rm Rb06})$ of Eq.~(\ref{alphaCs})
and $\alpha^{-1}({\rm Cs06})$ of Eq.~(\ref{alphaRb}),
whose uncertainties are about 7 ppb.
     

 The organization of the paper is as follows. 
In Sec. II, we briefly overview the ``old'' and ``new'' approaches
to the numerical calculation of the electron $g\! - \! 2$ in QED. 
In Sec. III, the diagrams of Group V of the eighth-order term are 
discussed.  We compared the results of the ``old'' and ``new'' 
calculations and found
an unaccountable difference in the results of the diagram $M_{18}$.
In Sec. IV, the diagram $M_{16}$ is closely examined instead of $M_{18}$. 
This is because   
$M_{16}$ has a similar structure to $M_{18}$,  but  somewhat simpler.
We found a source of the discrepancy between the ``old'' and ``new'' 
results and the errors in the ``old'' calculation of $M_{16}$
and $M_{18}$ are corrected.
Sec. V gives the summary of the updated  value of the eighth-order
contribution to the electron $g\! - \! 2$.

Appendix A presents the tests of the automation system {\sc gencode}{\it N}
for the fourth-order  and sixth-order $g\! - \! 2$'s.  
Appendix B gives our renormalization scheme of the magnetic moment amplitude
in the ``new'' approach.
Similarly, Appendix C gives the renormalization scheme of 
the renormalization constants in the ``new'' approach.


\section{Old {\it vs} new approach}
\label{sec:automation}

The purpose of this paper is the presentation of
the results of evaluation of $A_1^{(8)}$ by two independent
methods.
Although these methods started from the same Feynman-parametric
representation of $A_1^{(8)}$, they took different approaches,
in particular, in the handling of the self-energy subdiagrams
and associated infrared (IR) divergences.
Furthermore, the ``new" approach was instrumental in discovering
an error in the handling of infrared divergence in the old method \cite{Kinoshita:2005zr}.
After correcting this error, we now have two independent evaluations
of $A_1^{(8)}$, enhancing substantially the credibility of the calculation.


\subsection{Common starting point}
\label{commonstartpoint}

The anomalous magnetic moment $a_e$ is given by the static limit 
of the magnetic form factor that is related to the proper vertex part 
$\Gamma^\nu$. 
Throughout this paper our attention is focused on the $q$-type diagrams,
namely, proper vertex diagrams that have no
closed lepton loops.
In both old and ``new" formulations, we use a relation derived from the 
Ward-Takahashi identity \cite{Cvitanovic:1974uf, Kinoshita:1990}
\begin{equation}
  \Lambda^\nu(p,q) 
  \simeq 
  - q^\mu \left[
    \frac{\partial\Lambda_\mu(p,q)}{\partial q_\nu}
  \right]_{q\to 0}
  - \frac{\partial\Sigma(p)}{\partial p_\nu}
\label{wtidentity}
\end{equation}
between the self-energy part $\Sigma(p)$ and 
the sum of vertex parts $\Lambda^\nu(p,q)$ obtained by 
inserting an external vertex in the lepton lines of $\Sigma$ in 
all possible ways. 
Here, the momentum of the incoming lepton is $p-\frac{1}{2}q$ 
and that of the outgoing lepton is $p+\frac{1}{2}q$. 
By means of Eq.~(\ref{wtidentity}) a set of vertex diagrams are amalgamated 
into a single self-energy-like diagram, which reduces the
number of independent integrals substantially. For the
eighth-order $q$-type diagrams, the number of Feynman diagrams is reduced from 
518 to 74. Taking into account the time-reversal symmetry, the number is
further reduced from 74 to 47.

The amplitude of the magnetic moment contribution of 
a diagram is obtained by applying Feynman-Dyson rules 
of QED in the momentum space. 
Carrying out the momentum integration analytically, 
we can express the amplitude of $2n$th-order  diagram $\mathcal{G}$ 
as an integral over the Feynman parameters ${z_i}$: 
\begin{multline}
  M_\mathcal{G}^{(2n)} 
  = 
  \left(-\frac{1}{4}\right)^{n} (n-1)! 
  \int (dz)_\mathcal{G}
  \left[
    \frac{1}{n-1}\left(
    \frac{E_0 + C_0}{U^2 V^{n-1}} + \frac{E_1 + C_1}{U^3 V^{n-2}} + \cdots
    \right)
\right. \\
    +
\left .
    \left(
    \frac{N_0 + Z_0}{U^2 V^{n}} + \frac{N_1 + Z_1}{U^3 V^{n-1}} + \cdots
    \right)
  \right],
\label{parametricint}
\end{multline}
where $(dz)_\mathcal{G} = \prod_i d z_i \delta(1-\sum_i z_i)$. 
The factor  $(\alpha/\pi)^n$ is omitted for simplicity.

The quantities $E_k$, $C_k$, $N_k$, and $Z_k$ are polynomials 
of symbols called building blocks $B_{ij}$, $A_i$, and $C_{ij}$ 
\cite{Cvitanovic:1974uf}. 
The symbols $B_{ij}$ and $U$ are homogeneous polynomials of Feynman 
parameters, related to the flow of loop momenta in the diagram. 
The symbol $A_i$ is called scalar current that is associated 
with the flow of external momenta. 
They are functions of $B_{ij}$, $U$, and $z_i$. 
The symbol $C_{ij}$ is given by $z_i, B_{ij}$ and $U$. 
The denominator function $V$ is defined by 
\begin{equation}
  V = \sum_i z_i - G, \qquad G = \sum_i z_i A_i,
\label{denominatorV}
\end{equation}
where the summation is over the electron lines only, and
the electron mass is chosen to unity  for simplicity.


\subsection{Different structure of integrand}

Although the ``old" and ``new" methods have the common starting point,
they have an important difference in practice. In the ``old" version
of the programs, the size of the integrand was reduced 
by taking  symmetries of a diagram into account. 
One type of modifications was applied to the integrand by using 
8 ``junction laws" and 4 ``loop laws" satisfied by the scalar currents $A_i$
(where Feynman parameters $z_i$ play the role of resistance) 
\cite{Cvitanovic:1974uf,Kinoshita:1990}. 
Another type of modification was to reduce the number of integration variables by 
exploiting the fact that in some diagrams the integrand depends only on a particular 
combination of Feynman parameters. 
These resulted in substantial reduction in the size of integrands and the amount of 
computing time required to achieve desired precision. 
In the ``new'' version, those modifications were not employed at all because 
they are diagram-specific and not suitable for automation. 
As a result, the size of FORTRAN source code for $M_{01}$ (see Fig.~\ref{fig:EighthOrderSetV}), which 
requires only vertex renormalization, is about 515KB in the ``new" version 
in contrast to 316KB of the old version. 
A more notable difference is seen for the diagram $M_{12}$ which  requires IR 
subtraction.
The ``new" $M_{12}$ occupies 630KB while the old $M_{12}$ occupies only 21KB. 
As a consequence, the old and ``new" integrals have much different forms so that they can be 
regarded to be independent of each other as far as numerical integration is concerned. 

\subsection{Ultraviolet (UV) divergence}
\label{UVsubtraction}

The amplitude $M_{\mathcal G}$ has (logarithmic) UV-divergences in general. 
Suppose we want to find out whether $M_{\mathcal G}$ diverges when all loop momenta
of a subdiagram ${\mathcal S}$ consisting of $N_{\mathcal S}$ lines and $n_{\mathcal S}$ 
closed loops go to infinity.
In the parametric formulation this limit corresponds to 
the vanishing of $U$ when all $z_i$ for $i \in {\mathcal S}$ vanish simultaneously.
To find the criterion for the UV divergence from ${\mathcal S}$,
consider the part of the integration domain where $z_i$ for $i \in S$
satisfy $\sum_{i \in {\mathcal S} } z_i \leq \epsilon$.
In the limit $\epsilon \rightarrow 0$ one finds \cite{Cvitanovic:1974sv,Kinoshita:1990}
 \begin{align}
  V &= \mathcal{O}(1),
  \qquad
  U = \mathcal{O}(\epsilon^{n_\mathcal{S}}) \nonumber \\
  B_{ij} &=
  \begin{cases}
    \mathcal{O}(\epsilon^{n_\mathcal{S}-1}) 
    & \text{for $i,j \in \mathcal{S}$,} \\
    \mathcal{O}(\epsilon^{n_\mathcal{S}}) 
    & \text{otherwise}.
  \end{cases}
\label{Eq17}
\end{align}

The UV-divergent part can be identified by the following procedure 
called {\it K}-operation: 
\begin{enumerate}[(a)]
\item \label{kop:step1}
  In the limit (\ref{Eq17}) keep only terms with the lowest
  power of $\epsilon$ in $U$, $B_{ij}$, and $A_i$. 
  In this limit $U$ factorizes as
  $U_\mathcal{S} U_{\mathcal{G}/\mathcal{S}}$
  where $\mathcal{G}/\mathcal{S}$ is obtained from $\mathcal{G}$
  by shrinking $\mathcal{S}$ to a point in $\mathcal{G}$. 
  $B_{ij}$ factorizes similarly. 
  $V$ is reduced to $V_{\mathcal{G}/\mathcal{S}}$,
  where $V_{\mathcal{G}/\mathcal{S}}$ is
  the $V$ function defined on $\mathcal{G}/\mathcal{S}$. 
\item \label{kop:step2}
  Replace $V_{\mathcal{G}/\mathcal{S}}$ by
  $V_{\mathcal{G}/\mathcal{S}}+V_\mathcal{S}$. 
\item \label{kop:step3}
  Rewrite the integrand of $M_\mathcal{G}$ in terms of parametric
  functions redefined in (\ref{kop:step1}) and (\ref{kop:step2}),
  and drop all terms except those with the largest number of
  {\it contractions} \cite{Cvitanovic:1974uf} within $\mathcal{S}$. 
  The result is denoted by $\mathbb{K}_\mathcal{S} M_\mathcal{G}$,
  in which $\mathbb{K}_\mathcal{S}$ stands for an operator
  acting on $M_\mathcal{G}$. 
\end{enumerate}
By construction, $\mathbb{K}_\mathcal{S} M_\mathcal{G}$ has the same UV divergence as $M_\mathcal{G}$ in the same integration domain. 
Therefore it can be used as a pointwise subtraction term in the subtractive renormalization.

%
%
%
%

An important feature of {\it K}-operation 
is that the resulting integral can be factorized 
exactly into a product or a sum of products 
of lower-order quantities that consists of 
a leading UV-divergent part of the renormalization constant and 
the magnetic moment part. 
The {\it K}-operation associated with a UV-divergent subdiagram 
$\mathcal{S}$ produces, when $\mathcal{S}$ is of vertex type,  
the subtraction term of the form \cite{Cvitanovic:1974sv,Kinoshita:1990}
\begin{equation}
  \mathbb{K}_{\mathcal{S}} M_\mathcal{G}
  =
  L_{\mathcal{S}}^{\rm UV} M_{\mathcal{G}/\mathcal{S}}
  \,,
\end{equation}
where $L_{\mathcal{S}}^{\rm UV}$ is the leading UV-divergent part 
of the vertex renormalization constant $L_\mathcal{S}$ and 
$M_{\mathcal{G}/\mathcal{S}}$ is the magnetic moment part of 
the reduced diagram $\mathcal{G}/\mathcal{S}$. 
When $\mathcal{S}$ is a self-energy subdiagram, the {\it K}-operation 
yields  \cite{Cvitanovic:1974sv,Kinoshita:1990}
\begin{equation}
  \mathbb{K}_{\mathcal{S}} M_\mathcal{G}
  =
  {\delta m}_{\mathcal{S}}^{\rm UV} M_{\mathcal{G}/\mathcal{S}\,(i^\star)}
  +
  B_{\mathcal{S}}^{\rm UV} M_{\mathcal{G}/[\mathcal{S},i]}
  \,,
\end{equation}
where ${\delta m}_{\mathcal{S}}^{\rm UV}$ is the leading UV-divergent 
part of the mass-renormalization constant ${\delta m}_{\mathcal{S}}$, 
$B_{\mathcal{S}}^{\rm UV}$ is the leading UV-divergent part of 
the wave-function renormalization constant $B_{\mathcal{S}}$, 
and the reduced diagram $\mathcal{G}/[\mathcal{S},i]$ is obtained 
from $\mathcal{G}$ by removing $\mathcal{S}$ and a lepton line $i$ 
adjacent to $\mathcal{S}$. 

The whole UV-divergent structure of the amplitude $M_\mathcal{G}$ 
can be recognized by Zimmermann's forest formula \cite{Aoyama:2005kf}. 
A forest is a set of UV-divergent subdiagrams in which any pair 
of subdiagrams is either disjoint 
(they do not share lines or vertices) or inclusive 
(one subdiagram is a subgraph of the other subdiagram). 
Each subtraction term corresponds to a forest.
In our formulation, the subtraction term 
is obtained by successive application of {\it K}-operations 
for every element of the forest. 
The UV-finite amplitude $\underline{M}_\mathcal{G}$ created  
by {\it K}-operation is thus expressed in the form 
\begin{equation}
  \underline{M}_\mathcal{G}
  =
  M_\mathcal{G}
  +
  \sum_{f} \prod_{\mathcal{S}\in f} \left(-\mathbb{K}_{\mathcal{S}}\right)
  M_\mathcal{G}
  \,,
\label{uvrenormalized}
\end{equation}
where the summation is taken over the {\it normal forests} of the diagram 
$\mathcal{G}$ that do not include $\mathcal{G}$ itself as an element. 

{\it N. B. In both old and ``new" approaches UV divergence is treated 
by the same {\it K}-operation.}


\subsection{Infrared (IR) divergence}
\label{IRsubtraction}

A diagram may have an IR singularity when some of the internal
photon momenta vanish.
In order that this singularity becomes actually divergent, however, 
it must be enhanced by vanishing of denominators of two or more electron 
propagators (called {\it enhancers}) due to kinematical constraints. 
Such a situation occurs in diagrams that have self-energy-like subdiagrams.
In Eq. (\ref{parametricint}) this corresponds to the vanishing of $V$-function
of the denominators in the integration domain characterized 
by \cite{Cvitanovic:1974sv, Kinoshita:1990}
\begin{equation}
  z_i = \begin{cases}
    \mathcal{O}(\delta) 
    & \text{if $i$ is an electron line in $\mathcal{R}$,} \\
    \mathcal{O}(1)
    & \text{if $i$ is a photon line in $\mathcal{R}$,} \\
    \mathcal{O}(\epsilon), \quad \epsilon\sim\delta^2, \quad
    & \text{if $i \in \mathcal{S}$,}
  \end{cases}
\label{irlimit}
\end{equation}
where ${\mathcal R}={\mathcal G}/{\mathcal S}$.  

This enables us to obtain a simple IR power-counting rule for identifying
IR divergent terms.
When there are two enhancers, the amplitude shows a logarithmic
IR divergence. We can identify and construct the corresponding
subtraction term by the following procedure called
$\mathfrak{I}$-operation \cite{Cvitanovic:1974sv,Kinoshita:1990}:
\begin{enumerate}[(a)]
\item  In the limit (\ref{irlimit}) keep only terms with lowest power of $\epsilon$
and $\delta$ in $U, B_{ij}, A_i$. The numerator  then factorizes to the product
\begin{equation}
F \rightarrow    F_{\mathcal R} F_{\mathcal{S}}~, 
\end{equation}
where $F_{\mathcal R}$ is a numerical factor obtained by replacing 
all scalar currents $A_i$ in the
diagram $\mathcal{R}$  by one.

\item  Make the following replacements:
\begin{equation}
U \rightarrow U_{\mathcal S} U_{\mathcal R},~~~V \rightarrow V_{\mathcal S} + V_{\mathcal R}, ~~~ F \rightarrow F_0 [L_{\mathcal R}] F_{\mathcal S},
\end{equation}
where $F_0 [L_{\mathcal R}]$ is the no-contraction part of the vertex renormalization
constant defined in ${\mathcal R}$. The difference between $F_0[L_{\mathcal R}]$
and $F_{\mathcal R}$ causes a finite difference of the integration.  

\item  Rewrite the integrand of $M_{\mathcal G}$ in terms of redefined parametric 
functions, keeping only the IR-divergent terms.
\end{enumerate}
In the ``old" method all logarithmic IR divergences 
have been subtracted by means of the $\mathfrak{I}$-operation.
However, the case involving linear IR divergence, which has three enhancers,
was handled by an {\it ad hoc} manner
instead of a systematic approach.
Actually, the cause of linear IR-divergence is easy to identify.
It is caused by our treatment of self-energy subdiagram
by means of {\it K}-operation which subtracts only the UV-divergent part
of the self-mass.
The unsubtracted part of self-mass keeps the number of enhancers unchanged,
except in second-order case where the {\it K}-operation subtracts
the self-mass term completely.

In the ``new" approach, a systematic method is developed to handle
the linear IR divergence.  To remove the finite remnant of 
self-mass term completely,  an {\it R}-subtraction operation \cite{Aoyama:2007IR}
is newly  introduced. 
After the {\it R}-subtraction operation is carried out, which decreases the number of
enhancers to two, only logarithmic 
IR divergences remain, which can be handled by the {\it I}-subtraction operation
similar to, but different  in detail 
from, the $\mathfrak{I}$-operation of the ``old'' method.

For a formal treatment, we introduce two operators for these subtractions. 
The {\it R}-subtraction operator $\mathbb{R}_\mathcal{S}$ acts as 
\begin{equation}
  \mathbb{R}_\mathcal{S} \underline{M}_\mathcal{G}
  =
  {\delta m}_{\mathcal{S}}^{\rm R}
  \underline{M}_{\mathcal{G}/\mathcal{S}\,(i^\star)} 
  \,,
\end{equation}
where ${\delta m}_{\mathcal{S}}^{\rm R}$ is the residual part 
of the mass renormalization constant defined by 
\begin{equation}
  {\delta m}_{\mathcal{S}}^{\rm R}
  =
  {\delta m}_{\mathcal{S}}
  -
  {\delta m}_{\mathcal{S}}^{\rm UV}
  +
  \sum_{f} \prod_{\mathcal{S}^\prime\in f}
  \left(-\mathbb{K}_{\mathcal{S}^\prime}\right)
  \widetilde{\delta m}_{\mathcal{S}}
\label{dmR}
\end{equation}
in which 
the leading UV-divergent part ${\delta m}_{\mathcal{S}}^{\rm UV}$ 
and the subdivergent parts associated with the forests 
$\prod_{\mathcal{S}^\prime\in f} \left(-\mathbb{K}_{\mathcal{S}^\prime}\right) \widetilde{\delta m}_{\mathcal{S}}$
are subtracted away, where
$\widetilde{\delta m} \equiv {\delta m} - {\delta m}^{\rm UV}$. 

The {\it I}-subtraction operator $\mathbb{I}_\mathcal{S}$ acts 
on the UV-renormalized amplitude $\underline{M}_\mathcal{G}$ as 
\begin{equation}
  \mathbb{I}_\mathcal{S} \underline{M}_\mathcal{G}
  =
  {L}_{\mathcal{G}/\mathcal{S}}^{\rm R} \underline{M}_\mathcal{S}
  \,,
\end{equation}
where ${L}_{\mathcal{G}/\mathcal{S}}^{\rm R}$ is the residual part 
of the vertex renormalization constant defined by 
\begin{equation}
  {L}_{\mathcal{G}/\mathcal{S}}^{\rm R}
  =
  {L}_{\mathcal{G}/\mathcal{S}}
  -
  {L}_{\mathcal{G}/\mathcal{S}}^{\rm UV}
  +
  \sum_{f} \prod_{\mathcal{S}^\prime\in f}
  \left(-\mathbb{K}_{\mathcal{S}^\prime}\right)
  \widetilde{L}_{\mathcal{G}/\mathcal{S}}
\label{LR}
\end{equation}
in which 
the leading UV-divergent part ${L}_{\mathcal{G}/\mathcal{S}}^{\rm UV}$ 
and the subdivergent parts associated with the forests 
$\prod_{\mathcal{S}^\prime\in f} \left(-\mathbb{K}_{\mathcal{S}^\prime}\right) \widetilde{L}_{\mathcal{G}/\mathcal{S}}$
are subtracted away, where 
$\widetilde{L} \equiv L - L^{\rm UV}$. 

{\it N. B.  The IR power counting rule identifies only IR-divergent terms.
It does not specify how to handle IR-finite term.
The ``new" {\it I}-subtraction operation handles the IR-finite terms differently 
from the ``old" $\mathfrak{I}$-operation. 
The {\it I}-subtraction operation needs not  deal with the IR divergence
associated with a vertex subdiagram of the self-energy-like diagram, 
while the $\mathfrak{I}$-operation directly acts on the vertex subdiagram. }


The whole set of IR subtraction terms can be obtained by 
the combinations of these two operations, both of which belong to {\it annotated forests}
\cite{Aoyama:2007IR}. 
An annotated forest is a set of self-energy-like subdiagrams, to 
each element of which the distinct operation of {\it I}-subtraction 
or {\it R}-subtraction is assigned. 
The IR-subtraction term associated with an annotated forest is 
constructed by successively applying operators $\mathbb{I}$ or 
$\mathbb{R}$, and takes the form 
\begin{equation}
  (-\mathbb{I}_{\mathcal{S}_i})\dots
  (-\mathbb{R}_{\mathcal{S}_j})\dots
  \underline{M}_\mathcal{G}
\end{equation}
where the annotated forest $\widetilde{f}$ consists of the subdiagrams 
$\mathcal{S}_i$, \dots and $\mathcal{S}_j$, \dots.

{\it  N. B.  The IR divergence is treated differently in the old and ``new" approaches.
This difference plays an important part in ensuring the independence of two calculations.}


\subsection{Residual renormalization}

Because of difference in the handling of IR divergences in the ``old'' and ``new''
methods,
we obtain different forms of residual renormalization.
Since the ``old'' residual renormalization is described in Refs.~\cite{Cvitanovic:1974sv,Kinoshita:1990},
let us consider here only the ``new'' residual renormalization.

In the ``new'' approach the
UV- and IR-finite amplitude has the form 
\begin{equation}
  {\Delta M}_\mathcal{G}
  =
  M_\mathcal{G}
  +
  \sum_{f} \prod_{\mathcal{S}\in f} (-\mathbb{K}_{\mathcal{S}} )
  M_\mathcal{G}
  +
  \sum_{ \widetilde{f}} 
     ( -\mathbb{I}_{\mathcal{S}_i} )\cdots
     ( -\mathbb{R}_{\mathcal{S}_j}  )\cdots
  \underline{M}_\mathcal{G}
  \,,
\label{deltamg}
\end{equation}
where $\underline{M}_\mathcal{G}$ is 
the UV-finite  quantity defined by Eq. (\ref{uvrenormalized}).
$\Delta M_\mathcal{G}$ can be readily turned into a numerical
integration code by {\sc gencode}{\it N}
\cite{Aoyama:2005kf,Aoyama:2007IR} and is to be evaluated by numerical means.

This procedure is different from the standard on-shell renormalization which is 
defined by the on-shell quantities. 
The difference between the on-shell quantities and the quantity 
evaluated by Eq. (\ref{deltamg}) must be compensated by 
products of known lower-order quantities. 
We call this step the residual renormalization. 
See Appendix B for details.

\AllFiguresOfEighthOrderSetV

\section{Eighth-order terms}
\label{Sec:EighthOrderTerms}
 
 The eighth-order term  $A_1^{(8)}$ 
receives contributions from 891 Feynman diagrams.
The 373 of them have closed lepton loops and
had been evaluated by two or more  
independent methods \cite{Kinoshita:2005zr}.
The remaining 518 diagrams of $q$-type 
form one gauge-invariant set (Group V). 
In our approach they are represented by 47 independent diagrams
shown in Fig.\ref{fig:EighthOrderSetV} by using the relation
derived from Ward-Takahashi identity and the time-reversal symmetry. 
Thus far, there is only one complete evaluation
of the eighth-order term, which was performed by numerical means 
 \cite{Kinoshita:2005zr}. 
 Some of these diagrams have linear IR divergence, 
which was treated by an {\it ad hoc} subtraction method.
In contrast  {\sc gencode}{\it N} is capable of dealing with 
such hard IR divergence
in a systematic fashion \cite{Aoyama:2007IR}. 
The application of {\sc gencode}{\it N} to 
the calculation of the eighth-order $q$-type diagrams 
provides us the opportunity 
not only to test if it works properly, 
but also to check the previous result. 

 Even in the eighth-order case  {\sc gencode}{\it N} creates FORTRAN
programs very rapidly. 
 The entire 47 program sets are generated in less than ten minutes 
on {\sl hp}'s Alpha.
 The numerical evaluation is, however, quite non-trivial and requires 
a huge computational resource.
 For the preliminary evaluation we have used 64 to 256 Xeon CPU's 
{\it per} diagram and run the programs 
over a few months.  
To our surprise it uncovered an inconsistency in the treatment
of IR subtraction terms in the old calculation. In Secs.~\ref{Sec:EighthOrderTerms}
and \ref{m16m18}  we describe 
how this inconsistency was uncovered
by a detailed comparison of the old code and the code 
generated by {\sc gencode}{\it N}.  

\subsection{IR treatments of the eighth-order diagrams}
 
The treatments of IR subtraction terms are
different in {\sc gencode}{\it N} and the ``old" approach. 
The difference of IR subtraction terms leads 
to the difference of the finite part of the amplitude
$\Delta M_i$ ($i = 01,\,\cdots,\,47$). 
The difference 
$\Delta M_{i}^{\rm old} - \Delta M_{i}^{\rm new}$ 
between the old amplitude $\Delta M_{i}^{\rm old}$ 
and the ``new" one $\Delta M_{i}^{\rm new}$
is finite and can be expressed analytically 
in terms of finite lower-order quantities. 
We will see if this difference is numerically reproduced
by substituting the numerical values calculated separately for
these lower-order quantities. If the numerical discrepancy is found,
there must be something wrong in either the ``old" or ``new" calculation. 
This is what we tried to find out.

\MsixteenMeighteenFigure


We noted in Sec. \ref{UVsubtraction} that 
the subtraction of UV divergences 
is achieved by the same {\it K}-operation 
in both {\sc gencode}{\it N} and  ``old" calculation. 
 Therefore, 
the difference between $\Delta M_i^{\rm old}$ 
and $\Delta M_i^{\rm new}$, 
if it exists, 
comes  exclusively from the difference 
of IR subtraction procedures between the ``new" and ``old" calculations. 
 To examine this difference more closely 
let us begin by considering relatively simple diagrams which contain only 
one second-order self-energy subdiagram.
Diagrams belonging to this class are 
$M_{02}$, $M_{03}$, $M_{09}$, $M_{13}$, $M_{14}$, 
$M_{15}$, $M_{23}$, $M_{24}$, 
$M_{27}$, $M_{43}$, $M_{44}$.
 As an example  let us consider the diagram $M_{14}$.

 The IR divergence occurs in $M_{14}$ 
from the second-order self-energy subdiagram 
which consists of an electron line ``2'' and a photon line ``b'' 
in Fig.~\ref{fig:M14andL6g5}(a).
 In the W-T summed diagram this subdiagram plays dual roles.
 One part of this subdiagram behaves as a genuine self-mass term 
and the associated UV singularity is removed completely 
by the ${\mathbb K}_2$-operation.
 Another part works as the second-order magnetic moment $M_2$, 
and the residual diagram surrounding $M_2$ behaves 
like a sixth-order vertex diagram $L_{6g5}$
of Fig. \ref{fig:M14andL6g5}(b),
which is IR-divergent.

\MfourteenFigure

 In the ``old'' approach, 
the finite contribution $\Delta M_{14}$ was defined by
\begin{equation}
  \Delta M_{14}^{\rm old} 
  \equiv 
  M_{14} 
  + \sum_f  \prod_{{\mathcal S} \in f} 
  (- {\mathbb K}_\mathcal{S} ) M_{14} 
  - (I_{6g5}  + I_2 \Delta L_{4l} ) M_2.
\end{equation}
 Here the second term on the right-hand side 
is the sum of UV subtraction terms given 
by the {\it K}-operation.
 The last two terms beginning with the letter ``$I$'' 
are the IR subtraction terms generated 
by the $\mathfrak{I}$-operations 
$\mathfrak{I}_{134567}$ and $\mathfrak{I}_{13}(1-\mathfrak{I}_{134567})$ 
in the ``old" approach. 
 They arise from the ``magnetic-moment part'' 
of the self-energy-like subdiagram mentioned above. 
 Note that they are exactly identical 
with the IR-divergent parts 
of $L_{6g5}$:
\begin{equation}
  L_{6g5} 
  \equiv 
  I_{6g5} + I_2 \Delta L_{4l} + \Delta L_{6g5} 
  - \sum_f  \prod_{\mathcal{S} \in f }
  (-{\mathbb K}_\mathcal{S}) \widetilde{L}_{6g5} 
  + L_{6g5}^{\rm UV}
  \,.
\label{L6g5}
\end{equation} 
 Here the sum appearing on the right-hand side 
denotes all the UV subdivergences contained in $L_{6g5}$  
(whose explicit form is 
$-L^{UV}_2 \widetilde{L}_{4c} - L^{UV}_{4l} \widetilde{L}_2 + (L^{UV}_2)^2 
\widetilde{L}_2$), 
and the last term is the overall UV divergence 
of $L_{6g5}$. 

In the ``new'' (or {\sc gencode}{\it N}) approach, we introduce a 
term $L_{6g5}^{\rm R}$ defined by
\begin{align}
  L_{6g5}^{\rm R}
  & \equiv 
  L_{6g5} - L_{6g5}^{\rm UV} 
  + \sum_f \prod_{\mathcal{S} \in f} 
  (-{\mathbb K}_\mathcal{S}) 
  \widetilde{L}_{6g5} 
\label{L6g5R}
\end{align}
and $\Delta M_{14}^{\rm new}$ by
\begin{equation}
\Delta M_{14}^{\rm new} = \underline{M}_{14} -{\mathbb I}_{2} ~\underline{M}_{14},
\label{II2M14}
\end{equation}
where ${\mathbb I}_2$ is an $I$-subtraction operation associated with the 
self-energy-like subdiagram $\mathcal{S}=\{ 2,b \}$, and yields
\begin{equation}
{\mathbb I}_{2} ~\underline{M}_{14}=L_{6g5}^{\rm R} ~M_2.
\label{L6g5M2}
\end{equation}
From Eqs.(\ref{L6g5}) and (\ref{L6g5R}) we obtain
\begin{equation}
  L_{6g5}^{\rm R}  
  = 
  I_{6g5}+I_2 \Delta L_{4l} + \Delta L_{6g5}.
\end{equation} 
This is UV-finite and consists of not only IR divergent terms but also a completely finite term $\Delta L_{6g5}$.
 It follows that $\Delta M_{14}^{\rm old}$ and $\Delta M_{14}^{\rm new}$ 
differ by
\begin{equation}
  \Delta M_{14}^{\rm old}-\Delta M_{14}^{\rm new}
  = 
  \Delta L_{6g5} M_2 
  \,.
\label{diff_M14} 
\end{equation}
 Since $\Delta L_{6g5}$ is UV- and IR-finite,
it can be computed without encountering with UV or IR divergence. 
 This is true 
for every $\Delta M_i^{\rm old} - \Delta M_i^{\rm new}$ 
as it originates from the choice of 
finite pieces that accompany the singular terms.  
 The choice adopted in $\Delta M_{14}^{\rm new}$ turned 
out to be preferred since 
it leads to a simpler formula and can be readily extended 
to other cases. 

 All eleven diagrams listed above can be analyzed in the same manner.
 The diagrams $M_{04}$, $M_{11}$, $M_{12}$, $M_{17}$, $M_{29}$, $M_{30}$, 
which contain two or three {\it second-order}
self-energy-like subdiagrams, are slightly more complicated, but can be 
treated in a similar manner.
 Evaluation of the diagrams 
with one self-energy subdiagram of fourth- or sixth-order 
such as $M_{08}$, $M_{10}$, $M_{26}$, $M_{38}$, $M_{40}$, $M_{41}$ 
is more complicated 
and needs the residual self-mass renormalization, the $R$-subtraction, 
as well as the $I$-subtraction. 
 But they do not present particular difficulty 
as far as IR subtraction is concerned.
(See Appendix B for more information on these diagrams.)

The diagrams $M_{28}$, $M_{42}$, $M_{45}$, $M_{46}$, $M_{47}$ 
are even more complicated due to nested structure, but they 
can also be handled by slight extensions. (See Appendix B.) 

 The most difficult of the eighth-order $q$-type diagrams 
are those containing one second-order self-energy-like subdiagram and one fourth-order
self-energy-like subdiagram, namely, $M_{16}$ and $M_{18}$ 
of  Fig.~\ref{fig:M16andM18}. 
 The difficulty originates from the fact 
that these diagrams have linear IR divergence.
 Detailed analysis of these diagrams is deferred to Sec.~\ref{m16m18}.

\ComparFirstTable

\ComparSecondTable

\ComparThirdTable

\subsection{Numerical result of eighth-order calculation}

 We present the results of our numerical study 
for $\Delta M_i^{\rm old} - \Delta M_i^{\rm new}$
in Tables \ref{Table:M01M15}, \ref{Table:M16M30} 
and \ref{Table:M31M47}. 
 In these tables, 
the second columns list 
the analytic expression of 
$\Delta M_i^{\rm old} - \Delta M_i^{\rm new}$ 
in terms of finite pieces 
of lower-order renormalization constants 
and magnetic moment amplitudes
multiplied by the multiplicity,
which is 1 for the time-reversal-symmetric diagram and 2 otherwise.
 Each value in the third columns, 
called ``value A'', 
is obtained by substituting
the values of these renormalization constants, etc., listed in Table IV, 
for the corresponding expression in the second columns. 

 In contrast to value A, each value in the fourth columns, 
called ``value B'', is obtained 
by taking the difference between the numerical value 
$\Delta M_i^{\rm old}$ quoted from the literature \cite{Kinoshita:2005zr} and 
the one $\Delta M_i^{\rm new}$ newly calculated 
via {\sc gencode}{\it N} according to the ``new" IR subtraction procedure 
\cite{Aoyama:2007IR}. 
 The fifth columns list up 
the difference of value A and value B 
for each $i$, denoted by $A - B$. 
 It must be zero within numerical precision 
if the whole calculation has been done correctly. 
If value A and value B are different, there are two possible sources.
One possibility is that the program used for a numerical calculation has a bug.
It means that either $\Delta M_i^{\rm old}$ or $\Delta M_i^{\rm new}$ is wrong, or both
are wrong.
 The other possibility is that we incorrectly identified
the analytic difference between the ``old'' and ``new'' methods.

 For a diagram $M_i$ without any self-energy-like subdiagrams, 
the analytic expression of 
$\Delta M_i^{\rm old} - \Delta M_i^{\rm new}$ 
is trivially zero, as it does not have IR divergence. 
 We  can see from the corresponding values B 
in Tables~\ref{Table:M01M15}, 
\ref{Table:M16M30} and \ref{Table:M31M47} 
that this is confirmed within the numerical precision employed. 

\RconstSixthTable

\RconstFourthTable

 The diagrams containing self-energy-like subdiagrams
suffer from IR divergence. 
 Tables~\ref{Table:M01M15}, 
\ref{Table:M16M30} and \ref{Table:M31M47} 
show that  ``old'' and ``new'' calculations are 
in good agreement for most of these diagrams. 
 However, a large discrepancy $-0.221~(21)$ 
is found for the diagram $M_{18}$. 
 Though no detectable discrepancy is found for $M_{16}$, 
it has a structure similar to $M_{18}$
and is somewhat simpler to analyze. 
 In Section \ref{m16m18} 
we thus look for the origin of such a discrepancy 
through a detailed investigation of $M_{16}$.

\section{Detailed examination of $M_{16}$}
\label{m16m18}



 In the ``old'' approach 
the finite contribution $\Delta M_{16}$ was given by \cite{Kinoshita:1981wm, Kinoshita:1990}
\begin{align}
  \Delta M_{16}^{\rm old} 
  &\equiv 
  M_{16} 
  + {\sum_{f}} \prod_{\mathcal{S}\in f} 
  (-{\mathbb K}_\mathcal{S}) M_{16} 
\nonumber \\
  &- I_{6c1} M_2 
  - \frac{1}{2} J_{6c} M_2 
  - I_{4s} \Delta M_{4a} 
  - \Delta \delta m_{4a} I_{4b(1^\star)}
  + I_{2^\star} \Delta \delta m_{4a} M_2
  \,,
\end{align}
while the ``new'' version is given by
\begin{align}
  \Delta M_{16}^{\rm new} 
  &\equiv
  M_{16} 
  + {\sum_{f}} \prod_{\mathcal{S} \in f} 
  (-{\mathbb K}_\mathcal{S}) M_{16} 
\nonumber \\
  &- L_{6c1}^{\rm R} M_2 
  - L_{4s}^{\rm R} \Delta M_{4a} 
  - \Delta \delta m_{4a} \underline{M}_{4b(1^\star)}
  + L_{2^\star}^{\rm R} \Delta \delta m_{4a} M_2
  \,,
\end{align}
where $L_{2^\star}^{\rm R} = I_{2^\star}$. 
 Note that ``$2^\star$'' denotes 
the second-order diagram  with 
a two-point vertex inserted into the internal lepton line. 
 ``$4b(1^\star)$'' denotes 
the diagram obtained from the fourth-order diagram $4b$ by inserting 
a two-point vertex into the lepton line $1$ .  

\subsection{Unrenormalized amplitude and UV subtraction terms of $M_{16}$}

We began our examination by comparing the unrenormalized amplitude $M_{16}$
and its UV subtraction terms 
in the ``old'' and ``new'' programs.  
For this purpose we used the ``spot-check'' method,
by which the values of ``old'' and ``new''
integrands are compared at the same set of numerical values
of integration variables.
The integrand  of $M_{16}$ is  defined in the Feynman 
parameter space that  spans a  hyperplane in 11-dimensional space
satisfying
\begin{equation}
z_1+z_2+ \cdots +z_7 + z_a + z_b + z_c+z_d=1~.
\end{equation} 
In the ``new" version, 
this hyperplane is mapped onto a unit 10-dimensional hypercube.
On the other hand,  in 
the ``old'' version the integration space is mapped onto an 8-dimensional
hypercube, since the integrand depends only on the combination of Feynman 
parameters $z_{137}=z_1+z_3+z_7$.
To carry out the spot check, we must use the same mappings, so 
we changed the mapping of the ``new'' integrand to the ``old" one
defining $z_1=z_3=z_7=(1/3) z_{137}$.
In practice the set of input parameters is chosen  from the neighborhood  of the
singular point of interest where numerical disagreement is likely to be magnified.
But points too close to the singular point are avoided, where the noise due
to round-off error obscures the meaningful information.
The ``old'' integrals and ``new'' integrals  
of the unrenormalized term and UV subtraction terms 
should be algebraically equivalent but
have different forms because of extensive simplification of the ``old'' integrands 
by means of various relations among scalar currents.
The ``spot-check'' comparison of ``old'' and ``new''  unrenormalized and 
UV integrands proves unambiguously
that they have nevertheless the same values within the precision of numerical evaluation. (Typically
more than 10 digits in 14 digits precision.)

\subsection{IR subtraction terms of $M_{16}$}

The ``spot-check'' method, however, is not directly applicable for comparison of 
``old'' and ``new'' IR subtraction terms,
because they are algebraically different by construction.
For this purpose we need to understand precisely
the analytic structure of the IR subtraction terms in both ``old'' and
``new'' methods. Thus we follow an alternative approach by which we can 
identify how they differ from each other in the analytic form.
 
In the ``old'' method, the IR singularities of $M_{16}$ are isolated by the 
$\mathfrak{I}_\mathcal{R}$-operations, where $\mathcal{R}$ = $\{1,3,7,a\}, \{1,2,3,7,a,b\}, $ or $\{1,3,4,5,6,7,a,c,d\}$. 
Thus, the formal expression 
of the IR-free contribution of $M_{16}$ is given by
\cite{Cvitanovic:1974sv} 
\begin{equation}
\overline{\Delta M_{16}} \equiv (1-\mathfrak{I}_{137})(1-\mathfrak{I}_{134567})(1-\mathfrak{I}_{1237}) \underline{M}_{16},
\label{irfree}
\end{equation}
where $\underline{M}_{16}$ is the UV-finite amplitude obtained 
by the {\it K}-operations.
The product $\mathfrak{I}_{134567} \mathfrak{I}_{1237}$ gives no contribution, 
since they overlap each other
and cannot take these IR limits simultaneously. 
 
Following the ``old'' prescription  in Ref. \cite{Cvitanovic:1974sv}, 
individual IR subtraction terms of Eq. (\ref{irfree}) can be written as follows:
\begin{align}
& -\mathfrak{I}_{134567}\underline{M}_{16} = -I_{6c1} M_2 ,   
        \label{I134567M16} \\
& -\mathfrak{I}_{1237}\underline{M}_{16} = -I_{4c}\Delta M_{4a} 
                              - M_{4b(1^\star)}[I] \Delta \delta m_{4a}, 
      \label{I1237M16}  \\
& -\mathfrak{I}_{137}(1-\mathfrak{I}_{134567})(1-\mathfrak{I}_{1237})\underline{M}_{16}
                            = + I_{2^\star} \Delta \delta m_{4a} M_2 
      \label{I137XXM16} ~.
\end{align}
$I_{6c1}$ and $I_{4c}$ are the no-contraction terms of the vertex renormalization 
constant $L_{6c1}$ and $L_{4c}$, respectively.  $M_{4b(1^\star)}$ is the magnetic
moment amplitude, which is obtained from the fourth-order diagram $M_{4b}$ with
the two-point vertex inserted into the fermion line $1$. Its argument $[I]$ implies 
that the numerator of the no-contraction term of $M_{4b(1^\star)}$ is replaced by that 
of the vertex renormalization constant $L_{4s}$, 
while discarding the contraction terms.

If there were only
logarithmic IR divergence, $\overline{\Delta M}_{16}$ defined in (\ref{irfree}) would be
IR-finite, but it is not.  
The problem here is that $I_{6c1}$ and $M_{4b(1^\star)}[I]$ in Eqs. (\ref{I134567M16})
and (\ref{I1237M16}) have linear IR divergence. 
The $\mathfrak{I}$-operation prescription is constructed so that it only deals with 
the leading  IR singularity.  
Of all  eighth-order $q$-type diagrams, 
$M_{16}$ and $M_{18}$ have the linear IR divergence. Since these are the only cases, 
we chose to deal with their next-to-leading-order IR divergences
by an $ad~ hoc$ method rather than
constructing a general rule.

In the $\mathfrak{I}_{134567}$ limit of Eq. (\ref{I134567M16}), 
the diagram $M_{16}$ decouples into
the vertex diagram $L_{6c1}$ which consists of lepton lines $1,3,4,5,6,7$ and 
photon lines $a,c,d$ and the magnetic moment part $M_2$ which consists of lepton line 
$2$ and photon line $b$.  All IR singularities originate from 
the vertex diagram $L_{6c1}$. The no-contraction term $L_{6c1}[F_0]$, 
namely $I_{6c1}$, includes the leading linear 
IR singularity as well as the next-to-leading logarithmic singularity.

The logarithmic IR singularity also arises from 
the $\mathfrak{I}_{137}$-limit of the one-contraction term $L_{6c1}[F_1]$. 
To deal with this, we constructed the quantity $J_{6c}^{\rm unrenorm.} $ 
in which the numerator is 
the $\mathfrak{I}_{137}$ limit
of $L_{6c1}[F_1]$, but the denominator $V$ and $U$ is the same as $L_{6c1}$.
The UV divergences of $J_{6c}^{\rm unrenorm.}$ are removed by the ${\mathbb K}_{456}, {\mathbb K}_{56}, {\mathbb K}_{45}$-operations:
\begin{align}
  \frac{1}{2} J_{6c}& = (1-{\mathbb K}_{456})(1-{\mathbb K}_{45})(1-{\mathbb K}_{56})
 \left ( -\frac{1}{32}  \int (dz)_{G/S} \frac{f_1}{U^3 V^2} \right ), \nonumber \\
 & f_1 = -16 [ B_{45}(2-A_6) + 2 B_{46}(1-2A_5) + B_{56} ( 2-A_4) ]~~,
\end{align}
where $S=\{2,b\}$.

Next we consider the  $\mathfrak{I}_{1237}$-limit of Eq. (\ref{I1237M16}). 
In this limit, the
self-energy-like subdiagram consisting of lepton lines $4,5,6$ and 
photon lines $c,d$ plays dual roles. When this fourth-order
self-energy-like subdiagram behaves as a magnetic moment $M_{4a}$, the residual
diagram resembles a vertex diagram $L_{4s}$. Its singularity is 
logarithmic, so $\mathfrak{I}_{1237}$-operation properly works for this part.

The problem arises when the self-energy-like subdiagram acts as the self-mass 
$\delta m_{4a}$, which is the second one of its dual roles.
The residual
diagram is the magnetic moment amplitude with one two-point vertex
inserted, namely, $M_{4b(1^\star)}$. The power counting shows that it has a linear
IR singularity. Thus, the $\mathfrak{I}$-operation is not enough to remove
the IR singularity of this term. $M_{4b(1^\star)}[I]$ is not sufficient
to remove all IR singularities arising in the $\mathfrak{I}_{1237}$ limit.
The IR structure of $M_{4b(1^\star)}$ is more closely
scrutinized in the next subsection, where 
$I_{4b(1^\star)}$ is constructed to
include both linear and logarithmic IR singularities of the magnetic moment
amplitude $M_{4b(1^\star)}$.
(A similar subtraction method works also for $M_{18}$.)

Taking these considerations into account, we replace
the IR subtraction terms of $M_{16}$ in the ``old'' method listed in Eqs. (\ref{I134567M16}) and (\ref{I1237M16}) with \cite{Kinoshita:1981wm, Kinoshita:1990}
\begin{align}
&-\mathfrak{I}_{134567}^{\prime} \underline{M}_{16} = -(I_{6c1} +\frac{1}{2} J_{6c})M_2  
\label{oldI134567M16} \\
&-\mathfrak{I}_{1237}^{\prime} \underline{M}_{16} = -I_{4c}\Delta M_{4a}  - I_{4b(1^\star)} \Delta \delta m_{4a}~
\label{oldI1237M16} 
\end{align}
which are more convenient for comparison with the ``new" approach.
Note that Eq.(\ref{I137XXM16}) is unchanged.

Now, let us look at  the ``new'' approach.  
All IR singularities, both linear and logarithmic,  
are  subtracted by using  the general  rule applicable to any order of the
perturbation theory. 
The $R$- and $I$-subtractions, and their combinations determine the
IR subtraction terms of $M_{16}$ as follows: 
\begin{eqnarray}
&&-{\mathbb I}_{2}{\underline M}_{16} = - L_{6c1}^{\rm R}~ M_2~, \nonumber \\
&&-{\mathbb I}_{456}{ \underline M}_{16} = -L_{4s}^{\rm R}~ {\underline M}_{4a},
    \nonumber \\
&&-{\mathbb R}_{456}{ \underline M}_{16} = 
       -\delta m_{4a}^{\rm R}~ { \underline M}_{4b(1^\star)}~, \nonumber \\
&&+{\mathbb I}_{2}{\mathbb R}_{456}~{\underline M}_{16} =
                          +L_{2^\star}^{\rm R}~ \delta m_{4a}^{\rm R} ~M_2~~.
\label{newIRtermsofM16}
\end{eqnarray}
By definition given in Eqs. (\ref{dmR}) and (\ref{LR}),  
the residual quantities are explicitly given by 
\begin{align}
L_{6c1}^{\rm R} &= (1-{\mathbb K}_{456})(1-{\mathbb K}_{45})(1-{\mathbb K}_{56})
                                      (L_{6c1} -L_{6c1}^{UV}) \nonumber \\
                &=
                         L_{6c1} -L_{6c1}^{UV} 
                         -(B_{4a}^{UV} \widetilde{L}_2^{'} + \delta m_{4a}^{UV} L_{2^\star})
                         -2 L_2^{UV}\widetilde{L}_{4s}
                         + 2 L_2^{UV} (\delta m_2 L_{2^\star} + B_2^{UV} \widetilde{L}_2^{'} )
                             , \nonumber \\
L_{4s}^{\rm R} &        = (1-{\mathbb K}_{2}) (L_{4s}-L_{4s}^{UV}) 
                         = L_{4s}-L_{4s}^{UV}-(B_2^{UV}\widetilde{L}_2^{'} + \delta m_2 L_{2^\star} ), \nonumber \\ 
\underline{M}_{4a}&         =(1-{\mathbb K}_{45})(1-{\mathbb K}_{56})M_{4a} 
                         = M_{4a}- 2 L_2^{UV} M_2 ~,\nonumber \\ 
 \delta m_{4a}^{\rm R} &= \delta m_{4a}-\delta  m_{4a}^{UV} \nonumber \\
\underline{M}_{4b(1^\star)} &  = (1-{\mathbb K}_2) M_{4b(1^\star)} 
              = M_{4b(1^\star)} - (B_2^{UV} M_{2^\star}+\delta m_2 M_{2^{\star \star}})~,
\end{align}
where $\widetilde{L} = L - L^{UV}$. (See below Eq. (\ref{dmR}).)
In terms of the ``old'' expression of the unrenormalized amplitude 
and renormalization constants, the residual quantities
are related to   the 
IR divergent and finite pieces of the ``old'' method by the following relations: 
\begin{align}
&L_{6c1}^{\rm R} = I_{6c1} + \frac{1}{2} J_{6c} + \Delta L_{6c1}~, \nonumber \\
&L_{4s}^{\rm R} = I_{4s} + \Delta L_{4s}, \nonumber \\ 
&\underline{M}_{4a}=\Delta M_{4a},\nonumber \\
& \delta m_{4a}^{\rm R} =  \Delta \delta m_{4a}, \nonumber \\
& \underline{M}_{4b(1^\star )} = I_{4b(1^\star )} +\Delta M_{4b(1^\star )}~~. 
\end{align}

We are now ready to compare the IR subtraction terms of ``old'' and ``new'' method side by side:
\begin{equation}
\begin{array}{lllll}
   &~~~~~~&\multicolumn{1}{c}{\rm old}&~~~~~~~&\multicolumn{1}{c}{\rm new}   \\
(a)&&  - (I_{6c1}+\frac{1}{2}J_{6c} ) M_2       && - (I_{6c1}+\frac{1}{2}J_{6c}+\Delta L_{6c1}) M_2     \\
(b)&&  - I_{4s} \Delta M_{4a}                  && - (I_{4s}+\Delta L_{4s}) \Delta M_{4a} \\
(c)&&  - I_{4b(1^\star)} \Delta \delta m_{4a}       &&- (I_{4b(1^\star)}+\Delta M_{4b(1^\star)})\Delta \delta m_{4a} \\
(d)&&  + I_{2^\star} \Delta \delta m_{4a} M_2       && + I_{2^\star} \Delta \delta m_{4a} M_2 ~.
\end{array}
\label{comparIRofM16}
\end{equation}
Actually, instead of examining the IR subtraction terms
of the ``old'' method themselves 
we reconstructed them from the ``new''  programs 
by dropping 
finite terms (eg. $\Delta L_{4s}$ ) from the ``Residual'' term (eg. $L_{4s}^{\rm R}$),
and compared them with the terms in the ``old'' programs by the spot-check method.     
To obtain $I_{6c1}$ and $I_{4s}$, we only need to comment out 
the contraction terms (equivalently drop the terms proportional to $B_{ij}$) 
of $L_{6c1}^{\rm R}$
and $L_{4s}^{\rm R}$ of the ``new'' programs.
In this way,
we found that the {\it reconstructed} IR subtraction terms from the ``new" programs
are identical with ``old'' ones for (a), (b), and (d). 

However, 
the ``old'' IR subtraction term (c) $I_{4b(1^\star)} \Delta \delta m_{4a}$ 
cannot be constructed by such a simple recipe 
from the ``new'' programs generated by {\sc gencode}{\it N}.
Dropping the finite terms in $\underline{M}_{4b(1^\star)} \Delta  \delta m_{4a}$ is not 
enough to reproduce $I_{4b(1^\star)} \Delta \delta m_{4a}$.
Therefore, we reconstructed the subtraction term $I_{4b(1^\star)} \Delta \delta 
m_{4a}$ from the scratch
using the definitions of the fourth-order quantities $I_{4b(1^\star)}$ and 
$\Delta \delta m_{4a}$.
Then, the result is compared with the integrand in the ``old" program of 
$\Delta M_{16}^{\rm old}$.

\subsection{$I_{4b(1^\star)}\Delta \delta m_{4a}$ by the ``old"  $\mathfrak{I}$-operation}

Let us first explain how $I_{4b(1^\star)} \Delta \delta m_{4a}$ is obtained 
in the ``old" program.
In the 
old approach, the IR subtraction term $I_{4b(1^\star)} \Delta \delta m_{4a}$
originates from the $\mathfrak{I}_{1237}$-operation. 
In addition to this term, $\mathfrak{I}_{1237}$ 
operation yields the term $ I_{4s} \Delta M_{4a}$.   

The IR-limit associated with 
the operator $\mathfrak{I}_{1237}$ is given by
\begin{equation}
  {z_a+z_b = 1-\mathcal{O}(\delta)}, 
  \quad
  {z_1, z_2, z_3, z_7 = \mathcal{O}(\delta)}, 
  \quad
  {z_4, z_5, z_6, z_c, z_d = \mathcal{O}(\epsilon)}, 
  \quad
  \epsilon \sim \delta^2,
  \quad
  \delta \rightarrow 0.
\end{equation}
In the neighborhood of this limit we have 
$A_1 = 1 - \mathcal{O} (\delta)$, 
$A_2 = 1 - \mathcal{O}(\delta)$, 
and $V={\mathcal{O}} (\delta^2)$. 
As is discussed in the previous section, the result of $\mathfrak{I}_{1237}$-operation 
includes both  linear and logarithmic IR divergences. 
In particular, the linear divergence is associated 
with $\mathfrak{I}_{137} \mathfrak{I}_{1237}$ limit.
If we apply the 
$\mathfrak{I}_{137} \mathfrak{I}_{1237}$ operation, however, 
it subtracts the linear divergence correctly, but not the logarithmic
divergence. 
Thus, we chose an {\it ad hoc} method
in which the piece including linear divergence is separated out 
from the result of $\mathfrak{I}_{1237}$-operation and put aside for a while. 
The remainder that contains only logarithmic divergence is named $f_k$. 
The linear divergent piece is redefined so that 
it is defined on the subdiagram $\{ 1, 2, 3, 7, a, b \}$ without decomposing 
it into two subdiagrams $\{ 2, b \}$ and $\{ 1, 3, 7, a \}$, which occurs
in the na\"ive $\mathfrak{I}_{137} \mathfrak{I}_{1237}$ limit. 
This term is named $f_l$. 
The explicit forms of $f_k$ and $f_l$ in the old FORTRAN program of $\Delta M_{16}$ read: 
\begin{equation}
  f_k 
  = 
  \int(dz)_\mathcal{G}\, 
  \frac{1}{4 U^2 } L_{4s}[F_0] 
  \left(  
    \frac{ E_0 + C_0 + \delta m_{4a}[f_0] + g_\mathcal{S} F_1 + Y_1}{ V^3 } 
    + \frac{ 3(g_\mathcal{S}-V_t) \delta m_{4a}[f_0] + Y_0 }{ V^4 } 
  \right) 
  \,, 
\end{equation} 
\begin{equation}
  f_l 
  = 
  -\frac{3}{2} 
  \int (dz)_\mathcal{G}\,
  \frac{ \delta m_{4a}[f_0] }{ U^2V^4 } 
  z_2(1-A_2)^2 (-1+6A_1-3A_1^2+2A_1^3) 
  \,,
\end{equation}
%
%
where
\begin{gather}
  L_{4s}[F_0]=(4A_1-2A_2)(1-A_1+A_1^2)+(-2+A_1A_2)(1-4A_1+A_1^2) \nonumber \\
  F_1=B_{45}(2-A_6)/U+2B_{46}(1-2A_5)/U+B_{56}(2-A_4)/U \nonumber \\
  \begin{aligned}
  Y_1 =
    & -z_4(B_{45}(1-A_6)+B_{46}+B_{56}A_4)/U \nonumber \\
    & +z_5(B_{45}(1-A_6)-4B_{46}A_5+B_{56}(1-A_4))/U \nonumber \\
    & -z_6(B_{45}A_6+B_{46}+B_{56}(1-A_4))/U
  \end{aligned}
\nonumber\\ 
  V_t=z_{137}(1-A_1)-z_2(1-A_2) \nonumber \\
  E_0=2A_4A_5A_6-A_4A_5-A_4A_6-A_5A_6 \nonumber \\
  C_0= -3z_cz_d/U_\mathcal{S} \nonumber 
\end{gather}
\vspace{-0.5cm}
\begin{gather}
  \delta m_{4a}[f_0]=E_0+1-2A_5 \nonumber \\
  \begin{aligned}
  Y_0 = 
    & z_4(-A_4+A_5+A_6+A_4A_5+A_4A_6-A_5A_6) \nonumber \\
    & +z_5(1-A_4A_5+A_4A_6-A_5A_6+2A_4A_5A_6) \nonumber \\
    & +z_6(A_4+A_5-A_6-A_4A_5+A_4A_6+A_5A_6)
  \end{aligned}
\nonumber \\
  g_\mathcal{S} = z_4 A_4+z_5 A_5+z_6 A_6
  \,.
\label{miscdef}
\end{gather}
The building blocks $U, V, A_i, B_{ij}$ of the above integrands are 
obtained from those for $M_{16}$ 
by taking the IR-limit associated with the $\mathfrak{I}_{1237}$-operation. 
 Recall that in the IR-limit 
the subdiagram $\mathcal{S}$ consisting  of  the fermion lines $4,5,6$ and
photon lines $c,d$, and the reduced diagram $\mathcal{G}/\mathcal{S}$ 
consisting of the fermion lines $1,\,2,\,3,\,7$ 
and the photon lines $a,b$
decouple from each other.  
 Thus, the building blocks are actually the same as those 
obtained by taking the UV-limit associated 
with the ${\mathbb K}_{456}$ limit. 
 Their explicit forms are 
\begin{gather}
  U = U_\mathcal{S}~U_{\mathcal{G}/\mathcal{S}},
\nonumber \\
  U_\mathcal{S} = z_{46cd} z_{5} + z_{4c}z_{6d},
  \qquad
  U_{\mathcal{G}/\mathcal{S}} = z_{137a} z_{2b}+z_2z_b
\nonumber \\
  V = V_\mathcal{S} + V_{\mathcal{G}/\mathcal{S}},
\nonumber \\
  V_\mathcal{S} = z_{456}-z_4 A_4-z_5 A_5 - z_6 A_6 + \lambda^2 z_{cd},
  \qquad
  V_{\mathcal{G}/\mathcal{S}} 
  = z_{1237} - z_{137}A_1 - z_2 A_2 + \lambda^2 z_{ab},
\nonumber \\
  A_i = 1 - \sum_{j=1}^7 z_i B_{ij}/U,
  \quad
  i=1,\cdots 7,
\nonumber \\
  B_{11}=B_{13}=B_{17}=B_{33}=B_{37}=B_{37}= z_{2b} U_\mathcal{S},
\nonumber \\
  B_{45} = z_{6d}~U_{\mathcal{G}/\mathcal{S}},
  \qquad
  B_{46}= -z_5~U_{\mathcal{G}/\mathcal{S}},
  \qquad
  B_{56}=z_{4c}~U_{\mathcal{G}/\mathcal{S}},
\nonumber \\
  B_{ij} =0 
  \quad
  \text{for $i \in \mathcal{S}$ and $j \in \mathcal{G}/\mathcal{S}$.}
\end{gather}
In the above $z_{i_1 i_2 \cdots}$ stands for $z_{i_1}+z_{i_2}+\cdots$
and the electron mass is taken as a unit of mass scale 
(i.e., 1) and the photon mass is $\lambda$. 
 In the leading order of the $\mathfrak{I}_{1237}$ limit 
$L_{4s} [F_0]$ tends to $4$.
 The actual form of $L_{4s} [F_0]$ in Eq.~(\ref{miscdef}) was chosen 
so that  the integral $f_k$ decouples into  {\it known} lower-order
quantities. 
 This difference for $L_{4s}[F_0]$ is IR-finite.
 Note that $\delta m_{4a}[f_0]$ is related to the integrand of
$\Delta \delta m_{4a}$, namely, the UV- and IR-finite
part of the mass renormalization constant $\delta m_{4a}$ 
given later in Eq.~(\ref{dm4a}).

 In order to clarify the structure of $f_k$ and $f_l$, let us split 
$f_k$ into the three parts $f_{k1}, f_{k2}, f_{k3}$, 
\begin{align}
  f_{k1} 
  &=
  \int(dz)_\mathcal{G} \frac{ 1 }{ 4 U^2 } L_{4s}[F_0] 
  \left(  
    \frac{ E_0 + C_0 + g_{\mathcal{S}} F_1 + Y_1 }{ V^3 } 
    + \frac{ 3g_\mathcal{S} \delta m_{4a}[f_0] + Y_0 }{ V^4} 
  \right),
\nonumber \\                              
  f_{k2} 
  &=
  \int(dz)_\mathcal{G} \frac{ 1 }{ 4 U^2 } L_{4s}[F_0] \delta m_{4a}[f_0] 
  \left(
    \frac{ 1 }{ V^3 } - \frac{ 3z_{137}(1-A_1) }{ V^4 }
  \right) ,
\nonumber \\                              
  f_{k3}
  &=
  \int(dz)_\mathcal{G} \frac{ 1 }{ 4 U^2 } L_{4s}[F_0] \delta m_{4a}[f_0] 
  \frac{ 3 z_2 (1-A_2) }{ V^4 },
\end{align}
and $f_l$ into the two parts $f_{l1}, f_{l2}$ 
\begin{align}
  f_{l1} 
  &= 
  \frac{ 3 }{ 4 } \int (dz)_\mathcal{G} 
  \frac{ \delta m_{4a}[f_0] }{ U^2V^4 } 
  z_2 A_2 (1-A_2) (-1+6A_1-3A_1^2+2A_1^3) \,,
\nonumber \\                              
  f_{l2} 
  &= 
  - \frac{ 3 }{ 4 } \int (dz)_\mathcal{G} 
  \frac{ \delta m_{4a}[f_0] }{ U^2V^4 } 
  z_2 (1-A_2) (2-A_2) (-1+6A_1-3A_1^2+2A_1^3) \,.
\end{align}

The integrand $f_{k1}$ was compared with the integrand
$\mathbb{I}_{456} M_{16}$ generated by {\sc gencode}{\it N}
by the spot-check method.
We confirmed that the integral $f_{k1}$ minus 
its ${\mathbb K}_2$ limit, $f_{kv}$, 
is equal to $I_{4s} \Delta M_{4a}$, which is listed as (b) of
(\ref{comparIRofM16}). 
It turns out that 
$f_{k2}+f_{l1}$ is equal to $I_{4b(1^\star)} \Delta \delta m_{4a}$, 
which is reconstructed from the lower order quantities in the next subsection.
Thus, the difference between the ``old" calculation and the reconstructed
one is  confined in $f_{k3}+f_{l2}$.
It is IR-finite but contributes a nonzero value to $\Delta M_{16}^{\rm old}$.

\subsection{$I_{4b(1^\star)}\Delta \delta m_{4a}$ reconstruction from the 
lower-order quantities}
\label{reconst}

 In order to understand where this difference $f_{k3}+f_{l2}$ 
came from,
let us examine IR divergence structure of mass-inserted magnetic moment 
amplitude $M_{4b(1^\star)}$ in the ``old" approach.
$I_{4b(1^\star)}$ is defined from $M_{4b(1^\star)}$ 
as follows\cite{Kinoshita:1981wm, Kinoshita:1990}:
\begin{align}
  M_{4b(1^\star)} 
  & =
  (1 - {\mathbb K}_2) (1-{\mathbb I}_{all}) (M_{4b(1^\star)} - M_{4b(1^\star)}[f] ) 
  + {\mathbb K}_2 M_{4b(1^\star)} 
\nonumber \\
  & \quad 
  + {\mathbb I}_{all}(1-{\mathbb K}_2)M_{4b(1^\star)}[N_0 + Z_0 - f, E_0 + C_0] 
  + M_{4b(1^\star)}[f] 
\nonumber \\
  &\equiv 
  \Delta M_{4b(1^\star)} 
  + (\delta m_2 M_{2^{\star \star}}+ B_2^{UV} M_{2^\star}) 
  + I_{4b(1^\star)}
  \,.
\end{align}
The two terms in the second line of the r.h.s. define   $I_{4b(1^\star)}$:  
\begin{equation}
  I_{4b(1^\star)} 
  \equiv 
  {\mathbb I}_{all}(1-{\mathbb K}_2) M_{4b(1^\star)}[N_0 + Z_0 - f, E_0 + C_0] 
  + M_{4b(1^\star)}[f],
\label{I4b1s}
\end{equation}
where the function $f$  is introduced 
in the first term
in an {\it ad hoc} manner to subtract out the 
linear IR divergence
coming from $N_0 + Z_0$.
The linear IR divergence is confined to the second term $M_{4b(1^\star)}[f]$,
which has the form
\begin{gather}
  M_{4b(1^\star)}[f] = 
  - \frac{ 1 }{ 8 } \int(dy)_{\mathcal{G}/\mathcal{S}} 
  \frac{ f }{ U^2_{\mathcal{G}/\mathcal{S}} V^3_{\mathcal{G}/\mathcal{S}} }
  \,, 
\nonumber \\
  f =
  -8 y_2 A_2 (1-A_2) (-1 +6 A_1 - 3 A_1^2 + 2 A_1^3 )~.
\label{def_f}
\end{gather}
The explicit form of the first term of Eq.(\ref{I4b1s}) is
\begin{equation}
  {\mathbb I}_{all} M_{4b(1^\star)}[N_0 + Z_0-f, E_0+C_0] 
  = 
  \lim_{\lambda \rightarrow 0} \int (dy)_{\mathcal{G}/\mathcal{S}} 
  \frac{ L_{4s}[F_0] }{ U^2_{\mathcal{G}/\mathcal{S}} } 
  \left(
    \frac{ 1 }{ 2 V_{\mathcal{G}/\mathcal{S}}^2 } 
    - \frac{ y_{137} (1-A_1) }{ V^3_{\mathcal{G}/\mathcal{S}} } 
  \right),
\end{equation}
but i
where
\begin{gather}
  V_{\mathcal{G}/\mathcal{S}}
  =
  y_{137}(1-A_1)+y_2(1-A_2) + \lambda^2 (y_a + y_b),
\nonumber \\
  (dy)_{\mathcal{G}/\mathcal{S}} 
  =
  dy_1dy_2dy_3 dy_7 dy_a dy_b\,\delta(1-y_{1237}-y_{ab}). 
\end{gather}
The finite part of the mass renormalization constant $\Delta \delta m_{4a}$ is
\begin{equation}
  \Delta \delta m_{4a} = 
  \frac{ 1 }{ 4 } \int (dy)_\mathcal{S} 
  \frac{ \delta m_{4a}[f_0] }{ U_\mathcal{S}^2 V_\mathcal{S} },
\label{dm4a}
\end{equation}
where $ \delta m_{4a}[f_0]$ is expressed in the same form in Eq. (\ref
{miscdef}) and 
\begin{equation}
  (dy)_\mathcal{S} = 
  dy_4 dy_5 dy_6 dy_c dy_d\,\delta(1-y_{456}-y_{cd}).
\end{equation}
Using the identity
\begin{equation}
  \frac{ 1 }{ V_\mathcal{S}^m V_{\mathcal{G}/\mathcal{S}}^n } 
  = 
  \frac{ \Gamma (m+ n) }{ \Gamma (m) \Gamma (n) }
  \int_0^1 dt \int_0^1 ds\, 
  \delta(1-t-s)\,
  \frac{ t^{m-1} s^{n-1} }
       { ( t ~V_\mathcal{S} ~ +~ s~V_{\mathcal{G}/\mathcal{S}}  )^{m+n} },
\end{equation}
we can express the product of $\Delta \delta m_{4a}$ and 
the first term of $I_{4b(1^\star)}$ defined in Eq.~(\ref{I4b1s}) in the
same Feynman parameter space 
as that for the original amplitude $M_{16}$ 
\begin{align}
  f_{k2}^{\rm rc} 
  & \equiv
  \Delta \delta m_{4a} 
  \times 
  {\mathbb I}_{all} M_{4b(1^\star)}[N_0 + Z_0 - f, E_0 + C_0]
\nonumber \\     
  & = 
  \int (dz)_\mathcal{G}\,\frac{ L_{4s}[F_0] \delta m_{4a}[f_0] }{ 4U^2 } 
  \left(
    \frac{ 1 }{ V^3 } 
    - \frac{ 3z_{137} (1-A_1) }{ V^4 } 
  \right),
\end{align}
which is identical with $f_{k2}$.
Similarly, the contribution of the product of $\Delta \delta m_{4a}$  
$M_{4b(1^\star)}[f]$  to $M_{16}$ is
\begin{align}
  f_{l1}^{\rm rc }
  & \equiv 
  \Delta \delta m_{4a} M_{4b(1^\star)}[f] 
\nonumber \\
  & =
  \frac{ 3 }{ 4 } \int (dz)_\mathcal{G}\,
  \frac{ \delta m_{4a}[f_0] }{ U^2 V^4} 
  z_2 A_2 (1-A_2) (-1+6A_1-3A_1^2+2A_1^3)
  \,, 
\end{align}
which is identical with $f_{l1}$.
%
%

Therefore we find that 
the combination, $-f_{k3}-f_{l2}$, is extra 
in the  ``old'' $\Delta M_{16}^{\rm old}$ 
so that the correction term
\begin{align}
  \Delta M_{16}^{\rm add} 
  & \equiv 
  2(f_{k3}+f_{l2}) 
\nonumber \\
  & = 
  -2 \times \frac{ 9 }{ 4 } \int (dz)_\mathcal{G} 
  \frac{ \delta m_{4a} [f_0] }{ U^2 V^4 }
  z_2 A_2 (1-A_1)^3 (1-A_2) 
  \,, 
\end{align}
where the overall factor 2 comes from time-reversal diagram, 
must be added to $\Delta M_{16}^{\rm old}$. 
Evaluating it numerically, we obtain   
$\Delta M_{16}^{\rm add}= 0.029~437~8~(98)$, which is smaller than the current 
uncertainty 
of value B for $M_{16}$ in Table \ref{Table:M16M30} 
and cannot be detected by the direct comparison of value A and value B
until the latter is evaluated more precisely.

The difference between $\Delta M_{18}^{\rm new}$ and $\Delta M_{18}^{\rm old}$
can be analyzed in the same manner. It is found that the difference is 
numerically not small for $M_{18}$:
\begin{align}
  \Delta M_{18}^{\rm add} 
  & \equiv 
  2 (1-{\mathbb K}_5) (f_{k3}+f_{l2})
\nonumber \\
  & = 
  -2 \times \frac{ 9 }{ 4 } \int (dz)_\mathcal{G} 
  (1- {\mathbb K}_5)
  \left\{
    \frac{ \delta m_{4b}[f_0] }{ U^2 V^4 }
    z_2 A_2 (1-A_1)^3 (1-A_2) 
  \right\}
\nonumber \\
  & = -0.215~542~(19)
  \,,
\end{align}
where $A_i,U,$ and $V$ are defined in the ${\mathfrak I}_{1237}$ limit of $M_{18}$.
Of course their explicit forms are different from those of $M_{16}$.
If we add  $\Delta M_{18}^{\rm add}$
 to $\Delta M_{18}^{\rm old}$, the value B for $M_{18}$ 
in Table \ref{Table:M16M30}  
becomes $16.974~(21)$ and the difference between value A and value B
is reduced to $-0.006~(21)$, which is consistent with zero within the precision 
of numerical calculation.

\section{conclusion}

The results described in this paper  are summarized as follows:
\begin{enumerate}[1)]
\item  There was an inconsistency between the ``old'' integrals $\Delta M_{16}^{\rm old}$ 
and $\Delta M_{18}^{\rm old}$ and their residual renormalization terms.
This inconsistency is resolved in this paper.

\item Other 45 integrals of Group V of the ``old'' calculation 
are in good agreement with 
the ``new'' ones.

\item Programs generated by {\sc gencode}{\it N} have no error for $N=8$. 
Namely, the automation scheme has cleared the eighth-order test without difficulty.
\end{enumerate}

The separation of the IR divergent 
and finite pieces in a given amplitude can be made arbitrarily. 
There is no overriding rule that dictates how to carry out such a separation.
We only have to keep a record of what is subtracted as 
an IR subtraction term. 
All IR subtraction terms are summed up in the end and 
the arbitrariness in the choice of IR divergent part
cancel out, leaving a finite
contribution as a part of the residual renormalization.

The important point is that the IR subtraction term prepared for the
numerical calculation and the one used to calculate the residual renormalization
must be the same. What  we found is that $I_{4b(1^\star)}$ used in the
numerical calculation  of $M_{16}$ and $M_{18}$
and $I_{4b(1^\star)}$ for the residual renormalization
constant $\Delta M_{4b(1^\star)}$ had different forms in the 
FORTRAN programs of the ``old" calculation.
This is the reason why $M_{16}$ and $M_{18}$ had IR-finite but redundant 
contributions.

The development of automatic code generator \cite{Aoyama:2005kf,Aoyama:2007IR}
was crucial in enabling us to discover
the existence of extra IR subtraction terms
in $M_{16}$ and $M_{18}$ on short notice.
Adding the correction terms $\Delta M_{16}^{\rm add}$
and $\Delta M_{18}^{\rm add}$ to
the ``old'' value, we find the entire contribution of Group V to be 
\begin{equation}
  A_{1}^{(8)}({\rm Group V}) = -2.179~16~(343),
\label{revisedA8V}
\end{equation}
which is in good agreement with the {\it still preliminary} value 
\begin{equation}
  A_{1}^{(8)}({\rm Group V}) = -2.219~(53),
\label{gencodeNA8V}
\end{equation}
obtained by the ``new" code generated by {\sc gencode}{\it N}.

Due to the different forms of IR subtraction terms, the forms of the residual 
renormalization are also different in the ``old" and ``new" calculations.
The residual renormalization terms 
in the ``old" IR procedure are given by 
\cite{Kinoshita:1981wm,Kinoshita:1990,Kinoshita:2005zr}
\begin{align}
  A_{1}^{(8)}({\rm Group V})^{\rm old}
  & =
  \Delta M^{(8) {\rm old}} 
  - 5 \Delta M^{(6) {\rm old}} \Delta B_2
\nonumber \\
  &
  - \Delta M^{(4)} \{ 4 \Delta L^{(4)} + 3 \Delta B^{(4)} - 9 (\Delta B_2)^2 \}
\nonumber \\
  &
  - M_2 \{ 2 \Delta L^{(6)} + \Delta B^{(6)} 
       - (10 \Delta L^{(4)} + 6 \Delta B^{(4)})\Delta B_2 + 5 (\Delta B_2)^3 \}
\nonumber \\
  &
  - \Delta M^{(4^\star)} \Delta \delta m^{(4)}
\nonumber \\
  &
  - (M_{2^\star}-M_{2^\star}[I]) \{ 
    \Delta \delta m^{(6)} 
    - \Delta \delta m^{(4)} ( 5 \Delta B_2+\Delta \delta m_{2^\star} ) \} 
\nonumber \\
  &
  + M_2 \Delta \delta m^{(4)} ( 4 \Delta L_{2^\star} + \Delta B_{2^\star}-B_{2^\star}[I])
  \,,
\end{align}
where 
\begin{gather}
  \Delta M^{(8) {\rm old}} = \sum_{i=01}^{47} \Delta M_i^{\rm old}, 
\nonumber \\
  \Delta M^{(6) {\rm old}} 
    = \sum_{x=a}^{h} \Delta M_{6x}^{\rm old} 
\nonumber \\
  \Delta M^{(4)} = \Delta M_{4a} + \Delta M_{4b}, 
\nonumber \\
  \Delta M^{(4^\star)} = 
    2\Delta M_{4a(1^\star)} + \Delta M_{4a(2^\star)}
    + 2 \Delta M_{4b(1^\star)} + \Delta M_{4b(2^\star)} 
\nonumber \\ 
  \Delta L^{(6)} = \sum_{x=a}^{h} \sum_{i=1}^{5} 
    \eta_x \Delta L_{6x i} 
\nonumber \\
  \Delta L^{(4)} = 
    2 \Delta L_{4c} + \Delta L_{4x} + 2\Delta L_{4s} + \Delta L_{4l}, 
\nonumber \\
  \Delta B^{(6)} = \sum_{x=a}^{h} \eta_x \Delta B_{6x}, 
\nonumber \\
  \Delta B^{(4)} = \Delta B_{4a} + \Delta B_{4b}, 
\nonumber \\ 
  \Delta \delta m^{(6)} 
    = \sum_{x=a}^{h} \eta_x \Delta \delta m_{6x}, 
\nonumber \\
  \Delta \delta m^{(4)} = \Delta \delta m_{4a} + \Delta \delta m_{4b} 
\nonumber \\
  \eta_x = 
    \begin{cases}
      1 & \text{for $x =a,b,c,e,f,h$} \\
      2 & \text{for $x =d,g$.}
    \end{cases}
\end{gather}
The numerical values of the finite renormalization constants are listed in Tables \ref{Table:L6} and  \ref{Table:L4}, and also in Appendix \ref{testM4M6}.
$B_{2^\star}[I]$  is obtained from the $\mathfrak{I}_{all}$-operation of the
wave function renormalization constant $B_{2^\star}$.

The formula of the residual renormalization for the ``new" calculation is
much simpler 
than that for the ``old" one. Since the mass renormalization is completed
within the numerical calculation, the mass renormalization constant should not
appear in the residual renormalization.  The exceptions are 
the vertex and wave-function
renormalization constants that have self-energy subdiagrams.  
The mass inserted vertex (wave-function) renormalization constant 
$L_{2^\star} (B_{2^\star})$ has no overall UV divergence. As a result,
the {\it K}-operation cannot pick up 
the renormalization terms proportional to $L_{2^\star}(B_{2^\star})$. It must be 
restored in the residual renormalization
in order to carry out the complete on-shell renormalization.
The residual renormalization formula 
in the ``new" approach is given by  
\begin{align}
  A_{1}^{(8)}({\rm Group V})^{\rm new}
  & =
  \Delta M^{(8) {\rm new}} - 5 \Delta M^{(6){\rm new}} \Delta B_2
\nonumber \\
  &
  - \Delta M^{(4)} \{ 3 \Delta L^{(4)} + 3 \Delta B^{(4)} - 9 (\Delta B_2)^3 \}
\nonumber \\
  &
  - M_2 \{ \Delta LB^{(6)} - 6(\Delta L^{(4)} 
    + \Delta B^{(4)}) \Delta B_2 + 5 (\Delta B_2)^2 \}
\nonumber \\
  &
  + M_2 \Delta \delta m^{(4)} ( 4 \Delta L_{2^\star} + \Delta B_{2^\star} )
  \,,
\label{a8_Residual_formula}
\end{align}
where
\begin{gather}
  \Delta LB^{(6)} = 
    \Delta L^{(6)} + \Delta B^{(6)} + \Delta L^{(4)} \Delta B_2 
    + \Delta \delta m^{(4)} B_{2^\star}[I],
\label{LB_6} \\
  \Delta M^{(6){\rm new}} = \Delta M^{(6){\rm old}} 
    - (M_{2^\star}-M_{2^\star}[I]) \Delta \delta m^{(4)} - M_2 \Delta L^{(4)}.
\label{M6new}
\end{gather}
In Eq.~(\ref{a8_Residual_formula}) 
the vertex renormalization
constant $\Delta L^{(n)}$ and the wave function renormalization constant 
$\Delta B^{(n)}$ appear in the same weight for each order of the perturbation.
It is because we have already subtracted 
one $\Delta L^{(n)}$ as an IR subtraction term.  Calculating  the
combination $\Delta L^{(n)}
+\Delta B^{(n)}$ is much easier than calculating each of them separately.
Because of
the Ward-Takahashi-identity for 
the renormalization constants $L^{(n)} + B^{(n)} = 0$, many cancellations
occur between two terms. Thus, we introduced a combined renormalization
constant $\Delta LB^{(6)}$. Its relation to the ``old" renormalization
constants $\Delta L^{(6)}$ and $\Delta B^{(6)}$ are given in 
Eq.~(\ref{LB_6}).
More detailed definitions of $\Delta LB_{6x}$ for each diagram are
given in Appendix \ref{appendixRconstsSixthorder}.
The left-hand side of Eq. (\ref{LB_6}), 
$\Delta LB^{(6)}$,  was 
directly calculated with the programs made by the automatic 
code generator for the residual renormalization constants 
\cite{Aoyama:2007RR} and obtained as 
\begin{equation}
  \Delta LB^{(6)} 
  = \sum_{x =a}^{h} \eta_x \Delta LB_{6x}
  = 0.100~86~(77)
  \,.
\end{equation}
This result was checked by comparing
with the right-hand side of Eq. (\ref{LB_6})
calculated using the residual renormalization constant 
for the ``old" calculation.
The sixth-order magnetic moment $\Delta M^{(6) {\rm new}}$ was calculated
with the programs generated by {\sc gencode}{\it N} and given  as
\begin{equation}
  \Delta M^{(6){\rm new}}  = 0.42610~(53).
\label{M6_new_update}
\end{equation} 
See Appendix \ref{SixthOrderTest} for the detail 
of $\Delta M^{(6){\rm new}}$.

As we have shown in  the paper,  
the two results (\ref{revisedA8V}) and (\ref{gencodeNA8V})
of the eighth-order contribution from Group V diagrams are 
obtained by means of the totally independent calculations.  
Further theory corrections to the eighth-order term of the electron $g\!-\!2$ 
is very unlikely.  The new theoretical prediction should be announced
when we  complete all the tenth-order calculation.


\begin{acknowledgments} 
 This work is partly supported by JSPS
Grant-in-Aid for Scientific Research (C)19540322. 
 M.~H.'s work is also supported in part 
by JSPS and the French Ministry of Foreign Affairs 
under the Japan-France Integrated Action Program 
(SAKURA). 
 T.~K.~'s work is supported by the U.~S.~National Science Foundation
under Grant PHY-0355005. T.~K.~ also thanks for JSPS Invitation Fellowship
for Research in Japan S-07165, 2007. 
Numerical calculations were 
conducted on the RIKEN Super Combined Cluster System (RSCC). 
\end{acknowledgments}


\appendix

\input M4M6.tex
\input M8app.tex


\bibliographystyle{apsrev}
\bibliography{b}

\end{document}

%% file: M4M6.tex
%
%
%
%
%
%
%


\section{Test of gencodeN by lower-order $a_e$}
\label{testM4M6}

Although {\sc gencode}{\it N} was developed primarily to deal with  
the tenth-order $q$-type diagrams, it can be readily applied
to the calculation of fourth-, sixth-, and eighth-order
$q$-type diagrams.
 Since these lower-order terms are also known from previous works,
this serves for debugging of {\sc gencode}{\it N}.


\subsection{Fourth-order $a_e$}
\FourthFigures

 The fourth-order case is the simplest nontrivial example. 
 The $g\!-\!2$ receives correction at this order 
from four types of vertex diagrams, 
$4c$, $4x$, $4s$ and $4l$ 
shown in Fig.~\ref{4th-figures}(b). 
Following the remark in Sec.~\ref{commonstartpoint},
the sum of $g\!-\!2$ from the vertex diagrams $2 M_{4c(s)}+M_{4x(l)}$ 
is expressed as a quantity associated with a single self-energy 
diagram $M_{4a(4b)}$ in Fig.~\ref{4th-figures} (a)
via the Ward-Takahashi identity (\ref{wtidentity}).  
(The factor $2$ is assigned to the diagram 
to account for the presence of the diagram 
which is related by reversing the orientation of the lepton line.) 
{\sc gencode}{\it N} creates two program sets for $M_{4a}$ and $M_{4b}$
within a few seconds on a generic Linux PC. 
These programs give the finite amplitudes 
$\Delta M_{4a (4b)}$ 
as the sum of the unrenormalized Ward-Takahashi summed $g\!-\!2$ 
amplitudes, also denoted by $M_{4a(4b)}$, 
and necessary UV and/or IR subtraction terms. 
 The relation of $\Delta M_{4a(4b)}$ 
to $M_{4a(4b)}$ is given in Appendix \ref{appendixM4}.
It took about 10 minutes each to carry out their numerical
integration by VEGAS \cite{vegas} with ten million sampling points 
per iteration for
50 iterations on {\sl hp}'s Alpha machine. 

 The values obtained in this way are 
\begin{align}
  \Delta M_{4a} &= 0.218~78~(35) \,, 
\nonumber \\
  \Delta M_{4b} &= -0.187~73~(40) \,. 
\label{deltam4}
\end{align}
The contribution from the fourth-order  
$q$-type diagrams is expressed as 
\begin{equation}
  A_1^{(4)}(\mbox {{\rm $q$-type}})
  =
  \Delta M_{4a} + \Delta M_{4b} - \Delta B_2 M_2 \, 
\label{a4_qtype}
\end{equation}
taking the residual renormalization
into account. $M_2=1/2$ is the second-order correction 
to $g\!-\!2$ 
and  $\Delta B_2 = 3/4$  is the finite part of  
the second-order wave function renormalization constant $B_2$. 
Substituting the numerical values (\ref{deltam4}) for the formal expression 
(\ref{a4_qtype}), we obtain
\begin{equation}
  A_1^{(4)}(\mbox {{\rm $q$-type}}) 
  = -0.343~95 (53),
\end{equation}
which is in good agreement with the analytic value $-0.344~166 \cdots$ 
\cite{Petermann:1957,Sommerfield:1957}.

%
%



\subsection{Sixth-order $a_e$}
\label{SixthOrderTest}

\SixthFigures
 
 The sixth-order diagrams can be evaluated in a similar manner 
and are found to give a result 
in good agreement with the numerically 
\cite{Kinoshita:1995} and analytically \cite{Laporta:1996mq} known
values as follows. 
 Fifty vertex diagrams of the sixth-order $q$-type
diagrams are reduced to eight self-energy-like diagrams shown 
in Fig.~\ref{6th-figures} 
by means of the Ward-Takahashi identity and the time-reversal symmetry.
 It takes just one minute to create all eight FORTRAN programs 
for $M_{6x}$ 
($x = a, b, \ldots, h$) by {\sc gencode}{\it N} on {\sl hp}'s Alpha machine. 
 Numerical evaluation was carried out on RIKEN's PC-cluster system (RSCC). 
 After computation of $2$ to $6$ wall-clock hours 
with 16 Xeon-CPU's for each diagram to carry out a
VEGAS integration with one hundred million sampling points per iteration for
450 iterations, we obtained 
\begin{equation}
  \Delta M^{(6){\rm new}} 
  = \sum_{x =a}^{h} \Delta M_{6x}^{\rm new} 
  = 0.426~0 ~(11).
\label{M_6new}
\end{equation} 
%
 After continuing computation with one billion sampling points per iteration 
for 200 iterations
for each diagram, we obtain the updated result $\Delta M^{(6){\rm new}}$ 
given in Eq. (\ref{M6_new_update}).
 The contribution of $q$-type diagrams to $A_1^{(6)}$
including the residual renormalization is given by
\begin{align}
  A_1^{(6)}(\mbox{{\rm $q$-type}}) 
  & =
  \Delta M^{(6) {\rm new}} -3 \Delta M^{(4)}\Delta B_2
  + M_2 
  \left\{ 
    -\Delta B^{(4)} 
    -\Delta L^{(4)} + 2(\Delta B_2)^2 
  \right\} 
\nonumber \\
  & = 0.905~26~(53),
\label{a6_result}
\end{align}
where $\Delta B(L)^{(4)}$ is the sum of the finite parts of
the fourth-order wave function (vertex) renormalization constants.
See Appendix \ref{appendixRconstsFourthorder} for their definitions. 
The formula of the residual renormalization (\ref{a6_result}) can be
obtained by using the definitions of the finite quantities 
$\Delta M_{4a(b)}$, $\Delta M_{6x}$, etc., 
in Appendixes  \ref{appendixM} and \ref{appendixRconst}.
 The values of finite quantities such as $\Delta L_{4c} \ldots$ are 
given in Table \ref{Table:L4}.
 The various finite pieces appearing in (\ref{a6_result}) are 
\begin{gather}
  \Delta B^{(4)} 
  = \Delta B_{4a} + \Delta B_{4b} 
  =-0.437094(21)
\\
  \Delta L^{(4)} 
  = 2\Delta L_{4c} + \Delta L_{4x} + 2 \Delta L_{4s} + \Delta L_{4l} 
  = 0.465~024~(17)
\\
  \Delta M^{(4)} 
  = \Delta M_{4a} + \Delta M_{4b} 
  = 0.030~804 \cdots ~~~~~~~(\mbox{known exactly })\,.
\end{gather}
Eq.~(\ref{a6_result}) again shows  good agreement with 
the analytic result $0.904~979\cdots$ 
by Laporta and Remiddi ~\cite{Laporta:1996mq}. 
                         

%% file: M8app.tex



\section{Divergence structure of the magnetic
moments}
\label{appendixM}

We briefly summarize our notation in Appendices \ref{appendixM} and \ref{appendixRconst}.
 The relation between  the unrenormalized  amplitude $M$
and the finite amplitude $\Delta M$ of  magnetic moment 
is listed in this appendix. 

A symbol with a prefix $\Delta$ means  a finite quantity.
A renormalization constant with a superscript ``UV", $A^{\rm UV}$,  
is the leading UV-divergent term of  the  on-shell renormalization constant $A$. 
$A$ can be $L$, $B$, and $\delta m$ according to a vertex-,  wave-function ,
and  mass-renormalization constant, respectively. 
$A^{\rm UV}$ is identical with $\hat{A}$ 
in Refs. \cite{Kinoshita:1990,Kinoshita:2005zr}. 
The subtraction terms proportional to a UV-renormalization term $A^{\rm UV}$ 
are generated by the $K$-operations.  
A renormalization constant with a superscript ``R",
such as $L_{4s}^{\rm R}$, is the residual term defined in  Eqs. (\ref{dmR}) and (\ref{LR}).
$A$ of $A^{\rm R}$ must be either $L$ or $\delta m$. $B^{\rm R}$ is also 
defined accordingly, but there appears no $B^{\rm R}$ term
in the definition of a finite magnetic moment amplitude.  
The subtraction terms involving  $A^{\rm R}$  are generated 
by the $R/I$-subtraction operations.
 
The subscript of $M$ or $A$ stands for the name of a diagram.
A self-energy-like diagram of the second-, fourth-, and sixth-orders are
called 
$2$, $4a$ and $4b$, and $6a, 6b, 6c, 6d, 6e, 6f, 6g,$ and $6h$, respectively.
(See Figs. \ref{4th-figures} and \ref{6th-figures}.)
The 47 independent self-energy-like diagrams of the eighth-order are named
from $01$ to $47$.  (See Fig. \ref{fig:EighthOrderSetV}.) 
The fermion lines are always named from 1 to $n-1$ from the
left to right, where $n$ is the order of the perturbation theory.

The name of a vertex diagram is determined based on the Ward-Takahashi related
self-energy diagram. When the vertex diagram is obtained by inserting the 
external photon into 
the fermion line $i$ of a self-energy diagram $nx$, this is called $nx i$.
Thus, we have $L_{4x i},~i=1,2,3,~x=a,b$, 
and $L_{6 x i},~i=1,\cdots 5, ~x=a,\cdots h$
for the fourth-, and sixth-order vertex renormalization constants, respectively.
The vertex renormalization
constant of the second-order is named $L_2$, since there is only one vertex
diagam of the second order.
In the early article \cite{Cvitanovic:1974um}, the fourth-order vertex diagrams are
given  other names. We follow the naming system Ref.{\cite{Cvitanovic:1974um} 
in this paper. 
The correspondence between two names of the fourth-order vertex diagrams
is that $4a1=4a3=4c$, $4a2 = 4x$, $4b1=4b3=4s$, and $4b2=4l$.
(See Fig. \ref{4th-figures}).

When a two-point vertex is inserted into the fermion line $i$ of a diagram
$nx$, the resulting diagram is called $nx(i^\star)$. Namely, $\star$ indicates
the two-point vertex.

The primed quantity, for example  $L_{4x(i^\prime )}$,   
is the derivative amplitude
obtained by applying $-z_i \frac{\partial}{\partial z_i} $ 
operation on the integrand, where $z_i$ is
the Feynman parameter assigned to the fermion line $i$. Note
that  $L_{4x(i^\prime)}$ is equal to $L_{4x}$, but its UV-divergent
part $L_{4x(i^\prime)}^{\rm UV}$ is not equal to $L_{4x}^{\rm UV}$.
Since second order quantities such as $B_2$ and $\delta m_2$
have only one electron line,
it is not really necessary to distinguish different electron lines.
We therefore use somewhat sloppy notations $B_{2^{\star}}$ and $B_{2^{\prime}}$ 
instead of $B_{2(1^{\star})}$ and $B_{2(1^{\prime})}$.

For $L_2$, which contains electron lines 1 and 2, it is sometimes
necessary to distinguish lines in which insertion is made.
$L_{2^{\star \star \dagger}}$  implies that two two-point vertices 
are inserted into the fermion 
line 1 of $L_{2}$, while $L_{2^{\star \dagger \star}}$ 
means that one two-point vertex is inserted into
the line 1 and another  into the line 2. 
$M_{4a}$ contains three electron lines 1,2,3 and $M_{4a(1^{\star \star})}$ means that two-point vertex 
insertion has been made twice in the electron line 1, and so on.

\subsection{Fourth-order magnetic moments}
\label{appendixM4}
The fourth-order magnetic moments are the same for both
old and new approaches. The UV-finite amplitude are also given here.
\begin{align*}  
&M_{4a}=\Delta M_{4a}+2~L_2^{\rm UV}~M_2
\\
&\underline{M}_{4a}=\Delta M_{4a}
\\
&M_{4b}=\Delta M_{4b}+B_2^{\rm UV}~M_2+\delta m_2~M_{2^\star}+L_2^{\rm R}~M_2
\\
&\underline{M}_{4b}=\Delta M_{4b} + L_2^{\rm R}~M_2
\end{align*}

In the new approach  no explicit form of $M_{4^\star}$ is needed
because the mass renormalization is completed by the $R$-subtraction operation.
They are, however, listed here, since they are used in the old approach.
\begin{align*}  
&M_{4a(1^\star)}=\Delta M_{4a(1^\star)}+L_2^{\rm UV}~M_{2^\star}+I_{4a(1^\star)}
\\
&M_{4a(2^\star)}=\Delta M_{4a(2^\star)}+I_{4a(2^\star)}
\\  
&M_{4b(1^\star)}=\Delta M_{4b(1^\star)}
           +(\delta m_2~M_{2^{\star \star}}+B_2^{\rm UV}~M_{2^\star})+I_{4b(1^\star)}
\\  
&M_{4b(2^\star)}=\Delta M_{4b(2^\star)}
         +\delta m_{2^\star}^{\rm UV}~M_{2^\star}+I_{4b(2^\star)}+L_2^{\rm R}~M_{2^\star}
         +M_{2^\star}[I]~\delta \widetilde{m}_{2^\star}-2~M_{2^\star}[I]~L_2^{\rm R}
\end{align*}
\vspace{-0.9cm}

\subsection{Sixth-order magnetic moments by {\sc gencode}{\it N}}

The finite amplitudes of the sixth-order  are given in the following.
For simplicity, we drop the superscript  ``new" from
$\Delta M_{6 x}^{\rm new}$.

\begin{align*}
M_{6a} &=\Delta M_{6a}
           +2~\delta m_2~M_{4b(1^\star)}+2~B_2^{\rm UV}~M_{4b}
           -\delta m_2~(\delta m_2~M_{2^{\star \star}}+B_2^{\rm UV}~M_{2^{\star}})
          \\ & -B_2^{\rm UV}~(\delta m_2~M_{2^{\star}}+B_2^{\rm UV}~M_2)
           \\ &
           +2~L_{4s}^{\rm R}~M_2
\end{align*}
\vspace{-0.9cm}
\begin{align*}           
M_{6b} &=\Delta M_{6b}
          +\delta m_2~M_{4b(2^\star)}+B_2^{\rm UV}~M_{4b}+\delta m_{4b}^{\rm UV}~M_{2^{\star}}
          +B_{4b}^{\rm UV}~M_2
          -\delta m_2~\delta m_{2^\star}^{\rm UV}~M_{2^{\star}} 
          \\  & -B_2^{\rm UV}~(\delta m_{2'}^{\rm UV}~M_{2^{\star}}+B_{2'}^{\rm UV}~M_2)
          \\& +L_2^{\rm R}~\Delta M_{4b} +L_{4l}^{\rm R}~M_2+M_{2^{\star}}~\Delta \delta m_{4b} 
\end{align*}
\vspace{-0.9cm}
\begin{align*}
M_{6c}& =\Delta M_{6c}
         +2~L_2^{\rm UV}~M_{4b}+\delta m_{4a}^{\rm UV}~M_{2^{\star}}+B_{4a}^{\rm UV}~M_2
         -2~L_2^{\rm UV}~(\delta m_2~M_{2^{\star}}+B_2^{\rm UV}~M_2)
         \\ &
         +L_2^{\rm R}~\Delta M_{4a} +M_{2^{\star}}~\Delta \delta m_{4a} 
\end{align*}
\vspace{-0.9cm}
\begin{align*}
M_{6d}& =\Delta M_{6d}
         +L_{4s}^{\rm UV}~M_2+\delta m_2~M_{4a(1^\star)}+B_2^{\rm UV}~M_{4a}+L_2^{\rm UV}~M_{4b}
         -B_2^{\rm UV}~L_{2'}^{\rm UV}~M_2
        \\ & - L_2^{\rm UV}~(\delta m_2~M_{2^{\star}}+B_2^{\rm UV}~M_2)
         \\  &
        +L_{4c}^{\rm R}~M_2
\end{align*}
\vspace{-0.9cm}
\begin{align*}
M_{6e}& =\Delta M_{6e}+2~L_{4s}^{\rm UV}~M_2+\delta m_2~M_{4a(2^\star)}+B_2^{\rm UV}~M_{4a}-2~L_{2'}^{\rm UV}~B_2^{\rm UV}~M_2
         \\  & 
        +L_{4x}^{\rm R}~M_2
\end{align*}
\vspace{-0.9cm}
\begin{align*}
M_{6f}& =\Delta M_{6f}+2~L_{4c}^{\rm UV}~M_2
              +2~L_2^{\rm UV}~M_{4a}-3~L_2^{\rm UV}~L_2^{\rm UV}~M_2
\end{align*}
\vspace{-0.9cm}
\begin{align*}
M_{6g}& =\Delta M_{6g}+L_{4c}^{\rm UV}~M_2+L_{4l}^{\rm UV}~M_2
              +L_2^{\rm UV}~M_{4a}-2~L_2^{\rm UV}~L_2^{\rm UV}~M_2
\end{align*}
\vspace{-0.9cm}
\begin{align*}
M_{6h}& =\Delta M_{6h}+2~L_{4x}^{\rm UV}~M_2
\end{align*}

\subsection{eighth-order magnetic moments by {\sc gencode}{\it N}}

The finite amplitudes of the eighth-order  are given in the following.
For simplicity, we drop the superscript ``new'' from
$\Delta M_i^{\rm new}$.

\begin{align*}
M_{01}&=\Delta M_{01}+2~L_2^{\rm UV}~M_{6f}+2~L_{4c}^{\rm UV}~M_{4a}
              +2~L_{6f1}^{\rm UV}~M_2-3~(L_2^{\rm UV})^2~M_{4a}
\\ &
              -6~L_2^{\rm UV}~L_{4c}^{\rm UV}~M_2+4~(L_2^{\rm UV})^3~M_2
\end{align*}
\vspace{-0.9cm}
\begin{align*}
M_{02}&=\Delta M_{02}+\delta m_2~M_{6f(1^\star)}
             +B_2^{\rm UV}~M_{6f}+L_2^{\rm UV}~M_{6d}+L_{4s}^{\rm UV}~M_{4a}
         +L_{4c}^{\rm UV}~M_{4b}+L_{6d5}^{\rm UV}~M_2
         \\ &
      -L_2^{\rm UV}~(\delta m_2~M_{4a(1^\star)}+B_2^{\rm UV}~M_{4a})
      -B_2^{\rm UV}~L_{2'}^{\rm UV}~M_{4a}
      -L_{4c}^{\rm UV}~(\delta m_2~M_{2^\star}+B_2^{\rm UV}~M_2)
      \\ &
     -B_2^{\rm UV}~L_{4c(3')}^{\rm UV}~M_2
      -2~L_2^{\rm UV}~L_{4s}^{\rm UV}~M_2
      -(L_2^{\rm UV})^2~M_{4b}+2~L_2^{\rm UV}~B_2^{\rm UV}~L_{2'}^{\rm UV}~M_2
      \\ &
      +(L_2^{\rm UV})^2~(\delta m_2~M_{2^\star}+B_2^{\rm UV}~M_2)
      \\ & +L_{6f1}^{\rm R}~M_2 
\end{align*}
\vspace{-0.9cm}
\begin{align*}
M_{03}&=\Delta M_{03}+2~L_2^{\rm UV}~M_{6d}
+\delta m_2~M_{6f(3^\star)}+B_2^{\rm UV}~M_{6f}+2~L_{6d1}^{\rm UV}~M_2
\\ &
      -2~L_2^{\rm UV}~(\delta m_2~M_{4a(1^\star)}
      +B_2^{\rm UV}~M_{4a})
      -(L_2^{\rm UV})^2~M_{4b}
      -2~L_2^{\rm UV}~L_{4s}^{\rm UV}~M_2 \\ & -2~B_2^{\rm UV}~L_{4c(1'')}^{\rm UV}~M_2
    +(L_2^{\rm UV})^2~(\delta m_2~M_{2^\star}+B_2^{\rm UV}~M_2)
      +2~L_2^{\rm UV}~B_2^{\rm UV}~L_{2'}^{\rm UV}~M_2
\\ &        +L_{6f3}^{\rm R}~M_2
\end{align*}
\vspace{-0.9cm}
\begin{align*}
M_{04}&=\Delta M_{04}+\delta m_2~M_{6d(3^\star)}+B_2^{\rm UV}~M_{6d}
        +\delta m_2~M_{6d(1^\star)}+B_2^{\rm UV}~M_{6d}+L_2^{\rm UV}~M_{6a}
        +L_{6a1}^{\rm UV}~M_2
        \\ &
      -\delta m_2~(\delta m_2~M_{4a(1^{\star\star})}+B_2^{\rm UV}~M_{4a(1^\star)} )
          -B_2^{\rm UV}~(\delta m_2~M_{4a(1^\star)}+B_2^{\rm UV}~M_{4a})
      \\ &    -2~L_2^{\rm UV}~(\delta m_2~M_{4b(1^\star)}+B_2^{\rm UV}~M_{4b})
         -B_2^{\rm UV}~(L_{4s(3')}^{\rm UV}+L_{4s(1'')}^{\rm UV})~M_2
      \\ & +L_2^{\rm UV}~\delta m_2~(\delta m_2~M_{2^{\star\star}}+B_2^{\rm UV}~M_{2^\star})
             +L_2^{\rm UV}~B_2^{\rm UV}~(\delta m_2~M_{2^\star}+B_2^{\rm UV}~M_2)
       \\&  +(B_2^{\rm UV})^2~L_{2''}^{\rm UV}~M_2
        \\ &    +L_{6d3}^{\rm R}~M_2+L_{6d1}^{\rm R}~M_2
\end{align*}
\vspace{-0.9cm}
\begin{align*}
M_{05}&=\Delta M_{05}+L_2^{\rm UV}~M_{6h}+L_{4x}^{\rm UV}~M_{4a}+L_{6f2}^{\rm UV}~M_2
           +L_{6h1}^{\rm UV}~M_2-3~L_2^{\rm UV}~L_{4x}^{\rm UV}~M_2
\end{align*}
\vspace{-0.9cm}
\begin{align*}
M_{06}&=\Delta M_{06}+L_2^{\rm UV}~M_{6g}+L_2^{\rm UV}~M_{6f}
       +L_{4l}^{\rm UV}~M_{4a}+L_{6f3}^{\rm UV}~M_2
       +L_{6g5}^{\rm UV}~M_2
       \\ &    -2~(L_2^{\rm UV})^2~M_{4a}-2~L_2^{\rm UV}~L_{4l}^{\rm UV}~M_2
       -3~L_2^{\rm UV}~L_{4c}^{\rm UV}~M_2+3~(L_2^{\rm UV})^3~M_2
\end{align*}
\vspace{-0.9cm}
\begin{align*}
M_{07}&=\Delta M_{07}+L_2^{\rm UV}~M_{6g}+L_2^{\rm UV}~M_{6f}
        +L_{4c}^{\rm UV}~M_{4a}+L_{6d2}^{\rm UV}~M_2
         +L_{6g1}^{\rm UV}~M_2
         \\ &    -2~(L_2^{\rm UV})^2~M_{4a}-4~L_2^{\rm UV}~L_{4c}^{\rm UV}~M_2
         -L_2^{\rm UV}~L_{4l}^{\rm UV}~M_2+3~(L_2^{\rm UV})^3~M_2
\end{align*}
\vspace{-0.9cm}
\begin{align*}
M_{08}&=\Delta M_{08}+L_2^{\rm UV}~M_{6c}+2~L_2^{\rm UV}~M_{6d}
       +\delta m_{4a}^{\rm UV}~M_{4a(1^\star)}+B_{4a}^{\rm UV}~M_{4a}
        +L_{6c1}^{\rm UV}~M_2
      \\ & -2~(L_2^{\rm UV})^2~M_{4b}
         -L_2^{\rm UV}~(\delta m_{4a}^{\rm UV}~M_{2^\star}+B_{4a}^{\rm UV}~M_2)
       -2~L_2^{\rm UV}~(\delta m_2~M_{4a(1^\star)}+B_2^{\rm UV}~M_{4a})
      \\ & -2~L_2^{\rm UV}~L_{4s}^{\rm UV}~M_2
      -B_{4a}^{\rm UV}~L_{2'}^{\rm UV}~M_2
          +2~(L_2^{\rm UV})^2~(\delta m_2~M_{2^\star}+B_2^{\rm UV}~M_2)
      \\ &  +2~L_2^{\rm UV}~B_2^{\rm UV}~L_{2'}^{\rm UV}~M_2
      \\ &    +L_{4c}^{\rm R}~\Delta M_{4a}+\underline{M}_{4a(1^\star)}~\Delta \delta m_{4a}
\end{align*}
\vspace{-0.9cm}
\begin{align*}
M_{09}&=\Delta M_{09}+\delta m_2~M_{6f(2^\star)}+B_2^{\rm UV}~M_{6f}
       +L_2^{\rm UV}~M_{6e}+L_{4s}^{\rm UV}~M_{4a}
        +L_{6e1}^{\rm UV}~M_2+L_{6d3}^{\rm UV}~M_2
      \\ &    -L_2^{\rm UV}~(\delta m_2~M_{4a(2^{\star\star})}+B_2^{\rm UV}~M_{4a})
   -B_2^{\rm UV}~L_{2'}^{\rm UV}~M_{4a}-B_2^{\rm UV}~L_{4c(2')}^{\rm UV}~M_2
      \\ &    -B_2^{\rm UV}~L_{4c(1')}^{\rm UV}~M_2
     -3~L_2^{\rm UV}~L_{4s}^{\rm UV}~M_2+3~L_2^{\rm UV}~B_2^{\rm UV}~L_{2'}^{\rm UV}~M_2
     \\ &
          +L_{6f2}^{\rm R}~M_2
\end{align*}
\vspace{-0.9cm}
\begin{align*} 
M_{10}&=\Delta M_{10}+\delta m_2~M_{6d(2^\star)}+B_2^{\rm UV}~M_{6d}
            +\delta m_{4b}^{\rm UV}~M_{4a(1^\star)}+B_{4b}^{\rm UV}~M_{4a}
     \\ &     +L_2^{\rm UV}~M_{6b}+L_{6b1}^{\rm UV}~M_2
        -\delta m_2~\delta m_{2^\star}^{\rm UV}~M_{4a(1^\star)}-B_2^{\rm UV}~(
       \delta m_{2'}^{\rm UV}~M_{4a(1^\star)}
+B_{2'}^{\rm UV}~M_{4a})
      \\ &    -L_2^{\rm UV}~(\delta m_2~M_{4b(2^\star)}
      +B_2^{\rm UV}~M_{4b})
      -B_2^{\rm UV}~L_{4s(2')}^{\rm UV}~M_2
     \\ & -L_2^{\rm UV}~(\delta m_{4b}^{\rm UV}~M_{2^\star}
             +B_{4b}^{\rm UV}~M_2)- B_{4b}^{\rm UV}~L_{2'}^{\rm UV}~M_2
      +L_2^{\rm UV}~\delta m_2~\delta m_{2^\star}^{\rm UV}~M_{2^\star}
     \\ &   +L_2^{\rm UV}~B_2^{\rm UV}~(\delta m_{2'}^{\rm UV}~M_{2^\star}
    +B_{2'}^{\rm UV}~M_2 )+B_2^{\rm UV}~B_{2'}^{\rm UV}~L_{2'}^{\rm UV}~M_2
\\ & 
+L_{4c}^{\rm R}~(\Delta M_{4b}+L_2^{\rm R}~M_2)
        -L_{4c}^{\rm R}~L_2^{\rm R}~M_2+L_{6d2}^{\rm R}~M_2+\underline{M}_{4a(1^\star)}~\Delta \delta m_{4b}
\end{align*}
\vspace{-0.9cm}
\begin{align*}
M_{11}&=\Delta M_{11}+2~\delta m_2~M_{6d(5^\star)}
              +2~B_2^{\rm UV}~M_{6d}+2~L_{4s}^{\rm UV}~M_{4b}
      \\ & -\delta m_2~(\delta m_2~M_{4a(1^\star 3^\star)}+B_2^{\rm UV}~M_{4a(1^\star)} )
            -B_2^{\rm UV}~(\delta m_2~M_{4a(1^\star)}+B_2^{\rm UV}~M_{4a})
      \\ & -2~L_{4s}^{\rm UV}~(\delta m_2~M_{2^\star}+B_2^{\rm UV}~M_2)
          -2~B_2^{\rm UV}~L_{2'}^{\rm UV}~M_{4b}
      \\ &  +2~B_2^{\rm UV}~L_{2'}^{\rm UV}~(\delta m_2~M_{2^\star}+B_2^{\rm UV}~M_2)
      \\ &     +2~L_{6d5}^{\rm R}~M_2
\end{align*}
\vspace{-0.9cm}
\begin{align*}
M_{12}&=\Delta M_{12}+2~\delta m_2~M_{6a(1^\star)}+2~B_2^{\rm UV}~M_{6a}
  +\delta m_2~M_{6a(3^\star)}+B_2^{\rm UV}~M_{6a} 
      \\ & -2~\delta m_2~(\delta m_2~M_{4b(1^{\star \star})} 
         +B_2^{\rm UV}~M_{4b(1^\star)} )
       -2~B_2^{\rm UV}~(\delta m_2~M_{4b(1^\star)}+B_2^{\rm UV}~M_{4b})
      \\ & -\delta m_2~(\delta m_2~M_{4b(1^\star 3^\star)}+B_2^{\rm UV}~M_{4b(1^\star)} )
         -B_2^{\rm UV}~(\delta m_2~M_{4b(1^\star)}+B_2^{\rm UV}~M_{4b})
       \\ & +(\delta m_2)^2~(\delta m_2~M_{2^{\star \star \star}}+B_2^{\rm UV}~M_{2^{\star\star}})
        +2~\delta m_2~B_2^{\rm UV}~(\delta m_2~M_{2^{\star \star}}+B_2^{\rm UV}~M_{2^\star})
      \\ & +(B_2^{\rm UV})^2~(\delta m_2~M_{2^\star}+B_2^{\rm UV}~M_2)
      \\ &    +2~L_{6a1}^{\rm R}~M_2+L_{6a3}^{\rm R}~M_2
\end{align*}
\vspace{-0.9cm}
\begin{align*}
M_{13}&=\Delta M_{13}+\delta m_2~M_{6h(1^\star)}
+B_2^{\rm UV}~M_{6h}+L_{4x}^{\rm UV}~M_{4b}+L_{6d4}^{\rm UV}~M_2
        \\ &      -L_{4x}^{\rm UV}~(\delta m_2~M_{2^\star}+B_2^{\rm UV}~M_2)
          -B_2^{\rm UV}~L_{4x(1')}^{\rm UV}~M_2
   \\ &   +L_{6h1}^{\rm R}~M_2
\end{align*}
\vspace{-0.9cm}
\begin{align*}
M_{14}&=\Delta M_{14}+\delta m_2~M_{6g(5^\star)}+B_2^{\rm UV}~M_{6g}+L_2^{\rm UV}~M_{6d}+L_{4l}^{\rm UV}~M_{4b}
+L_{6d3}^{\rm UV}~M_2
      \\ &    -L_2^{\rm UV}~(\delta m_2~M_{4a(1^\star)}
           +B_2^{\rm UV}~M_{4a})
      -L_{4l}^{\rm UV}~(\delta m_2~M_{2^\star}+B_2^{\rm UV}~M_2)
        -B_2^{\rm UV}~L_{4c(1')}^{\rm UV}~M_2
      \\ &    -(L_2^{\rm UV})^2~M_{4b}-L_2^{\rm UV}~L_{4s}^{\rm UV}~M_2
       +(L_2^{\rm UV})^2~(\delta m_2~M_{2^\star}
       +B_2^{\rm UV}~M_2)
       +L_2^{\rm UV}~B_2^{\rm UV}~L_{2'}^{\rm UV}~M_2
     \\ &    +L_{6g5}^{\rm R}~M_2
\end{align*}
\vspace{-0.9cm}
\begin{align*}
M_{15}&=\Delta M_{15}+\delta m_2~M_{6g(1^\star)}
       +B_2^{\rm UV}~M_{6g}+L_2^{\rm UV}~M_{6d}+L_{4c}^{\rm UV}~M_{4b}
        +L_{6a2}^{\rm UV}~M_2
      \\ &    -L_2^{\rm UV}~(\delta m_2~M_{4a(1^\star)}+B_2^{\rm UV}~M_{4a})
     -L_{4c}^{\rm UV}~(\delta m_2~M_{2^\star}+B_2^{\rm UV}~M_2)
        -B_2^{\rm UV}~L_{4l(1')}^{\rm UV}~M_2
      \\ & -(L_2^{\rm UV})^2~M_{4b}-L_2^{\rm UV}~L_{4s}^{\rm UV}~M_2
      \\&    +(L_2^{\rm UV})^2~(\delta m_2~M_{2^\star}
      +B_2^{\rm UV}~M_2)+L_2^{\rm UV}~B_2^{\rm UV}~L_{2'}^{\rm UV}~M_2
     \\ &     +L_{6g1}^{\rm R}~M_2
\end{align*}
\vspace{-0.9cm}
\begin{align*}
M_{16}&=\Delta M_{16}+\delta m_2~M_{6c(1^\star)}
      +B_2^{\rm UV}~M_{6c}+\delta m_{4a}^{\rm UV}~M_{4b(1^\star)}
    +B_{4a}^{\rm UV}~M_{4b}
+2~L_2^{\rm UV}~M_{6a}
    \\ &      -\delta m_2~(\delta m_{4a}^{\rm UV}~M_{2^{\star\star}}
   +B_{4a}^{\rm UV}~M_{2^\star})
         -B_2^{\rm UV}~(\delta m_{4a}^{\rm UV}~M_{2^\star}+B_{4a}^{\rm UV}~M_2)
     \\ &  -4~L_2^{\rm UV}~(\delta m_2~M_{4b(1^\star)}+B_2^{\rm UV}~M_{4b})
     \\ &     +2~L_2^{\rm UV}~\delta m_2~(\delta m_2~M_{2^{\star\star}}+B_2^{\rm UV}~M_{2^\star})
      +2~B_2^{\rm UV}~L_2^{\rm UV}~(\delta m_2~M_{2^\star}+B_2^{\rm UV}~M_2)
     \\ &     +L_{6c1}^{\rm R}~M_2+L_{4s}^{\rm R}~\Delta M_{4a}+\underline{M}_{4b(1^\star)}~
     \Delta \delta m_{4a}-L_{2^\star}^{\rm R}~\Delta \delta m_{4a}~M_2
\end{align*}
\vspace{-0.9cm}
\begin{align*}
M_{17}&=\Delta M_{17}+\delta m_2~M_{6e(1^\star)}+B_2^{\rm UV}~M_{6e}
 +\delta m_2~M_{6d(4^\star)}+B_2^{\rm UV}~M_{6d}+L_{4s}^{\rm UV}~M_{4b}
   \\ & +L_{6a3}^{\rm UV}~M_2
         -\delta m_2~(\delta m_2~M_{4a(1^\star 2^\star)}+B_2^{\rm UV}~M_{4a(1^\star)})
     \\ & -B_2^{\rm UV}~(\delta m_2~M_{4a(2^\star)}+B_2^{\rm UV}~M_{4a})
      -L_{4s}^{\rm UV}~(\delta m_2~M_{2^\star}
          +B_2^{\rm UV}~M_2)-2~B_2^{\rm UV}~L_{4s(1')}^{\rm UV}~M_2
     \\ &    -B_2^{\rm UV}~L_{2'}^{\rm UV}~M_{4b}
       +\delta m_2~B_2^{\rm UV}~L_{2'}^{\rm UV}~M_{2^\star}
      +(B_2^{\rm UV})^2~(L_{2'}^{\rm UV}+L_{2''}^{\rm UV})~M_2
     \\ &     +L_{6e1}^{\rm R}~M_2+L_{6d4}^{\rm R}~M_2
\end{align*}
\vspace{-0.9cm}
\begin{align*}
M_{18}&=\Delta M_{18}+\delta m_2~M_{6b(1^\star)}+B_2^{\rm UV}~M_{6b}
+\delta m_2~M_{6a(2^\star)}+B_2^{\rm UV}~M_{6a}
      +\delta m_{4b}^{\rm UV}~M_{4b(1^\star)}
     \\ & +B_{4b}^{\rm UV}~M_{4b}
         -\delta m_2~(\delta m_2~M_{4b(1^\star 2^\star)}
      +B_2^{\rm UV}~M_{4b(1^\star)})
        -B_2^{\rm UV}~(\delta m_2~M_{4b(2^\star)}+B_2^{\rm UV}~M_{4b})
        \\ & -\delta m_2~(\delta m_{4b}^{\rm UV}~M_{2^{\star \star}}
     +B_{4b}^{\rm UV}~M_{2^\star})-B_2^{\rm UV}~(\delta m_{4b}^{\rm UV}~M_{2^\star}
     +B_{4b}^{\rm UV}~M_2)
       \\ &  -\delta m_2~\delta m_{2^\star}^{\rm UV}~M_{4b(1^\star)}
       -B_2^{\rm UV}~(\delta m_{2'}^{\rm UV}~M_{4b(1^\star)}+B_{2'}^{\rm UV}~M_{4b})
       \\ &  +\delta m_2~\delta m_{2^\star}^{\rm UV}~
        (\delta m_2~M_{2^{\star \star}}+B_2^{\rm UV}~M_{2^\star})
         +B_2^{\rm UV}~\delta m_{2'}^{\rm UV}~(\delta m_2~M_{2^{\star \star}}
         +B_2^{\rm UV}~M_{2^\star})
      \\ & +B_2^{\rm UV}~B_{2'}^{\rm UV}~(\delta m_2~M_{2^\star}
      +B_2^{\rm UV}~M_2)
      \\ &
      +L_{6b1}^{\rm R}~M_2+L_{4s}^{\rm R}~(\Delta M_{4b}+L_2^{\rm R}~M_2)
     +L_{6a2}^{\rm R}~M_2-L_{4s}^{\rm R}~L_2^{\rm R}~M_2
      \\ & +(\underline{M}_{4b(1^\star)} -L_{2^\star}^{\rm R}~M_2)~\Delta \delta m_{4b}
\end{align*}
\vspace{-0.9cm}
\begin{align*}
M_{19}&=\Delta M_{19}+2~L_{6h2}^{\rm UV}~M_2
\end{align*}
\vspace{-0.9cm}
\begin{align*}
M_{20}&=\Delta M_{20}+L_2^{\rm UV}~M_{6h}+L_{6f2}^{\rm UV}~M_2+L_{6g4}^{\rm UV}~M_2
-2~L_2^{\rm UV}~L_{4x}^{\rm UV}~M_2
\end{align*}
\vspace{-0.9cm}
\begin{align*}
M_{21}&=\Delta M_{21}+2~L_{6g2}^{\rm UV}~M_2
\end{align*}
\vspace{-0.9cm}
\begin{align*}
M_{22}&=\Delta M_{22}+L_2^{\rm UV}~M_{6g}+L_{4c}^{\rm UV}~M_{4a}+L_{6f1}^{\rm UV}~M_2
+L_{6c2}^{\rm UV}~M_2
       \\ & -(L_2^{\rm UV})^2~M_{4a}
-3~L_2^{\rm UV}~L_{4c}^{\rm UV}~M_2-L_2^{\rm UV}~L_{4l}^{\rm UV}~M_2+2~(L_2^{\rm UV})^3~M_2
\end{align*}
\vspace{-0.9cm}
\begin{align*}
M_{23}&=\Delta M_{23}+\delta m_2~M_{6h(2^\star)}
   +B_2^{\rm UV}~M_{6h}+L_{6e2}^{\rm UV}~M_2 \\ & +L_{6d4}^{\rm UV}~M_2
    -B_2^{\rm UV}~(L_{4x(1')}^{\rm UV}+L_{4x(2')}^{\rm UV})~M_2
         \\ & +L_{6h2}^{\rm R}~M_2
\end{align*}
\vspace{-0.9cm}
\begin{align*}
M_{24}&=\Delta M_{24}+\delta m_2~M_{6g(2^\star)}+B_2^{\rm UV}~M_{6g}+L_{4s}^{\rm UV}~M_{4a}+L_{6b2}^{\rm UV}~M_2
\\ & +L_{6d5}^{\rm UV}~M_2
      -B_2^{\rm UV}~L_{2'}^{\rm UV}~M_{4a}
     -B_2^{\rm UV}~L_{4l{2'}}^{\rm UV}~M_2
    \\ &  -2~L_{4s}^{\rm UV}~L_2^{\rm UV}~M_2
-B_2^{\rm UV}~L_{4c(3')}^{\rm UV}~M_2
+2~L_2^{\rm UV}~B_2^{\rm UV}~L_{2'}^{\rm UV}~M_2 
     \\&   +L_{6g2}^{\rm R}~M_2
\end{align*}
\vspace{-0.9cm}
\begin{align*}
M_{25}&=\Delta M_{25}+2~L_2^{\rm UV}~M_{6g}+2~L_{6d2}^{\rm UV}~M_2
  -(L_2^{\rm UV})^2~M_{4a}-2~L_2^{\rm UV}~L_{4l}^{\rm UV}~M_2
    \\ & -2~L_2^{\rm UV}~L_{4c}^{\rm UV}~M_2
+2~(L_2^{\rm UV})^3~M_2
\end{align*}
\vspace{-0.9cm}
\begin{align*}
M_{26}&=\Delta M_{26}+2~L_2^{\rm UV}~M_{6c}+2~L_{4c}^{\rm UV}~M_{4b}
+\delta m_{6f}^{\rm UV}~M_{2^\star}+B_{6f}^{\rm UV}~M_2-3~(L_2^{\rm UV})^2~M_{4b}
   \\ &     -2~L_2^{\rm UV}~(\delta m_{4a}^{\rm UV}~M_{2^\star}+B_{4a}^{\rm UV}~M_2)
       -2~L_{4c}^{\rm UV}~(\delta m_2~M_{2^\star}+B_2^{\rm UV}~M_2)
   \\ &   +3~(L_2^{\rm UV})^2~(\delta m_2~M_{2^\star}+B_2^{\rm UV}~M_2)
      \\ &
     +L_2^{\rm R}~\Delta M_{6f}+M_{2^\star}~\Delta \delta m_{6f}
\end{align*}
\vspace{-0.9cm}
\begin{align*}
M_{27}&=\Delta M_{27}+\delta m_2~M_{6g(4^\star)}+B_2^{\rm UV}~M_{6g}+L_2^{\rm UV}~M_{6e}+L_{6d1}^{\rm UV}~M_2
+L_{6a2}^{\rm UV}~M_2
\\ & -B_2^{\rm UV}~L_{4l(1')}^{\rm UV}~M_2
     -B_2^{\rm UV}~L_{4c(1'')}^{\rm UV}~M_2-2~L_2^{\rm UV}~L_{4s}^{\rm UV}~M_2
             -L_2^{\rm UV}~(\delta m_2~M_{4a(2^\star)}
\\ & +B_2^{\rm UV}~M_{4a})+2~L_2^{\rm UV}~B_2^{\rm UV}~L_{2'}^{\rm UV}~M_2
    \\ &   +L_{6g4}^{\rm R}~M_2
\end{align*}
\vspace{-0.9cm}
\begin{align*}
M_{28}&=\Delta M_{28}+\delta m_2~M_{6c(2^\star)}+B_2^{\rm UV}~M_{6c}+L_2^{\rm UV}~M_{6b}
+L_{4s}^{\rm UV}~M_{4b}+\delta m_{6d}^{\rm UV}~M_{2^\star}+B_{6d}^{\rm UV}~M_2
 \\ &     -L_2^{\rm UV}~(\delta m_2~M_{4b(2^\star)}+B_2^{\rm UV}~M_{4b})
-B_2^{\rm UV}~L_{2'}^{\rm UV}~M_{4b}
-\delta m_2~\delta m_{4a(1^\star)}^{\rm UV}~M_{2^\star}
\\&      -B_2^{\rm UV}~(\delta m_{4a(1')}^{\rm UV}~M_{2^\star}+B_{4a(1')}^{\rm UV}~M_2)
    -L_2^{\rm UV}~(\delta m_{4b}^{\rm UV}~M_{2^\star}+B_{4b}^{\rm UV}~M_2)
\\ &     -L_{4s}^{\rm UV}~(\delta m_2~M_{2^\star}+B_2^{\rm UV}~M_2)
   +L_2^{\rm UV}~\delta m_2~\delta m_{2^\star}^{\rm UV}~M_{2^\star}     
\\ &  +L_2^{\rm UV}~B_2^{\rm UV}~(\delta m_{2'}^{\rm UV}~M_{2^\star}
     +B_{2'}^{\rm UV}~M_2)
       +B_2^{\rm UV}~L_{2'}^{\rm UV}~(\delta m_2~M_{2^\star}+B_2^{\rm UV}~M_2)
     \\ & 
     +L_{6c2}^{\rm R}~M_2-L_2^{\rm R}~L_{4c}^{\rm R}~M_2+L_2^{\rm R}~(\Delta M_{6d}+L_{4c}^{\rm R}~M_2)
     +M_{2^\star}~\Delta \delta m_{6d}
\end{align*}
\vspace{-0.9cm}
\begin{align*}
M_{29}&=\Delta M_{29}
 +2~\delta m_2~M_{6e(2^\star)}+2~B_2^{\rm UV}~M_{6e}+2~L_{6a1}^{\rm UV}~M_2
     -\delta m_2~(\delta m_2~M_{4a(2^{\star \star})}+B_2^{\rm UV}~M_{4a(2^\star)})
  \\ &  -B_2^{\rm UV}~(\delta m_2~M_{4a(2^\star )}+B_2^{\rm UV}~M_{4a})
     -2~B_2^{\rm UV}~(L_{4s(3')}^{\rm UV}+L_{4s(1'')}^{\rm UV})~M_2
      +2~(B_2^{\rm UV})^2~L_{2''}^{\rm UV}~M_2
 \\ &    +2~L_{6e2}^{\rm R}~M_2
\end{align*}
\vspace{-0.9cm}
\begin{align*}
M_{30}&=\Delta M_{30}+2~\delta m_2~M_{6b(2^\star)}+2~B_2^{\rm UV}~M_{6b}
+\delta m_{6a}^{\rm UV}~M_{2^\star}+B_{6a}^{\rm UV}~M_2
 \\ &     -\delta m_2~(\delta m_2~M_{4b(2^{\star \star})}+B_2^{\rm UV}~M_{4b(2^\star)})
       -B_2^{\rm UV}~(\delta m_2~M_{4b(2^\star)}+B_2^{\rm UV}~M_{4b})
 \\&     -2~\delta m_2~\delta m_{4b(1^\star)}^{\rm UV}~M_{2^\star}
        -2~B_2^{\rm UV}~(\delta m_{4b(1')}^{\rm UV}~M_{2^\star}
         +B_{4b(1')}^{\rm UV}~M_2)
  \\&    +2~\delta m_2~B_2^{\rm UV}~\delta m_{2'^\star}^{\rm UV}~M_{2^\star}
      +(B_2^{\rm UV})^2~(\delta m_{2''}^{\rm UV}~M_{2^\star}+B_{2''}^{\rm UV}~M_2)
      \\ &
      +2~L_{6b2}^{\rm R}~M_2-2~L_2^{\rm R}~L_{4s}^{\rm R}~M_2+L_2^{\rm R}~(\Delta  
       M_{6a}+2~L_{4s}^{\rm R}~M_2)+M_{2^\star}~\Delta \delta m_{6a}
\end{align*}
\vspace{-0.9cm}
\begin{align*}
M_{31}&=\Delta M_{31}+2~L_{6h3}^{\rm UV}~M_2
\end{align*}
\vspace{-0.9cm}
\begin{align*}
M_{32}&=\Delta M_{32}+L_{6g3}^{\rm UV}~M_2+L_{6h2}^{\rm UV}~M_2
\end{align*}
\vspace{-0.9cm}
\begin{align*}
M_{33}&=\Delta M_{33}+2~L_{6g3}^{\rm UV}~M_2
\end{align*}
\vspace{-0.9cm}
\begin{align*}
M_{34}&=\Delta M_{34}+L_{4x}^{\rm UV}~M_{4a}+L_{6c3}^{\rm UV}~M_2
+L_{6h1}^{\rm UV}~M_2-2~L_{4x}^{\rm UV}~L_2^{\rm UV}~M_2
\end{align*}
\vspace{-0.9cm}
\begin{align*}
M_{35}&=\Delta M_{35}+L_2^{\rm UV}~M_{6h}+L_{6e3}^{\rm UV}~M_2+L_{6g4}^{\rm UV}~M_2
-2~L_{4x}^{\rm UV}~L_2^{\rm UV}~M_2
\end{align*}
\vspace{-0.9cm}
\begin{align*}
M_{36}&=\Delta M_{36}+L_2^{\rm UV}~M_{6g}+L_{4l}^{\rm UV}~M_{4a}+L_{6b3}^{\rm UV}~M_2
+L_{6g5}^{\rm UV}~M_2
     -(L_2^{\rm UV})^2~M_{4a}
    \\ & -3~L_2^{\rm UV}~L_{4l}^{\rm UV}~M_2-L_2^{\rm UV}~L_{4c}^{\rm UV}~M_2
+2~(L_2^{\rm UV})^3~M_2
\end{align*}
\vspace{-0.9cm}
\begin{align*}
M_{37}&=\Delta M_{37}+2~L_{6g2}^{\rm UV}~M_2
\end{align*}
\vspace{-0.9cm}
\begin{align*}
M_{38}&=\Delta M_{38}+2~L_{4x}^{\rm UV}~M_{4b}+\delta m_{6h}^{\rm UV}~M_{2^\star}+B_{6h}^{\rm UV}~M_2
-2~L_{4x}^{\rm UV}~(\delta m_2~M_{2^\star}+B_2^{\rm UV}~M_2)
      \\ & +L_2^{\rm R}~\Delta M_{6h}+M_{2^\star}~\Delta \delta m_{6h}
\end{align*}
\vspace{-0.9cm}
\begin{align*}
M_{39}&=\Delta M_{39}+L_2^{\rm UV}~M_{6g}+L_{4c}^{\rm UV}~M_{4a}
         +L_{6g1}^{\rm UV}~M_2+L_{6c2}^{\rm UV}~M_2
      -(L_2^{\rm UV})^2~M_{4a}
     \\ & -3~L_2^{\rm UV}~L_{4c}^{\rm UV}~M_2
      -L_2^{\rm UV}~L_{4l}^{\rm UV}~M_2+2~(L_2^{\rm UV})^3~M_2
\end{align*}
\vspace{-0.9cm}
\begin{align*}
M_{40}&=\Delta M_{40}+L_2^{\rm UV}~M_{6c}+L_{4l}^{\rm UV}~M_{4b}+L_{4c}^{\rm UV}~M_{4b}+
\delta m_{6g}^{\rm UV}~M_{2^\star}+B_{6g}^{\rm UV}~M_2
     \\ &  -2~(L_2^{\rm UV})^2~M_{4b}-L_2^{\rm UV}~(\delta m_{4a}^{\rm UV}~M_{2^\star}
+B_{4a}^{\rm UV}~M_2)-L_{4l}^{\rm UV}~(\delta m_2~M_{2^\star}+B_2^{\rm UV}~M_2)
 \\ &     -L_{4c}^{\rm UV}~(\delta m_2~M_{2^\star}+B_2^{\rm UV}~M_2)
         +2~(L_2^{\rm UV})^2~(\delta m_2~M_{2^\star}+B_2^{\rm UV}~M_2)
     \\ & +L_2^{\rm R}~\Delta M_{6g}+M_{2^\star}~\Delta \delta m_{6g}
\end{align*}
\vspace{-0.9cm}
\begin{align*}
M_{41}&=\Delta M_{41}+2~L_2^{\rm UV}~M_{6e}
+\delta m_{4a}^{\rm UV}~M_{4a(2^\star)}+B_{4a}^{\rm UV}~M_{4a}+2~L_{6c1}^{\rm UV}~M_2
  \\ &    -2~L_2^{\rm UV}~(\delta m_2~M_{4a(2^\star)}+B_2^{\rm UV}~M_{4a})-4~L_2^{\rm UV}~L_{4s}^{\rm UV}~M_2
      -2~B_{4a}^{\rm UV}~L_{2'}^{\rm UV}~M_2
  \\ & +4~L_2^{\rm UV}~B_2^{\rm UV}~L_{2'}^{\rm UV}~M_2
  \\ &    +L_{4x}^{\rm R}~\Delta M_{4a}+\underline{M}_{4a(2^\star)}~\Delta \delta m_{4a} 
\end{align*}
\vspace{-0.9cm}
\begin{align*}
M_{42}&=\Delta M_{42}+2~L_2^{\rm UV}~M_{6b}+\delta m_{4a}^{\rm UV}~M_{4b(2^\star)}
        +B_{4a}^{\rm UV}~M_{4b}
+\delta m_{6c}^{\rm UV}~M_{2^\star}+B_{6c}^{\rm UV}~M_2
 \\ &     -2~L_2^{\rm UV}~(\delta m_2~M_{4b(2^\star)}+B_2^{\rm UV}~M_{4b})
        -2~L_2^{\rm UV}~(\delta m_{4b}^{\rm UV}~M_{2^\star}+B_{4b}^{\rm UV}~M_2)
 \\ &     -\delta m_{4a}^{\rm UV}~\delta m_{2^\star}^{\rm UV}~M_{2^\star}
    -B_{4a}^{\rm UV}~(\delta m_{2'}^{\rm UV}~M_{2^\star}
+B_{2'}^{\rm UV}~M_2)
  \\ & +2~L_2^{\rm UV}~\delta m_2~\delta m_{2^\star}^{\rm UV}~M_{2^\star}
      +2~L_2^{\rm UV}~B_2^{\rm UV}~(\delta m_{2'}^{\rm UV}~M_{2^\star}+B_{2'}^{\rm UV}~M_2)
   \\ &
   +L_{4l}^{\rm R}~\Delta M_{4a}-(L_2^{\rm R})^2~\Delta M_{4a}
  +L_2^{\rm R}~(\Delta M_{6c}+L_2^{\rm R}~\Delta M_{4a}+M_{2^\star}~\Delta \delta m_{4a})
   \\ &    +M_{2^\star}~(\Delta \delta m_{6c}+L_2^{\rm R}~\delta m_{4a}^{\rm R})
      +\underline{M}_{4b(2^\star)}~\Delta \delta m_{4a}
    \\ & - \delta m_{2^\star}^{\rm R}~\Delta \delta m_{4a}~M_{2^\star}
      -\Delta \delta m_{4a}~L_2^{\rm R}~M_{2^\star}
\end{align*}
\vspace{-0.9cm}
\begin{align*}
M_{43}&=\Delta M_{43}+\delta m_2~M_{6h(3^\star)}+B_2^{\rm UV}~M_{6h}+2~L_{6e2}^{\rm UV}~M_2
-2~B_2^{\rm UV}~L_{4x(2')}^{\rm UV}~M_2
     \\&  +L_{6h3}^{\rm R}~M_2
\end{align*}
\vspace{-0.9cm}
\begin{align*}
M_{44}&=\Delta M_{44}+\delta m_2~M_{6g(3^\star)}+B_2^{\rm UV}~M_{6g}
+L_{4s}^{\rm UV}~M_{4a}
+L_{6b2}^{\rm UV}~M_2+L_{6e1}^{\rm UV}~M_2
\\&      -B_2^{\rm UV}~L_{2'}^{\rm UV}~M_{4a}
 -B_2^{\rm UV}~L_{4l(2')}^{\rm UV}~M_2
-B_2^{\rm UV}~L_{4c(2')}^{\rm UV}~M_2-2~L_{4s}^{\rm UV}~L_2^{\rm UV}~M_2
\\&     +2~L_2^{\rm UV}~B_2^{\rm UV}~L_{2'}^{\rm UV}~M_2
  \\ &   +L_{6g3}^{\rm R}~M_2
\end{align*}
\vspace{-0.9cm}
\begin{align*}
M_{45}&=\Delta M_{45}+\delta m_2~M_{6c(3^\star)}
    +B_2^{\rm UV}~M_{6c}+2~L_{4s}^{\rm UV}~M_{4b}
         +\delta m_{6e}^{\rm UV}~M_{2^\star}+B_{6e}^{\rm UV}~M_2
 \\ &     -2~B_2^{\rm UV}~L_{2'}^{\rm UV}~M_{4b}
        -\delta m_2~\delta m_{4a(2^\star)}^{\rm UV}~M_{2^\star}
       -B_2^{\rm UV}~(\delta m_{4a(2')}^{\rm UV}~M_{2^\star}+B_{4a(2')}^{\rm UV}~M_2)
  \\ &    -2~L_{4s}^{\rm UV}~(\delta m_2~M_{2^\star}+B_2^{\rm UV}~M_2)
          +2~B_2^{\rm UV}~L_{2'}^{\rm UV}~(\delta m_2~M_{2^\star}+B_2^{\rm UV}~M_2)
      \\ & +L_{6c3}^{\rm R}~M_2+L_2^{\rm R}~(\Delta M_{6e}+L_{4x}^{\rm R}~M_2)
      +M_{2^\star}~\Delta \delta m_{6e}-L_{4x}^{\rm R}~L_2^{\rm R}~M_2
\end{align*}
\vspace{-0.9cm}
\begin{align*}
M_{46}&=\Delta M_{46}+\delta m_2~M_{6e(3^\star)}+B_2^{\rm UV}~M_{6e}
   +\delta m_{4b}^{\rm UV}~M_{4a(2^\star)}+B_{4b}^{\rm UV}~M_{4a}
     +2~L_{6b1}^{\rm UV}~M_2
  \\ &    -\delta m_2~\delta m_{2^\star}^{\rm UV}~M_{4a(2^\star)}-B_2^{\rm UV}~(\delta m_{2'}^{\rm UV}~M_{4a(2^\star)}+B_{2'}^{\rm UV}~M_{4a})-2~B_2^{\rm UV}~L_{4s(2')}^{\rm UV}~M_2
 \\ &   -2~B_{4b}^{\rm UV}~L_{2'}^{\rm UV}~M_2+2~B_2^{\rm UV}~B_{2'}^{\rm UV}~L_{2'}^{\rm UV}~M_2
\\ &   +L_{6e3}^{\rm R}~M_2-L_{4x}^{\rm R}~L_2^{\rm R}~M_2+L_{4x}^{\rm R}~(\Delta M_{4b}+L_2^{\rm R}~M_2)+\underline{M}_{4a(2^\star)}~\Delta \delta m_{4b}
\end{align*}
\vspace{-0.9cm}
\begin{align*}
M_{47}&=\Delta M_{47}+\delta m_2~M_{6b(3^\star)}+B_2^{\rm UV}~M_{6b}
+\delta m_{4b}^{\rm UV}~M_{4b(2^\star)}+B_{4b}^{\rm UV}~M_{4b}
+\delta m_{6b}^{\rm UV}~M_{2^\star}
\\ & +B_{6b}^{\rm UV}~M_2
      -\delta m_2~\delta m_{2^\star}^{\rm UV}~M_{4b(2^\star)}
-B_2^{\rm UV}~(\delta m_{2'}^{\rm UV}~M_{4b(2^\star)}
+B_{2'}^{\rm UV}~M_{4b})
\\ & -\delta m_2~\delta m_{4b(2^\star)}^{\rm UV}~M_{2^\star}
       -B_2^{\rm UV}~(\delta m_{4b(2')}^{\rm UV}~M_{2^\star}+B_{4b(2')}^{\rm UV}~M_2)
       -\delta m_{4b}^{\rm UV}~\delta m_{2^\star}^{\rm UV}~M_{2^\star}
 \\ &     -B_{4b}^{\rm UV}~(\delta m_{2'}^{\rm UV}~M_{2^\star}+B_{2'}^{\rm UV}~M_2)
     +\delta m_2~(\delta m_{2^\star}^{\rm UV})^2~M_{2^\star}
       +B_2^{\rm UV}~\delta m_{2'}^{\rm UV}~\delta m_{2^\star}^{\rm UV}~M_{2^\star}
  \\ &    +B_2^{\rm UV}~B_{2'}^{\rm UV}~(\delta m_{2'}^{\rm UV}~M_{2^\star}+B_{2'}^{\rm UV}~M_2)
  \\ &    +M_{2^\star}~[ \Delta \delta m_{6b}+L_2^{\rm R}~\{ \delta m_{4b}^{\rm R}
         -(\delta m_2~\delta m_{2^\star}^{\rm R}+B_2^{\rm UV}~\delta m_{2'}^{\rm R}) \} ]
\\ &    
  +L_2^{\rm R}~(\Delta M_{6b}
        +M_{2^\star}~\Delta \delta m_{4b}
         +L_2^{\rm R}~\Delta M_{4b}+L_{4l}^{\rm R}~M_2)
      +M_{4b(2^\star)}^{\rm R}~\Delta \delta m_{4b}
\\&      +L_{4l}^{\rm R}~(\Delta M_{4b}+L_2^{\rm R}~M_2)
      +L_{6b3}^{\rm R}~M_2
    -\Delta \delta m_{4b}~\delta m_{2^\star}^{\rm R}~M_{2^\star}
      -L_2^{\rm R}~\Delta \delta m_{4b}~M_{2^\star}
\\ &      -(L_2^{\rm R})^2~(\Delta M_{4b}+L_2^{\rm R}~M_2)
      -2~L_{4l}^{\rm R}~L_2^{\rm R}~M_2
      +(L_2^{\rm R})^3~M_2
\end{align*}

\section{Divergence Structure of  the  renormalization
constants}
\label{appendixRconst}

\subsection{Second-order renormalization constants}
\vspace{-1cm}
\begin{align*}
&L_2 = L_2^{\rm UV} + \widetilde{L}_2
~,~~~~~~L_2^{\rm R}=\widetilde{L}_2 =I_2
\\
&B_2 = B_2^{\rm UV} + \widetilde{B}_2
~,~~~~~~B_2^{\rm R} = \widetilde{B}_2 = - I_2 + \Delta B_2
\\
&L_2^{\rm R}+B_2^{\rm R} = \Delta B_2
\\
&B_{2^\star} = - 2~L_{2^\star}~,~~~~~~L_{2^\star}=I_{2^\star}+\Delta L_{2^\star}
\\
&B_{2^{\star \star} } = -2 ~( 2~L_{2^{\star \star \dag}} + L_{2^{\star \dag \star}} )
\\
&\delta m_{2^\star} = \delta m_{2^\star}^{\rm UV} + I_2 + \Delta \delta m_{2^\star}
\end{align*}

\subsection{Fourth-order renormalization constants}
\label{appendixRconstsFourthorder}
\vspace{-1cm}
\begin{align*}  
&L_{4x}=L_{4x}^{\rm UV}+I_{4x}+\Delta L_{4x}
\\
&L_{4c}=L_{4c}^{\rm UV}+I_{4c}+\Delta L_{4c}+L_2^{\rm UV}~\widetilde{L}_2
\\
&B_{4a}=B_{4a}^{\rm UV}-I_{4x}+\Delta B_{4a}
                +2~L_2^{\rm UV}~\widetilde{B}_2-2~I_{4c}
\end{align*}
\vspace{-0.9cm}
\begin{align*}  
&L_{4l}=L_{4l}^{\rm UV}+I_{4l}+(L_2^{\rm R})^2
                +\Delta L_{4l}+\widetilde{L}_2~L_2^{\rm UV}
\\
&L_{4s}=L_{4s}^{\rm UV}+I_{4s}+\Delta L_{4s}
                +\delta m_2~L_{2^\star}+B_2^{\rm UV}~\widetilde{L}_{2'}
\\
&B_{4b}=B_{4b}^{\rm UV}+\Delta B_{4b}
                +\delta m_2~B_{2^\star}+B_2^{\rm UV}~\widetilde{B}_{2'}
                +L_2^{\rm R}~\widetilde{B}_2-2~I_{4s}-I_{4l}
\end{align*}
\vspace{-0.9cm}
\begin{align*}
&\Delta LB_{4a} = 2~L_{4c}^{\rm R} + L_{4x}^{\rm R} + B_{4a}^{\rm R}
                 =  2~\Delta L_{4c} + \Delta L_{4x} + \Delta B_{4a}
\\
&\Delta LB_{4b}  = 2~L_{4s}^{\rm R} + L_{4l}^{\rm R} + B_{4a}^{\rm R} - L_2^{\rm R}~\Delta B_2
                = 2~\Delta L_{4s} + \Delta L_{4l} + \Delta B_{4a}
\end{align*}
\vspace{-0.9cm}
\begin{align*}
\Delta LB^{(4)} = \Delta LB_{4a} + \Delta LB_{4b} = \Delta L^{(4)} + \Delta B^{(4)}
\end{align*}

\subsection{Sixth-order renormalization constants}
\vspace{-1cm}
\label{appendixRconstsSixthorder}

\begin{align*} 
L_{6a1}&=L_{6a1}^{\rm R}+2~\delta m_2~L_{4s(1^\star)}+2~B_2^{\rm UV}~\widetilde{L}_{4s(1')}
       -\delta m_2~(\delta m_2~L_{2^{\star \star \dag}}+B_2^{\rm UV}~L_{2'^\star})
 \\ &  -B_2^{\rm UV}~(\delta m_2~L_{2'^\star}+B_2^{\rm UV}~\widetilde{L}_{2''})+L_{6a1}^{\rm UV}
\end{align*}
\vspace{-0.9cm}
\begin{align*} L_{6a2}&=L_{6a2}^{\rm R}+L_2^{\rm UV}~\widetilde{L}_{4s}+\delta m_2~L_{4l(1^\star)}+B_2^{\rm UV}~\widetilde L_{4l(1')}
       -L_2^{\rm UV}~(\delta m_2~L_{2^\star}+B_2^{\rm UV}~\widetilde{L}_{2'})+L_{6a2}^{\rm UV}
\end{align*}
\vspace{-0.9cm}
\begin{align*} L_{6a3}&=L_{6a3}^{\rm R}+2~(\delta m_2~L_{4s(1^\star)}
         +B_2^{\rm UV}~\widetilde{L}_{4s(1')})
       -\delta m_2~(\delta m_2~L_{2^{\star \dag \star}}+B_2^{\rm UV}~L_{2'^\star})
    \\ &   -B_2^{\rm UV}~(\delta m_2~L_{2'^\star}+B_2^{\rm UV}~\widetilde{L}_{2''})+L_{6a3}^{\rm UV}
\end{align*}    
\vspace{-0.9cm}
\begin{align*} L_{6b1}&=L_{6b1}^{\rm R}+\delta m_2~L_{4s(2^\star)}
             +B_2^{\rm UV}~\widetilde{L}_{4s(2')}
              +\delta m_{4b}^{\rm UV}~L_{2^\star}+B_{4b}^{\rm UV}~\widetilde{L}_{2'}
        -\delta m_2~\delta m_{2^\star}^{\rm UV}~L_{2^\star}
  \\ &  -B_2^{\rm UV}~(\delta m_{2'}^{\rm UV}~L_{2^\star}
                +B_{2'}^{\rm UV}~\widetilde{L}_{2'})+L_{6b1}^{\rm UV}
\end{align*}
\vspace{-0.9cm}
\begin{align*} L_{6b2}&=L_{6b2}^{\rm R}+\delta m_2~L_{4l(2^\star)}
          +B_2^{\rm UV}~\widetilde{L}_{4l(2')}+L_{4s}^{\rm UV}~\widetilde{L}_2
       -B_2^{\rm UV}~L_{2'}^{\rm UV}~\widetilde{L}_2+L_{6b2}^{\rm UV}
\end{align*}
\vspace{-0.9cm}
\begin{align*} L_{6b3}&=L_{6b3}^{\rm R}+L_2^{\rm UV}~\widetilde{L}_{4l}+L_{4l}^{\rm UV}
        ~\widetilde{L}_2-(L_2^{\rm UV})^2~\widetilde{L}_2+ L_{6b3}^{\rm UV}
\end{align*}
\vspace{-0.9cm}
\begin{align*} L_{6c1}&=L_{6c1}^{\rm R}+2~L_2^{\rm UV}~\widetilde{L}_{4s}
    +\delta m_{4a}^{\rm UV}~L_{2^\star}+B_{4a}^{\rm UV}~\widetilde{L}_{2'}
    -2~L_2^{\rm UV}~(\delta m_2~L_{2^\star}+B_2^{\rm UV}~\widetilde{L}_{2'})+L_{6c1}^{\rm UV}
\end{align*}
\vspace{-0.9cm}
\begin{align*} L_{6c2}&=L_{6c2}^{\rm R}+L_2^{\rm UV}~\widetilde{L}_{4l}
      +L_{4c}^{\rm UV}~\widetilde{L}_2-(L_2^{\rm UV})^2~\widetilde{L}_2 +L_{6c2}^{\rm UV}
\end{align*}
\vspace{-0.9cm}
\begin{align*} L_{6c3}&=L_{6c3}^{\rm R}+L_{4x}^{\rm UV}~\widetilde{L}_2+L_{6c3}^{\rm UV}
\end{align*}
\vspace{-0.9cm}
\begin{align*} L_{6d1}&=L_{6d1}^{\rm R}+\delta m_2~L_{4c(1^\star)}
         +B_2^{\rm UV}~\widetilde{L}_{4c(1')}+L_2^{\rm UV}~\widetilde{L}_{4s}
       -L_2^{\rm UV}~(\delta m_2~L_{2^\star}+B_2^{\rm UV}~\widetilde{L}_{2'})+L_{6d1}^{\rm UV}
\end{align*}
\vspace{-0.9cm}
\begin{align*} L_{6d2}&=L_{6d2}^{\rm R}+L_2^{\rm UV}~\widetilde{L}_{4c}
   +L_2^{\rm UV}~\widetilde{L}_{4l}-(L_2^{\rm UV})^2~\widetilde{L}_2+L_{6d2}^{\rm UV}
\end{align*}
\vspace{-0.9cm}
\begin{align*} L_{6d3}&=L_{6d3}^{\rm R}+\delta m_2~L_{4c(1^\star)}
    +B_2^{\rm UV}~\widetilde{L}_{4c(1')}+L_2^{\rm UV}~\widetilde{L}_{4s}
       -L_2^{\rm UV}~(\delta m_2~L_{2^\star}+B_2^{\rm UV}~\widetilde{L}_{2'})+L_{6d3}^{\rm UV}
\end{align*}
\vspace{-0.9cm}
\begin{align*} L_{6d4}&=L_{6d4}^{\rm R}+\delta m_2~L_{4x(1^\star)}
+B_2^{\rm UV}~\widetilde{L}_{4x(1')}+L_{6d4}^{\rm UV}
\end{align*}
\vspace{-0.9cm}
\begin{align*} L_{6d5}&=L_{6d5}^{\rm R}
+\delta m_2~L_{4c(3^\star)}+B_2^{\rm UV}~\widetilde{L}_{4c(3')}
+L_{4s}^{\rm UV}~\widetilde{L}_2
       -B_2^{\rm UV}~L_{2'}^{\rm UV}~\widetilde{L}_2+L_{6d5}^{\rm UV}
\end{align*}
\vspace{-0.9cm}
\begin{align*} L_{6e1}&=L_{6e1}^{\rm R}+\delta m_2~L_{4c(2^\star)}
      +B_2^{\rm UV}~\widetilde{L}_{4c(2')}+L_{4s}^{\rm UV}~\widetilde{L}_2
       -B_2^{\rm UV}~L_{2'}^{\rm UV}~\widetilde{L}_2+L_{6e1}^{\rm UV}
\end{align*}
\vspace{-0.9cm}
\begin{align*} L_{6e2}&=L_{6e2}^{\rm R}+\delta m_2~L_{4x(2^\star)}
                        +B_2^{\rm UV}~\widetilde{L}_{4x(2')}+L_{6e2}^{\rm UV}
\end{align*}
\vspace{-0.9cm}
\begin{align*} L_{6e3}&=L_{6e3}^{\rm R}+L_2^{\rm UV}~\widetilde{L}_{4x}+L_{6e3}^{\rm UV}
\end{align*}
\vspace{-0.9cm}
\begin{align*} L_{6f1}&=L_{6f1}^{\rm R}+L_2^{\rm UV}~\widetilde{L}_{4c}
      +L_{4c}^{\rm UV}~\widetilde{L}_2-(L_2^{\rm UV})^2~\widetilde{L}_2+L_{6f1}^{\rm UV}
\end{align*}
\vspace{-0.9cm}
\begin{align*} L_{6f2}&=L_{6f2}^{\rm R}+L_2^{\rm UV}~\widetilde{L}_{4x}+L_{6f2}^{\rm UV}
\end{align*}
\vspace{-0.9cm}
\begin{align*} L_{6f3}&=L_{6f3}^{\rm R}+2~L_2^{\rm UV}~\widetilde{L}_{4c}
                  -(L_2^{\rm UV})^2~\widetilde{L}_2+L_{6f3}^{\rm UV}
\end{align*}
\vspace{-0.9cm}
\begin{align*} L_{6g1}&=L_{6g1}^{\rm R}+L_2^{\rm UV}~\widetilde{L}_{4c}
             +L_{4c}^{\rm UV}~\widetilde{L}_2-(L_2^{\rm UV})^2~\widetilde{L}_2+L_{6g1}^{\rm UV}
\end{align*}
\vspace{-0.9cm}
\begin{align*} L_{6g2}&=L_{6g2}^{\rm R}+L_{6g2}^{\rm UV}
\end{align*}
\vspace{-0.9cm}
\begin{align*} L_{6g3}&=L_{6g3}^{\rm R}+L_{6g3}^{\rm UV}
\end{align*}
\vspace{-0.9cm}
\begin{align*} L_{6g4}&=L_{6g4}^{\rm R}+L_2^{\rm UV}~\widetilde{L}_{4x}+L_{6g4}^{\rm UV}
\end{align*}
\vspace{-0.9cm}
\begin{align*} L_{6g5}&=L_{6g5}^{\rm R}+L_2^{\rm UV}~\widetilde{L}_{4c}
              +L_{4l}^{\rm UV}~\widetilde{L}_2-(L_2^{\rm UV})^2~\widetilde{L}_2+L_{6g5}^{\rm UV}
\end{align*}
\vspace{-0.9cm}
\begin{align*} L_{6h1}&=L_{6h1}^{\rm R}+L_{4x}^{\rm UV}~\widetilde{L}_2+L_{6h1}^{\rm UV}
\end{align*}
\vspace{-0.9cm}
\begin{align*} L_{6h2}&=L_{6h2}^{\rm R}+L_{6h2}^{\rm UV}
\end{align*}
\vspace{-0.9cm}
\begin{align*} L_{6h3}&=L_{6h3}^{\rm R}+L_{6h3}^{\rm UV}
\end{align*}

\begin{align*} 
  B_{6a}&=B_{6a}^{\rm R}+2~(\delta m_2~B_{4b(1^\star)}
       +B_2^{\rm UV}~\widetilde{B}_{4b(1')})
     -\delta m_2~(\delta m_2~B_{2^{\star \star}}+B_2^{\rm UV}~B_{2'^{\star}})
 \\ & -B_2^{\rm UV}~(\delta m_2~B_{2'^{\star}}+B_2^{\rm UV}~\widetilde{B}_{2''})+B_{6a}^{\rm UV}\end{align*}
\vspace{-0.9cm}
\begin{align*} B_{6b}&=B_{6b}^{\rm R}+\delta m_2~B_{4b(2^\star)}
        +B_2^{\rm UV}~\widetilde{B}_{4b(2')}
      +\delta m_{4b}^{\rm UV}~B_{2^\star}+B_{4b}^{\rm UV}~\widetilde{B}_{2'}
           -\delta m_2~\delta m_{2^\star}^{\rm UV}~B_{2^\star}
  \\ &   -B_2^{\rm UV}~(\delta m_{2'}^{\rm UV}~B_{2^\star}+B_{2'}^{\rm UV}~\widetilde{B}_{2'})
      +B_{6b}^{\rm UV}
\end{align*}
\vspace{-0.9cm}
\begin{align*} B_{6c}=B_{6c}^{\rm R}+2~L_2^{\rm UV}~\widetilde{B}_{4b}
      +\delta m_{4a}^{\rm UV}~B_{2^\star}+B_{4a}^{\rm UV}~\widetilde{B}_{2'}
       -2~L_2^{\rm UV}~(\delta m_2~B_{2^\star}+B_2^{\rm UV}~\widetilde{B}_{2'})
    +B_{6c}^{\rm UV}
\end{align*}
\vspace{-0.9cm}
\begin{align*} B_{6d}&=B_{6d}^{\rm R}+\delta m_2~B_{4a(1^\star)}
           +B_2^{\rm UV}~\widetilde{B}_{4a(1')}
       +L_2^{\rm UV}~\widetilde{B}_{4b}+L_{4s}^{\rm UV}~\widetilde{B}_2
      -L_2^{\rm UV}~(\delta m_2~B_{2^\star}+B_2^{\rm UV}~\widetilde{B}_{2'})
   \\ &    -B_2^{\rm UV}~L_{2'}^{\rm UV}~\widetilde{B}_2+B_{6d}^{\rm UV}
\end{align*}
\vspace{-0.9cm}
\begin{align*} B_{6e}=B_{6e}^{\rm R}+\delta m_2~B_{4a(2^\star)}
         +B_2^{\rm UV}~\widetilde{B}_{4a(2')}
      +2~L_{4s}^{\rm UV}~\widetilde{B}_2-2~B_2^{\rm UV}~L_{2'}^{\rm UV}~\widetilde{B}_2
     +B_{6e}^{\rm UV}
\end{align*}
\vspace{-0.9cm}
\begin{align*} B_{6f}=B_{6f}^{\rm R}+2~L_2^{\rm UV}~\widetilde{B}_{4a}
       +2~L_{4c}^{\rm UV}~\widetilde{B}_2
       -3~(L_2^{\rm UV})^2~\widetilde{B}_2+B_{6f}^{\rm UV}
\end{align*}
\vspace{-0.9cm}
\begin{align*} B_{6g}=B_{6g}^{\rm R}+L_2^{\rm UV}~\widetilde{B}_{4a}
       +L_{4c}^{\rm UV}~\widetilde{B}_2
      +L_{4l}^{\rm UV}~\widetilde{B}_2
       -2~(L_2^{\rm UV})^2~\widetilde{B}_2+B_{6g}^{\rm UV}
\end{align*}
\vspace{-0.9cm}
\begin{align*} B_{6h}=B_{6h}^{\rm R}+2~L_{4x}^{\rm UV}~\widetilde{B}_2+B_{6h}^{\rm UV}
\end{align*}

The residual renormalization constants $\Delta LB_{6x}$ for each diagram 
appearing in  Eq.~(\ref{LB_6}) are defined in the following equations. 

\begin{align*} \Delta LB_{6a}&=2~L_{6a1}^{\rm R}+2~L_{6a2}^{\rm R}+L_{6a3}^{\rm R}+B^{\rm R}_{6a}
                        \\&      -2~L_{4s}^{\rm R}~\Delta B_2
\end{align*}
\vspace{-0.9cm}
\begin{align*} \Delta LB_{6b}&=2~L_{6b1}^{\rm R}+2~L_{6b2}^{\rm R}+L_{6b3}^{\rm R}+B^{\rm R}_{6b}
                        \\&     -L_2^{\rm R}~(B_{4b}^{\rm R}+2~L_{4s}^{\rm R}+L_{4l}^{\rm R})
                              -L_{4l}^{\rm R}~\Delta B_2+(L_2^{\rm R})^2~\Delta B_2
\end{align*}
\vspace{-0.9cm}
\begin{align*} \Delta LB_{6c}&=2~L_{6c1}^{\rm R}+2~L_{6c2}^{\rm R}+L_{6c3}^{\rm R}+B^{\rm R}_{6c}
                      \\&        -L_2^{\rm R}~(B_{4a}^{\rm R}+2~L_{4c}^{\rm R}+L_{4x}^{\rm R})
\end{align*}
\vspace{-0.9cm}
\begin{align*} \Delta LB_{6d}&=L_{6d1}^{\rm R}+L_{6d2}^{\rm R}+L_{6d3}^{\rm R}
                          +L_{6d4}^{\rm R}+L_{6d5}^{\rm R}+B^{\rm R}_{6d}
                     \\ &         -L_{4c}^{\rm R}~\Delta B_2
\end{align*}
\vspace{-0.9cm}
\begin{align*} \Delta LB_{6e}&=2~L_{6e1}^{\rm R}+2~L_{6e2}^{\rm R}+L_{6e3}^{\rm R}+B^{\rm R}_{6e}
                      \\ &        -L_{4x}^{\rm R}~\Delta B_2
\end{align*}
\vspace{-0.9cm}
\begin{align*} \Delta LB_{6f}&=2~L_{6f1}^{\rm R}+2~L_{6f2}^{\rm R}+L_{6f3}^{\rm R}+B^{\rm R}_{6f}
\end{align*}
\vspace{-0.9cm}
\begin{align*} \Delta LB_{6g}=L_{6g1}^{\rm R}+L_{6g2}^{\rm R}+L_{6g3}^{\rm R}+L_{6g4}^{\rm R}
               +L_{6g5}^{\rm R}+B^{\rm R}_{6g}
\end{align*}
\vspace{-0.9cm}
\begin{align*} \Delta LB_{6h}=2~L_{6h1}^{\rm R}+2~L_{6h2}^{\rm R}+L_{6h3}^{\rm R}+B^{\rm R}_{6h}
\end{align*}
\vspace{-0.9cm}
\begin{align*}
\Delta LB^{(6)} = \sum_{x=a}^{h} \eta_x \Delta LB_{6x}
            = \Delta L^{(6)} + \Delta B^{(6)} 
                 +\Delta L^{(4)}~ \Delta B_2 + \Delta \delta m^{(4)} ~B_{2^\star}[I] 
\end{align*}
